\documentclass{pr-imfp03-new}

\begin{document}

\begin{flushright}
KEK-TH-928\\
December 2003\\
\end{flushright}

\vspace*{1.5cm}
\begin{center}
    {\baselineskip 25pt
    \Large{\bf

CKM Phenomenology and $B$-Meson Physics -\\
 Present Status and Current Issues

    }
    }

\vspace{1.2cm}
\centerline{\bf Ahmed Ali}

\vspace{.5cm}
\small{\it Theory Group, High Energy Accelerator Research Organization (KEK),
 Tsukuba, 305 -0801, Japan
\footnote{On leave of absence from Deutsches Elektronen-Synchrotron DESY,
 D-22603 Hamburg, FRG.
\\
E-mail: ahmed@post.kek.jp}}

    \vspace{.5cm}
    \today

    \vspace{1.5cm}
    {\bf Abstract}

\end{center}

\bigskip

We review the status of the
Cabibbo-Kobayashi-Maskawa (CKM) matrix elements and the CP-violating
phases in the CKM-unitarity triangle. The emphasis in these
lecture notes is on $B$-meson physics, 
though we also review the current status and
issues in the light quark sector of this matrix. Selected applications
of theoretical methods in QCD used in the interpretation of data are given
and some of the issues restricting the theoretical precision on the CKM
matrix elements discussed. The overall consistency of the 
CKM theory with the available
data in flavour physics is impressive and we quantify this consistency.
Current data also show some anomalies which, however, are not yet 
statistically significant.
They are discussed briefly. Some benchmark measurements that remain to be done
in experiments at the $B$-factories and hadron colliders are listed.
Together with the already achieved results, they will provide unprecedented
tests of the CKM theory and by the same token may lead to the discovery 
of new physics.

\vspace{3cm}
\begin{center}

To appear in the Proceedings of the International Meeting on Fundamental Physics,\\ Soto de 
Cangas (Asturias), Spain, February 23 - 28, 2003; Publishers: CIEMAT
Editorial Service (Madrid, Spain); J.~Cuevas and A.~Ruiz. (Eds.)

\end{center}

\newpage

\title{}

\newread\epsffilein 
\newif\ifepsffileok 
\newif\ifepsfbbfound 
\newif\ifepsfverbose 
\newdimen\epsfxsize 
\newdimen\epsfysize 
\newdimen\epsftsize 
\newdimen\epsfrsize 
\newdimen\epsftmp 
\newdimen\pspoints 
\pspoints=1bp 
\epsfxsize=0pt 
\epsfysize=0pt 
\def\epsfbox#1{\global\def\epsfllx{72}\global\def\epsflly{72}%
 \global\def\epsfurx{540}\global\def\epsfury{720}%
 \def\lbracket{[}\def\testit{#1}\ifx\testit\lbracket
 \let\next=\epsfgetlitbb\else\let\next=\epsfnormal\fi\next{#1}}%
\def\epsfgetlitbb#1#2 #3 #4 #5]#6{\epsfgrab #2 #3 #4 #5 .\\%
 \epsfsetgraph{#6}}%
\def\epsfnormal#1{\epsfgetbb{#1}\epsfsetgraph{#1}}%
\def\epsfgetbb#1{%
%
%
\openin\epsffilein=#1 
\ifeof\epsffilein\errmessage{I couldn't open #1, will ignore it}\else
%
%
 {\epsffileoktrue \chardef\other=12
 \def\do##1{\catcode`##1=\other}\dospecials \catcode`\ =10  
 \loop
 \read\epsffilein to \epsffileline
 \ifeof\epsffilein\epsffileokfalse\else
%
%
 \expandafter\epsfaux\epsffileline:. \\%
 \fi
 \ifepsffileok\repeat
 \ifepsfbbfound\else
 \ifepsfverbose\message{No bounding box comment in #1; using defaults}\fi\fi
 }\closein\epsffilein\fi}%
%
%
\def\epsfclipstring{}
\def\epsfclipon{\def\epsfclipstring{ clip}}%
\def\epsfclipoff{\def\epsfclipstring{}}%
\def\epsfsetgraph#1{%
 \epsfrsize=\epsfury\pspoints
 \advance\epsfrsize by-\epsflly\pspoints
 \epsftsize=\epsfurx\pspoints
 \advance\epsftsize by-\epsfllx\pspoints
%
%
 \epsfxsize\epsfsize\epsftsize\epsfrsize
 \ifnum\epsfxsize=0 \ifnum\epsfysize=0
 \epsfxsize=\epsftsize \epsfysize=\epsfrsize
 \epsfrsize=0pt
%
%
%
 \else\epsftmp=\epsftsize \divide\epsftmp\epsfrsize
 \epsfxsize=\epsfysize \multiply\epsfxsize\epsftmp
 \multiply\epsftmp\epsfrsize \advance\epsftsize-\epsftmp
 \epsftmp=\epsfysize
 \loop \advance\epsftsize\epsftsize \divide\epsftmp 2
 \ifnum\epsftmp>0
 \ifnum\epsftsize<\epsfrsize\else
 \advance\epsftsize-\epsfrsize \advance\epsfxsize\epsftmp \fi
 \repeat
 \epsfrsize=0pt
 \fi
 \else \ifnum\epsfysize=0
 \epsftmp=\epsfrsize \divide\epsftmp\epsftsize
 \epsfysize=\epsfxsize \multiply\epsfysize\epsftmp
 \multiply\epsftmp\epsftsize \advance\epsfrsize-\epsftmp
 \epsftmp=\epsfxsize
 \loop \advance\epsfrsize\epsfrsize \divide\epsftmp 2
 \ifnum\epsftmp>0
 \ifnum\epsfrsize<\epsftsize\else
 \advance\epsfrsize-\epsftsize \advance\epsfysize\epsftmp \fi
 \repeat
 \epsfrsize=0pt
 \else
 \epsfrsize=\epsfysize
 \fi
 \fi
%
%
 \ifepsfverbose\message{#1: width=\the\epsfxsize, height=\the\epsfysize}\fi
 \epsftmp=10\epsfxsize \divide\epsftmp\pspoints   
 \vbox to\epsfysize{\vfil\hbox to\epsfxsize{%
 \ifnum\epsfrsize=0\relax
 \includegraphics{#1}%
 \else
 \epsfrsize=10\epsfysize \divide\epsfrsize\pspoints
 \includegraphics{#1}%
 \fi
 \hfil}}%
\global\epsfxsize=0pt\global\epsfysize=0pt}%
%
%
 {\catcode`\%=12 \global\let\epsfpercent=
%
%
\long\def\epsfaux#1#2:#3\\{\ifx#1\epsfpercent
 \def\testit{#2}\ifx\testit\epsfbblit
 \epsfgrab #3 . . . \\%
 \epsffileokfalse
 \global\epsfbbfoundtrue
 \fi\else\ifx#1\par\else\epsffileokfalse\fi\fi}%
%
%
\def\epsfempty{}%
\def\epsfgrab #1 #2 #3 #4 #5\\{%
\global\def\epsfllx{#1}\ifx\epsfllx\epsfempty
 \epsfgrab #2 #3 #4 #5 .\\\else
 \global\def\epsflly{#2}%
 \global\def\epsfurx{#3}\global\def\epsfury{#4}\fi}%
%
%
\def\epsfsize#1#2{\epsfxsize}
%
%
\let\epsffile=\epsfbox
\def\sss{\scriptscriptstyle}
\def\barp{{\raise.35ex\hbox{${\sss (}$}}---{\raise.35ex\hbox{${\sss )}$}}}
\def\bdbarp{\hbox{$B_d$\kern-1.4em\raise1.4ex\hbox{\barp}}}
\def\bsbarp{\hbox{$B_s$\kern-1.4em\raise1.4ex\hbox{\barp}}}
\def\dbarp{\hbox{$D$\kern-1.1em\raise1.4ex\hbox{\barp}}}
\newcommand{\beq}{\begin{equation}}
\newcommand{\eeq}{\end{equation}}
\newcommand{\beqa}{\begin{eqnarray}}
\newcommand{\eeqa}{\end{eqnarray}}
\def\g{\gamma}
\def\ra{\rightarrow}
\def\etal{{\it et al.}}
\newcommand{\siz}{\mbox{$\sin^2\hat{\theta}_W(M_Z)\ $}}
\newcommand{\msb}{\mbox{$\overline{\rm{MS}}\ $}}
\def \branch{{\cal B}}
\newcommand{\mt}{\mbox{$m_t$}}
\newcommand{\mh}{\mbox{$M_H$}}
\newcommand{\mz}{\mbox{$M_Z$}}   
\newcommand{\mw}{\mbox{$M_W$}}
\newcommand{\alsz}{\mbox{$\alpha_s(M_Z)$}}
\newcommand{\als}{\mbox{$\alpha_s$}}
\newcommand{\con}[3]{{\bf #1} (19#3) #2}
\newcommand{\xpr}{{x^\prime}}    
\def\xbj          {x_{\rm Bj}}
\def\bsll{$b \rightarrow s \ell^+ \ell^- $ }
\def\bxsll{$B \rightarrow X_s \ell^+ \ell^- $ }
\def\bxsee{B \rightarrow X_s e^+ e^-  }
\def\bxsmm{B \rightarrow X_s \mu^+ \mu^-  }
\def\bxstt{B \rightarrow X_s \tau^+ \tau^- }
\def\bsee{$b \rightarrow s e^+ e^- $ }
\def\bxsg{$B \rightarrow X_s \gamma $ }
\def\s{\hat{s}}
\def\u{\hat{u}}
\def\z{v \cdot \hat{q}}
\def\dilep{l^+ l^-}
\def\dilpp{l^+ l^+}
\def\dilmm{l^- l^-}
\def\epsm{\epsilon_\mu}
\def\epsn{\epsilon_\nu}
\def\be{\begin{equation}}
\def\ee{\end{equation}}
\def\g{\gamma}
\newcommand{\BR}{{\cal B}}
\newcommand{\M}{{\cal M}}
\newcommand{\xg}{x}
\def\mt{m_t}
\def\mb{m_b}
\def\mc{m_c}
\def\bra{\langle}
\def\ket{\rangle}
\def\bea{\begin{eqnarray}}
\def\eea{\end{eqnarray}}
\def\be{\begin{equation}}
\def\ee{\end{equation}}
\def\a{\alpha}
\def\b{\beta}
\def\g{\gamma}
\def\d{\delta}
\def\e{\epsilon}
\def\p{\pi}
\def\ve{\varepsilon}
\def\ep{\varepsilon}
\def\et{\eta}
\def\l{\lambda}
\def\n{\nu}
\def\G{\Gamma}
\def\D{\Delta}
\def\L{\Lambda}
\def\as{\alpha _s}
\def\alfas{\alpha _s}
\def\etal{et al.}
\def\mt{m_t}  
\def\ml{m_\ell}
\def\Mw{M_W}
\def\dBR{{d^2 {\cal B}} \over {d {\hat s} dz}}
\def\Leff{L_{eff}}  
\def\bdull{ B_{d,u} \rightarrow  X_s + l^+ l^-}
\def\bduee{ B_{d,u} \rightarrow  X_s + e^+ e^-}
\def\bdumm{ B_{d,u} \rightarrow  X_s + \mu^+ \mu^-}
\newcommand{\bgamaxs}{$B \to X _{s} + \gamma$}
\newcommand{\brogam}{\ $B \to \rho+ \gamma$}
\newcommand{\bdgam}{\ $b \to d+ \gamma$}
\newcommand{\bsg}{\ $b \to s+ g$}
\newcommand{\bsdgam}{\ $b \to (s,d)+ \gamma$}
\newcommand{\bsdell}
   {$b \to (s,d)+ ~\ell \bar{\ell}$ ($\ell = e, \mu, \tau, \nu$)}
\newcommand{\bksell} 
   {$B \to K^*+ ~\ell \bar{\ell}$ ($\ell = e, \mu, \tau, \nu$)}
\newcommand{\ope}{operator product expansion}
\newcommand{\bsdnu}
   {$b \to (s,d)+ ~\nu \bar{\nu}$}
\newcommand{\broell}
   {$B \to \rho+ ~\ell \bar{\ell}$ ($\ell = e, \mu, \tau, \nu$)}
\newcommand{\bkell}
  {$B \to K+ ~\ell \bar{\ell}$ ($\ell = e, \mu, \tau, \nu$)}
\newcommand{\bpiell}
 {$B \to \pi+ ~\ell \bar{\ell}$ ($\ell = e, \mu, \tau, \nu$)}
\newcommand{\BGAMAXS}{B \ra X _{s} + \gamma}
\newcommand{\BGAMAXD}{B \ra X _{d} + \gamma}
\newcommand{\BBGAMAXS}{{\cal B}(B \ra  X _{s} + \gamma)}
\newcommand{\BBGAMAXD}{{\cal B}(B \ra  X _{d} + \gamma)}
\newcommand{\BBGAMARHO}{{\cal B}(B \ra  \rho + \gamma)}
\newcommand{\BBGAMAKSTAR}{{\cal B}(B \ra  K^{\star} + \gamma)}   
\newcommand{\BGAMARHO}{B \ra  \rho + \gamma}
\newcommand{\BGAMAKSTAR}{B \ra  K^{\star} + \gamma}
\newcommand{\GGAMAXD}{\Gamma(B \ra  X _{d} + \gamma)}
\newcommand{\BGAMAS}{b \ra s + \gamma}
\newcommand{\BGAMAD}{b \ra d + \gamma}
\newcommand{\BBGAMAS}{{\cal B}(b \ra s + \gamma)}
\newcommand{\BBKSTAR}{{\cal B}(B \ra K^\star + \gamma)}
\newcommand{\BKSTAR}{B \ra K^\star + \gamma}
\newcommand{\BBGAMAD}{{\cal B}(b \ra d + \gamma)}
\newcommand{\BGAMAGS}{ b \ra s  + g + \gamma}
\newcommand{\BGAMAGD}{ b \ra d  + g + \gamma}
\newcommand{\GGAMAXS}{\Gamma (B \ra  X _{s} + \gamma)}
\def\Vcd{V_{cd}}
\def\beq{\begin{equation}}
\def\eeq{\end{equation}}
\def\Vcdabs{\vert V_{cd} \vert}
\def\Vus{V_{us}}
\def\Vusabs{\vert V_{us} \vert}
\def\Vcs{V_{cs}}
\def\Vcsabs{\vert V_{cs} \vert}
\def\Vud{V_{ud}}
\def\Vudabs{\vert V_{ud} \vert}
\def\Vbc{V_{cb}}
\def\Vbcabs{\vert V_{cb} \vert}
\def\Vcbabs{\vert V_{cb} \vert}
\def\Vbu{V_{ub}}
\def\Vbuabs{\vert V_{ub}\vert}
\def\Vubabs{\vert V_{ub}\vert}
\def\Vtd{V_{td}}
\def\Vtdabs{\vert V_{td} \vert}
\def\Vts{V_{ts}}
\def\Vtsabs{\vert V_{ts} \vert}
\def\Vtb{V_{tb}}
\def\Vtbabs{\vert V_{tb}\vert}
\newcommand{\bd}{B_d^0}
\newcommand{\bdb}{\overline{B_d^0}}
\newcommand{\abseps}{\vert\epsilon\vert}
\newcommand{\bsdem}
   {$b \to (s,d)+ ~\ell \bar{\ell}$ ($\ell = e, \mu$)}
\def\MAT#1#2#3{<{#1}\vert {#2}\vert {#3}>}
\def\BZ{B_d^0}  
\def\BZB{\bar{B_d^0}}
\def\BS{B_s^0}
\def\BSB{\bar{B_s^0}}
\def\BP{B_u^+}
\def\BM{B_u^-}  
\def\UPS{\Upsilon}
\def\US{\Upsilon(1\hbox{S})}
\def\USS{\Upsilon(2\hbox{S})} 
\def\USSS{\Upsilon(3\hbox{S})}
\def\USSSS{\Upsilon(4\hbox{S})}
\def\pppi{p\bar{p}\pi^-}
\def\pppipi{p\bar{p}\pi^+\pi^-}
\newcommand{\fbb}{f^2_{B_d}B_{B_d}}
\newcommand{\fbbs}{f^2_{B_s}B_{B_s}}
\newcommand{\fbd}{f_{B_d}}
\newcommand{\fbs}{f_{B_s}}
\newcommand{\go}[1]{\gamma^{#1}}
\newcommand{\gu}[1]{\gamma_{#1}}
\newcommand{\xeta}{\eta \hspace{-6pt} / }
\newcommand{\xeps}{\epsilon \hspace{-5pt} / }
\newcommand{\xr}{r \hspace{-5pt} / }
\newcommand{\delmd}{\Delta M_{B_d}}
\newcommand{\delms}{\Delta M_{B_s}}
\newcommand{\ps}{10^{-12} s}
\def\lp{l^+}
\def\lm{l^-}
\def\sw{\sin{^2}\theta _{W}}   
\def\sh{\hat s}
\def\mh{\hat m}
\def\qbar{\overline q}
\def\sLbar{\overline s_L} 
\def\ubar{\overline u}
\def\sbar{\overline s}
\def\dbar{\overline d}
\def\bbar{\overline b}
\def\cbar{\overline c}
\def\bLbar{\overline b_L}
\newcommand{\bu}{B_u^\pm}
\def\qq{\qbar{{\lambda_a}\over 2} q}
\def\q5q{\qbar{{\lambda_a}\over 2} i\gamma_5 q}
\def\eLbar{\overline l_L}
\def\eRbar{\overline l_R}
\def\nulbar{\overline {\nu_\ell}}
\def\nul{\nu_\ell}
\def\gmu{\gamma_\mu}
\def\gmuu{\gamma^\mu}
\newcommand{\bsgam}{\ $b \to s+ \gamma$}
\newcommand{\bsdg}{\ $b \to (s,d) + g$}
\newcommand{\bsggam}{\ $b \to s+ \gamma+ g$}
\newcommand{\bdggam}{\ $b \to d+ \gamma+ g$}
\newcommand{\bsdggam}{\ $b \to (s,d)+ \gamma+ g$}
\newcommand{\btod}{$b \to d \gamma, d \gamma g $ }
\newcommand{\btos}{$b \to s \gamma, s \gamma g $ }
\newcommand{\bgamaxd}{$B \to X _{d} + \gamma$}
\newcommand{\brbgamaxd}{{\cal B}(B \to X _{d} + \gamma)}
\newcommand{\bksgam}{\ $B \to K^*+ \gamma$}
\def\to{\rightarrow}
\def\VA{(V-A)}
\def\ij{_{ij}}
\def\ji{_{ji}}
\def\dbrp{{dB^+\over d\sh}}
\def\dbrm{\int_0^1 dz {d^2 B \over {d\sh dz}} -
          \int_{-1}^0 dz {d^2 B \over {d\sh dz}} }
\def\ddb{{d^2 B \over d\sh dz}}
\def\mb{m_b}
\def\xt{x_t}
\def\xs{x_s}
\def\xd{x_d}
\newcommand{\kkbar}{$K^0$-${\overline{K^0}}$}
\newcommand{\bdbdbar}{$B_d^0$-${\overline{B_d^0}}$}
\newcommand{\bsbsbar}{$B_s^0$-${\overline{B_s^0}}$}
\def\Cz{{\overline C}^Z}
\def\Cb{{\overline C}^{Box}}
\def\Fs{{F_1}^{(s)}}
\def\Fl{{F_1}^{(l)}}
\def\Fm{{F_2}}
\def\as{\alpha _s}
\def\alfas{\alpha _s}
\def\etal{et al.}
\def\a{{\cal A}}
\def\b{{\cal B}}
\def\acp{{\cal A}_{\rm CP}}
\def\h{{\cal H}}
\def\he{{\cal H}^{\rm eff}}
\def\l{\lambda}
\def\p{\pi}
\def\e{{\rm e}}
\def\c{{\cal C}}
\def\brg{{B \rightarrow \rho \gamma}}
\def\cbdg{{B^\pm \rightarrow D^\pm \gamma}}
\def\cbkg{{B^\pm \rightarrow {K^\ast}^\pm \gamma}}
\def\cbrg{{B^\pm \rightarrow \rho^\pm \gamma}}
\def\nbkg{{B^0 \rightarrow {K^\ast}^0 \gamma}}
\def\nbrg{{B^0 \rightarrow \rho^0 \gamma}}
\def\bog{{B \rightarrow \omega \gamma}}
\def\brog{{B \rightarrow \rho(\omega) \gamma}}
\def\nbdg{{B^0 \rightarrow D^0 \gamma}}
\def\si{sin\theta_c}
\def\co{cos\theta_c}
\def\BDl{B \to D \ell \nu_\ell}
\def\BDSl{B \to D^* \ell \nu_\ell}
\def\BPIl{B \to \pi \ell \nu_\ell}
\def\BROl{B \to \rho \ell \nu_\ell}
\def\JMUl{J^\mu ^{lept.}}
\def\JMUh{J^\mu ^{hadr.}}
\def\vdvp{v \cdot v^\prime}
\def\xip{\xi_+(\vdvp )}
\def\xim{\xi_-(\vdvp )}
\def\xiv{\xi_V(\vdvp )}
\def\xiao{\xi_{A_1}(\vdvp )}
\def\xiaoo{\xi_{A_1}(\vdvp =1 )}
\def\xiat{\xi_{A_2}(\vdvp )}
\def\xiath{\xi_{A_3}(\vdvp )}
\def\FP{F_+ (q^2)}
\def\FM{F_- (q^2)}  
\def\VV{V (q^2)}
\def\Vbc{V_{cb}}
\def\Vbu{V_{ub}}
\def\Vtd{V_{td}}
\def\Vts{V_{ts}}
\def\Vtb{V_{tb}}
\def\pp{{\prime\prime}}  
\def \heff{H_{\mathrm{eff}}}
\def \eff{\hbox{eff}}  
\def \nn{\nonumber}
\def \dis{\displaystyle}
\def \tev{{\hbox{TeV}}}
\def \gev{{\hbox{GeV}}}
\def \kev{{\hbox{keV}}}
\def \mev{{\hbox{MeV}}}
\def \gh{\vphantom{$\fbox{\Big[}$}}
\def \B{\bar{B}}    
\def \cl#1{{#1\%\ \mathrm{C.L.}}}
\maketitle
\vspace*{-2.0cm}

\section{Introduction}
It is now forty years that Nicola Cabibbo formulated the notion of flavour 
mixing in the charged hadronic weak interactions~\cite{Cabibbo:yz}.  
The Cabibbo theory provides a consistent description of the muonic 
decay $\mu \to e \nu_e \nu_{\mu}$, the neutron 
$\beta$-decay $n \to p e \nu_e$, and the strangeness changing transitions, 
such  as the $K_{\ell 3}$ decays and the hyperon decays, in terms of a 
universal Fermi coupling constant $G_F$ and a mixing angle, the Cabibbo angle 
$\theta_C$. Thanks to dedicated experiments carried out well over 
four decades, and impressive theoretical progress, in particular in the
technology of the electroweak radiative corrections, we now have 
precise values for these fundamental parameters of nature~\cite{Hagiwara:fs}:
\begin{eqnarray}
G_F&=&1.16639(1)\times 10^{-5}~{\rm GeV}^{-2}\,, 
\nonumber\\[-1.5mm] 
\label{gftc}\\[-1.5mm]
\theta_c &=& 12.69(15)^\circ\,.
\nonumber
\end{eqnarray}
The Cabibbo theory~\cite{Cabibbo:yz}
describes in the quark language charged weak 
transitions $d \to u$ and $s \to u$ involving the three lightest quarks 
$u$, $d$, and $s$. However, it was not able to account for the flavour 
changing neutral current (FCNC) transition $s \to d$, such as the $K^0$ - 
$\overline{K^0}$ mass difference $\Delta M_K$. This outstanding 
problem with the Cabibbo theory was solved by the GIM 
mechanism~\cite{Glashow:gm} which required a fourth quark - the charm (c) 
quark. The  GIM-construction 
banished the FCNC transitions from the tree level, relegating them to 
loops (induced quantum effects) where they found their natural abode. 
Thus, in the Cabibbo-GIM theory, $\Delta M_K$,
as well as a number of $\vert \Delta S \vert = 
1, \Delta Q=0$ transitions, such as $K_L \to \mu^+ \mu^-$ and
$K_L \to \gamma \gamma$,  are well-accounted for  
in terms of  $G_F$ and $\theta_C$, and the
mass of the charm quark $m_c$~\cite{Gaillard:1974nj}, found later
to be in the right ball-park through the discovery of the $J/\psi,\psi^\prime,...$ 
resonances and charmed hadrons. Along with the GIM mechanism came also
a $(2\times 2)$ quark flavour mixing matrix characterized by the Cabibbo angle 
$\theta_C$.

 The final act in quark mixing
came through the seminal work of Kobayashi and Maskawa 
(KM)~\cite{Kobayashi:fv}, who enlarged the Cabibbo-GIM $(2\times 2$) quark 
mixing matrix to a $(3\times 3$) matrix by adding another doublet 
of heavier quarks $(t,b)$. This matrix which
relates the quarks in the weak interaction basis $(d^\prime, s^\prime, 
b^\prime)$ and the quark mass eigenstates $(d,s,b)$,
\beq
\left(\matrix{d^\prime, s^\prime,b^\prime}\right)
=V_{\rm CKM} \left(\matrix{d, s, b }\right)~,
\eeq  
is called the Cabibbo-Kobayashi-Maskawa matrix
$V_{\rm CKM}$, and  symbolically written as
\beq
V_{\rm CKM} \equiv \left(\matrix{
 V_{ud} & V_{us} &V_{ub} \cr
 V_{cd} & V_{cs} &V_{cb} \cr
 V_{td} & V_{ts} &V_{tb}}\right)\,.
\label{CKM}
\eeq
$V_{\rm CKM}$ is a unitary matrix, characterized by three 
independent rotation angles and a complex 
phase. The KM theory was formulated to incorporate the CP violation observed in
Kaon decays in 1964 by  Christenson {\it et al}.~\cite{Christenson:fg}.
In this theory CP symmetry is broken at the Lagrangian level in the charged 
current weak interactions and no where else. In principle, all the elements of the
matrix $V_{\rm CKM}$ are complex. In practice, only two of the matrix elements
have measurable phases. But, this is sufficient to anticipate CP violation in a  
large number of processes, some of which are now being measured 
with ever-increasing precision in the $K$- and $B$-meson decays.

In these lectures, I will summarize the current status of the magnitude 
of all the nine matrix elements $\vert V_{ij} \vert$ of the CKM matrix 
and the weak phases entering in these matrix elements. To discuss this, a 
parametrization of the CKM matrix is needed. It has become customary to discuss the 
CKM phenomenology using the Wolfenstein parametrization~\cite{Wolfenstein:1983yz}:
\begin{eqnarray}
V_{\rm CKM} &\simeq&
 \left(\matrix{
 1-{1\over 2}\lambda^2 & \lambda
 & A\lambda^3 \left( \rho - i\eta \right) \cr
 -\lambda ( 1 + i A^2 \lambda^4 \eta )
& 1-{1\over 2}\lambda^2 & A\lambda^2 \cr
 A\lambda^3\left(1 - \rho - i \eta\right) & -A\lambda^2\left(1+i \lambda^2 \eta\right) 
& 1 
\cr}\right)~,
\label{CKM-W}
\end{eqnarray}
where the four independent parameters are: $A$, $\lambda=\sin \theta_C$,
$\rho$ and $\eta$, of which $\eta$ is what makes this matrix complex and leads 
to CP Violation. Anticipating precise
data, a perturbatively improved Wolfenstein
parametrization~\cite{Buras:1994ec} with $\bar{\rho}=\rho(1-\lambda^2/2)$ and
$\bar{\eta}=\eta(1-\lambda^2/2)$ will be used.  This rescaling effects mainly the
matrix elements $V_{td}$, which now has the definition $V_{td} =A \lambda^3 (1- 
\bar{\rho} - i \bar{\eta})$, and $V_{ub}= A\lambda^3 (1+\lambda^2/2) \left( 
\bar{\rho} - i\bar{\eta }\right)$, and the other matrix elements remain essentially unchanged.

 As we shall see,
a quantitative determination of these matrix elements requires, apart from  dedicated 
experiments, reliable theoretical tools in the theory of strong 
interactions (QCD). These 
include, apart from the QCD-motivated quark models, chiral perturbation theory, QCD 
sum rules, Heavy Quark Effective  Theory (HQET) and Lattice QCD, combined with 
perturbative QCD. To illustrate their impact, I will discuss some  representative 
applications where a particular technique is the main 
theoretical workhorse. Further details and in-depth discussions can be
found, for example, in the proceedings of the CERN-CKM 
workshops~~\cite{Battaglia:2003in,ckm2003}.

These lecture notes are organized as follows: In section 2, 
I review the status of $\vert V_{ud} \vert$ and $\vert V_{us}\vert$
and the resulting test of unitarity of the CKM matrix.
In section 3, current status of $\vert V_{cd} \vert$ and $\vert V_{cs}\vert$
is briefly reviewed. Section 4 describes in considerable detail the current
measurements of the matrix elements $\vert V_{cb} \vert$ and $\vert V_{ub} \vert$
and the theoretical techniques used in arriving at the results. Status of the third
row of $V_{\rm CKM}$ is reviewed in Section 5.
Section 6 summarizes the current knowledge of  $\vert 
V_{ij}\vert$ and the Wolfenstein parameters, including the phases of the
unitarity triangle(s). This section also reports on the results of a global fit of
the CKM parameters using the CKM unitarity and knowledge of the various 
experimental and theoretical quantities. CP violation in $B$-meson decays is discussed in 
Section 7, and we restrict ourselves to the discussion of only the currently 
available results from the two B-factory experiments.
We conclude with a summary of the main results and some remarks in Section 8. 

\section{Current Status of $\vert V_{ud}\vert$ and $\vert V_{us}\vert$}
\subsection{Status of $\vert V_{ud} \vert$}
We start with the discussion of the matrix element $\vert V_{ud} \vert$.
The superallowed $0^+ \to 0^+$ Fermi transitions (SFT) have been measured 
so far in nine  nuclei $(^{10}{\rm C},~^{14}{\rm O},^{26}{\rm Al}, 
^{34}{\rm Cl}, ^{38}{\rm K}, ^{42}{\rm Sc}, ^{46}{\rm V}, ^{50}{\rm Mn}, 
^{54}{\rm Co})$, summarized by Towner and Hardy~\cite{Towner:1998qj}. As 
only the 
vector current contributes in the nuclear hadronic matrix element
$\langle p_f;0^+| \bar{u} \gamma_\mu d | p_i;0^+\rangle$, and the transitions involve 
members of a given isotriplet, the conserved vector current hypothesis 
helps greatly in reducing the
hadronic uncertainties. Radiative $(\delta_R$ and $\Delta_R$) and isospin breaking 
($\delta_C$) 
corrections have been calculated. Of these $\delta_C$ and $\delta_R$ are nucleus-dependent. 
As the $ft$ 
values ($f$ is the nuclear-dependent phase space and $t$ is the lifetime) of the nuclear 
transitions are also nucleus-dependent, one usually absorbs the nucleus-dependent radiative 
corrections by defining another quantity
\be
 {\cal F}t \equiv ft\, (1+\delta_R)\, (1-\delta_C)\,.
\ee
The ${\cal F}t$ values measured in the nine transitions are indeed 
consistent with each other, with their average having a value~\cite{Towner:1998qj}
\be
 \overline{{\cal F}t}=3072.3(9)~{\rm sec}\, \, [\chi^2/{\rm d.o.f.}=1.1]\,.
\ee
 The nucleus-independent radiative
correction $\Delta_R$ incorporates the short-distance contribution and
has been calculated by Marciano and Sirlin~\cite{Marciano:pd,Sirlin:qd}. 
The value of $\Delta_R$ depends on a parameter $m_A$ which enters in the process of 
matching the short-distance and long-distance contributions. Taking $m_A$  
in the range ${m_{A}}_1/2 \leq m_A \leq 2 {m_{A}}_1$, where ${m_{A}}_1$ 
is the $1^{+-}$ meson mass, one estimates: 
$\Delta_R=(2.40 \pm 0.08)\%$. With the precise determination  of ${\overline{\cal F}t}$
and $\Delta_R$, the matrix element $\vert V_{ud} \vert$ is derived from the 
expression:
\be
\vert V_{ud} \vert^2 = \frac{K}{2 G_F^2 {\overline{\cal 
F}t}\, (1+\Delta_R)}~,
\label{vudsftform}
\ee
where $K$ is a phase space factor, $K=2\pi^3 \ln 2/m_e^5$, with $m_e$
being the electron mass, and its value is known to a 
high accuracy: $K=(8120.271 \pm 0.012) \times 10^{-10}$ GeV$^{-4} sec$.
This yields a value~\cite{Towner:1998qj}:
\be
  \vert V_{ud} \vert_{\rm SFT} =0.9740 \pm 0.0005\,.
\label{vudsft}
\ee
The error shown here is dominated by theoretical 
errors, contributed mainly by the 
(somewhat arbitrary) choice of the low energy cutoff in estimating 
$\Delta_R$ and the nuclear-dependent 
isospin-breaking corrections $\delta_C$~\cite{Towner:1998qj,Saito:1995de}. 
The value listed by the PDG~\cite{Hagiwara:fs} 
from this method is: $\vert V_{ud} \vert 
=0.9740 \pm 0.0005 \pm 0.0005$, where the added error
reflects the PDG concern about the systematic uncertainty due to the 
nucleus-dependent radiative corrections. 

The other precise method of determining the matrix 
element $\vert V_{ud}\vert$ is through the polarized neutron beta-decay
$(n \to p e^- \overline{\nu_e})$. The currently attained precision owes 
itself 
to the enormous progress made in having highly polarized cold neutron beams.
For example, for the cold neutron beam at the High Flux Reactor at the 
Institut-Laue-Langevin, Grenoble, the degree of neutron polarization has been
measured to be $P=98.9(3)\%$ over the full cross-section of the 
beam~\cite{Abele:2002wc}.
Also, the neutron lifetime, $\tau_n=(885.7\pm 0.8)$ sec~\cite{Hagiwara:fs}, is 
now measured to an accuracy of one part in a thousand. The charged weak 
current has a $V-A$ structure and the hadronic matrix element  
can be parametrized as:
$\langle p |\bar{u} \gamma_\mu(1-\gamma_5)d|n \rangle =\bar{u}_p
\gamma_\mu(g_V+ g_A\gamma_5)u_n$, requiring the knowledge of $g_V$ and 
$g_A$. (We have neglected a small weak magnetism contribution $\propto 
(\mu_p -\mu_n)/(2 m_p) \sigma_{\mu \nu} q^\nu$, where $\mu_p$ 
and $\mu_n$ are the 
anomalous magnetic moments of the proton and neutron, respectively, 
with $q_\nu$  
and $m_p$ being the momentum transfer and the proton mass, respectively). 
However, in the neutron 
beta-decay, radiative corrections are under better  theoretical control.

The theoretical expression for determining $\vert V_{ud} \vert$ from the neutron 
lifetime is:
\be
\vert V_{ud} \vert^2= \frac{1}{C \tau_n (1+3 x^2) f^{\rm 
R}\, (1+\Delta_R)}~, 
\label{vudnbd}
\ee
where $C=G_F^2 m_e^5/(2\pi^3)$, $x=g_A/g_V$,  $f^{\rm R}=1.71482(15)$ 
is the phase 
space factor including model-dependent radiative corrections~\cite{Wilkinson:hu}, 
and the model-independent radiative correction $\Delta_R$ has been specified earlier. 
The high accuracy on
$f^{\rm R}$ owes itself to the Ademollo-Gatto theorem, which makes the
departure of $g_V$ from the symmetry limit tiny, with current estimates yielding $\delta 
g_V\equiv 1-g_V=O(10^{-5})$~\cite{Paver:gz,Donoghue:1990ti,Kaiser:2001yc}.
 
To extract $\vert V_{ud} \vert$ from the neutron lifetime, one has to know  
$g_A/g_V$. This can be determined from the electron ($\beta$) asymmetry or the
$e$-$\bar{\nu}$ correlation in the decay of a polarised neutron. 
For example, the probability that an electron is emitted with 
an angle $\theta$ with respect to the 
neutron spin polarization, denoted here by $P=\langle \sigma_z \rangle$, is
\be
W(\theta) =1+\beta P A \cos \theta~,
\label{betasymm}
\ee  
where $\beta$ is the electron velocity and the coefficient 
$A$ depends on $x=g_A/g_V$:
\be
A= -2\,\frac{x(1+x)}{1 + 3x^2}\,.
\ee
It is understood here that a small correction due to the 
weak magnetism has been included 
in extracting $g_A/g_V$ from $A$. 
Thus, the measurements of $\tau_n$ and $A$ determine
both $g_A/g_V$ and $\vert V_{ud}\vert$. However, currently
the two most precise measurements of this quantity, 
namely $g_A/g_V=-1.2594 \pm 0.0038$~\cite{Abele:2002wc} 
and $g_A/g_V=-1.2739 \pm 0.0019$~\cite{Erozolimsky:wi} differ 
by more than 3$\sigma$, and hence the experimental spread 
in the values of $g_A/g_V$ is currently the main uncertainty in the 
determination of $\vert V_{ud} \vert$ from the 
neutron $\beta$-decay. Of these, the  PERKEOII 
experiment~\cite{Abele:2002wc} yields a 
value $\vert V_{ud} \vert=0.9724  \pm 0.0013$, 
which, on using the PDG values for
$\vert V_{us}\vert$ ($=0.2196 \pm 0.0026$) and $\vert 
V_{ub}\vert$ ($=(3.6 \pm 0.7) \times 10^{-3}$) leads to

\vspace*{-2mm}

\be
\vert V_{ud} \vert^2 + \vert V_{us} \vert^2 + \vert V_{ub} \vert^2 
\equiv 1-\Delta_1 =0.9924 \pm 0.0028\,.
\label{deltaparkeo}
\ee

The value $\Delta_1 =(7.6 \pm 2.8) \times 10^{-3}$ differs from zero 
(the unitarity value) by $2.7 \sigma$. However, following the advice 
of the PDG and restricting to the experiments using neutron polarization 
of more than 90\%~\cite{Abele:2002wc,Liaud:vu,Mostovoi:ye}, 
a recent compilation of the experimental results 
yields~\cite{Isidori:2003a}:
\begin{eqnarray}
g_A/g_V &=&-1.2720 \pm 0.0022\,, 
\nonumber\\[-1.5mm] 
\label{Vudisidori}\\[-1.5mm]
\vert V_{ud} \vert_{\rm n~decay} &=& 0.9731 \pm 0.0015\,.
\nonumber
\end{eqnarray}
The result for $\vert V_{ud} \vert$ from the neutron $\beta$-decay 
is less precise than the one in (\ref{vudsft}), obtained  from the 
$0^+ \to 0^+$ SFTs, though the two values of $\vert V_{ud} \vert$ 
are completely consistent with each other. To improve the precision on 
$\vert V_{ud} \vert$ from the neutron beta-decay,
it is imperative to resolve the inconsistencies in the current measurements of 
$g_A/g_V$ and determine this ratio more accurately.

The third method for determining $\vert V_{ud}\vert$ is through the 
$\pi_{e3}$ decay: $\pi^+ \to \pi^0 e^+ \nu_e$. This decay is governed 
by the vector pion form factor $f_+^{\pi^\pm\pi^0}(t)$, where $t$ is the 
transfer momentum squared. In the isospin limit, 
$f_+^{\pi^\pm\pi^0}(0)=1$.  An updated analysis of the radiative 
corrections to the pionic beta-decay has been recently undertaken 
in an elegant paper by Cirigliano et al.~\cite{Cirigliano:2002ng}, 
including all electromagnetic corrections of order 
$e^2p^2$ (here $e^2p^2$ implies both corrections of order $e^2p^2$ and of order
$(m_u -m_d) p^2$), using the framework of chiral perturbation theory with 
virtual photons and leptons. Accounting for the isospin-breaking 
and radiative corrections,
$\vert V_{ud} \vert$ can be obtained from the following 
expression~\cite{Cirigliano:2002ng}:
\be
\vert V_{ud}\vert^2 =\frac{\Gamma_{\pi_{e3[\gamma]}}}
{ {\cal N}_\pi \, S_{\rm ew}  \, \vert 
f^{\pi^\pm\pi^0}(0) \vert^2 \,  I_{\pi}(\lambda_t,\alpha)}\,,
\label{cirigliano}
\ee
where
\bea
& & {\cal N}_\pi = G_F^2\, M_{\pi^\pm}^5/(64\pi^3)\,, 
\nonumber\\ [0.9mm]
& & I_\pi(\lambda_t,\alpha) = I_\pi(\lambda_t,0)\,
\left (1+ \Delta I_\pi(\lambda_t, \alpha)\right )\,, 
\label{betapirad}\\[0.9mm]
& & f_+^{\pi^\pm\pi^0}(0) = (1+\delta_{SU(2)}^\pi)\,(1+\delta_{e^2p^2}^\pi)\,.
\nonumber
\eea  
Here $\lambda_t$ is the slope parameter in the parametrization 
of the form factor
$f_+^{\pi^\pm\pi^0}(t)=f_+^{\pi^\pm\pi^0}(0)(1+\lambda_t t/M_\pi^2 + 
O(\lambda_t^2))$, $I_{\pi}(\lambda_t,\alpha)$ is a slope-dependent phase 
space integral, and $S_{\rm ew}$ is what was earlier called $\Delta_R$. 
Estimates of the
various quantities in these expressions are~\cite{Cirigliano:2002ng}:  
\begin{eqnarray}
&& \delta_{SU(2)}^\pi \sim  10^{-5}\,, 
\nonumber\\[0.9mm]
&& \delta_{e^2p^2}^\pi=(0.46\pm 0.05)\%\,, 
\label{cirigicf}\\[0.9mm]
&& \Delta I_\pi(\lambda_t,\alpha)= 0.1\%\,.
\nonumber
\end{eqnarray}
 The precision
on $\vert V_{ud} \vert$ is dominated by the precision on the quantity
$\Gamma_{\pi_{e3[\gamma]}}$ (i.e., the branching ratio ${\rm BR}(\pi^\pm \to \pi^0 e^\pm 
\nu_e [+ \gamma])$. The present preliminary result of the PIBETA 
collaboration~\cite{Pocanic:2003tp} is:
\be
 {\rm BR}(\pi^\pm \to \pi^0 e^\pm \nu_e)=(1.044 \pm 0.007({\rm stat.}) \pm 
0.009({\rm syst.}) )\times 10^{-8}\,,
\ee
which is significantly more precise than the earlier most accurate
measurement by 
McFarlane et al.~\cite{Mcfarlane:ry}: ${\rm BR}(\pi^\pm \to \pi^0 e^\pm 
\nu_e)=(1.026 \pm 0.030) \times 10^{-8}$. The PIBETA measurement yields a 
value~\cite{Cirigliano:2003yr}:
\be
\vert V_{ud} \vert_{\pi_{\ell 3}} =0.9771 \pm 0.0056\,.
\label{vudpibeta}
\ee
This measurement of $\vert V_{ud} \vert$ is almost an order of magnitude 
less precise than $\vert V_{ud} \vert$ determined  from the nuclear 
$0^+ \to 0^+$ SFTs. One expects a factor three improvement in the value 
of $\vert V_{ud} \vert$ at the end of the PIBETA experiment.

Taken the three determinations of $\vert V_{ud} \vert$ discussed here,
the current world average of this quantity,
\be
 \vert V_{ud} \vert_{\rm WA} =0.9739\pm 0.0005\,,
\label{vudwa}
\ee
is essentially the same as in (\ref{vudsft}) from the nuclear $0^+ \to
0^+$ transitions. The impact of the neutron beta decay experiments on $\vert 
V_{ud} \vert$ can be significantly enhanced if the experimental spread 
in $g_A/g_V$ 
is resolved, and the resulting accuracy on this quantity improved.  

\subsection{Status of $\vert V_{us} \vert$}

The determination of the matrix element $\vert V_{us} \vert$ from 
$K_{\ell 3}$ decays (with $\ell=e,\mu$) has been extensively reviewed 
recently~\cite{Isidori:2003a,Cirigliano:2003yr} to which we refer for 
further details. The value for $\vert V_{us} \vert$ quoted by the PDG is 
based essentially on the theoretical analysis of Leutwyler and 
Roos~\cite{Leutwyler:1984je}, done some twenty years ago,
which yields: $\vert V_{us} \vert=0.2196 \pm 0.0023$. During the last 
couple of years, new analytical calculations of the radiative corrections 
have been reported by Cirigliano et al.~\cite{Cirigliano:2002ng}, carried 
out in the context of the chiral perturbation theory, which were used in 
extracting $\vert V_{ud} \vert$ from the $\pi_{e 3}$ decay, discussed 
earlier. Also, the so-called long-distance part of 
the electromagnetic radiative corrections, calculated by 
Ginsberg long ago~\cite{Ginsberg:jh} and used in the Leutwyler-Roos 
analysis~\cite{Leutwyler:1984je} has been recently checked
(and corrected)~\cite{Cirigliano:2002ng,Bytev:2002nx}. 
Finally, two $O(p^6)$ chiral perturbation theory calculations of the 
isospin-conserving contribution to the 
$K_{\ell3}$ form factor have also been  
undertaken~\cite{Post:2000gk,Bijnens:2003uy}. In particular, it has been 
pointed out by Bijnens and  Talavera~\cite{Bijnens:2003uy} that the low 
energy constants (LEC's) which appear in order $p^6$ in the form factor 
$f_+(0)$  can be determined from $K_{\ell3}$ measurements via the slope 
and the curvature of the scalar form-factor $f_0(q^2)$. In fact, there is 
some model-dependence also in the order $p^4$ parameters which impacts 
on $\vert V_{us} \vert$, and one should firm up the existing 
phenomenological estimates by new measurements and/or 
calculations of the LEC's on the lattice. 

In addition to these theoretical developments, new experiments and/or 
analysis have been reported during this year by 
several groups. This includes a new, high statistics measurement of 
the $K^+ \to \pi^0 e^+ \nu$ $(K_{e3}^+)$ branching ratio by the BNL 
experiment E865~\cite{Sher:2003fb}, which impacts on the 
determination of $\vert V_{us} \vert$. New results in $K_{\ell 3}$ decays 
have been reported by the KLOE collaboration at 
DA$\Phi$NE~\cite{Sciascia:2003qp,Moulson:2003zu}, 
based on the measurements of the decays
$K_L \to \pi \mu \nu_\mu$, $K_L \to \pi e \nu_e$, and 
$K_S \to \pi e \nu_e$. In addition, semileptonic hyperon decays 
have been revisited  by Cabibbo {\it et al}.~\cite{Cabibbo:2003ea} to 
determine the Cabibbo angle (or $\vert V_{us}\vert$). Finally, a 
determination of $\vert V_{us} \vert$ has been undertaken from hadronic 
$\tau$-decays by Gamiz et al.~\cite{Gamiz:2002nu}. In this subsection, we 
summarize these results, some of which are new additions in this field 
since the CERN-CKM workshop.

The four $K_{\ell 3}$ decay widths for the decays $K_{e3}^+$, $K_{e3}^0$, 
$K_{\mu 3}^+$, and $K_{\mu 3}^0$ have been analyzed 
by Cirigliano~{\it et~al}.~\cite{Cirigliano:2001mk}. 
Normalizing the  decay widths in terms of the quantity $f_+^{K^0 \pi^-}(0)$, 
evaluated in the absence of the electromagnetic corrections, 
the following master formula is used to extract~$\vert V_{us} 
\vert f_+^{K^0 \pi^-}(0)$~\cite{Cirigliano:2001mk,Cirigliano:2003yr}:

\bea
\vert V_{us} \vert f_+^{K^0 \pi^-}(0) = 
\left( \frac{\Gamma_{n[\gamma]}}
{{\cal N}_n \, I_n(\lambda_t,0)}\right)^{1/2} \,
\left(\frac{1}{S_{\rm ew}} \right)^{1/2}  
\frac{1}{1+ \delta_{SU(2)}^n + \delta_{\rm EM}^n + \Delta 
I_n(\lambda_t,\alpha)/2}\,,
\label{kleall}
\eea
where the index $n$ runs over the four modes, 
${\cal N}_n=C_n^2 G_F^2 M_{n}^5/192 
\pi^3$, with $C_n=(1,1/\sqrt{2})$ for $(K^0,K^+)$. 
The various corrections and the compilation of the decay widths can be 
seen in the literature~\cite{Cirigliano:2003yr}.
This yields the following value:
\be
\vert V_{us} \, \vert f_+^{K^0 \pi^-}(0) =0.2115 \pm 0.0015\,.
\label{cirivusf}
\ee
The quantity $f_+^{K^0 \pi^-}(0)$ has been studied in the context of the 
chiral perturbation theory. The result up to the next-to-next-to-leading 
order is known~\cite{Leutwyler:1984je}:
\be
f_+^{K^0 \pi^-}(0) =1 + f_2 + f_4 +{\cal O}(p^6)\,,
\ee
with $f_2=-0.023$ and $f_4=-0.016 \pm 0.008$, yielding
the Leutwyler-Roos value $f_+^{K^0 \pi^-}(0)=0.961 \pm 0.008$. This
gives~\cite{Cirigliano:2003yr}:
\be
\vert V_{us} \, \vert_{K_{\ell 3}} = 0.2201 \pm 0.0024 \,.
\label{oldvus}
\ee 
Bijnens and Talavera~\cite{Bijnens:2003uy} have included the isospin-conserving part 
of the $O(p^6)$ corrections in the determination of $f_+^{K^0 \pi^-}(0)$, getting
$f_+^{K^0 \pi^-}(0)=0.9760 \pm 0.0102$, which, in turn, yields
$\vert V_{us} \vert_{K_{\ell 3}}=0.2175 \pm 0.0029$. However, as emphasized by these
authors, this result should be treated as preliminary since the isospin-breaking
$O(p^6)$ contributions are not yet included. Also, the effect of the curvature in
the form factor on the experimental value remains to be evaluated.

Recently, the E865 collaboration at Brookhaven~\cite{Sher:2003fb} has published a 
branching ratio for the 
decay $K_{e3[\gamma]}^+$: ${\rm BR}(K_{e3[\gamma]}^+)=(5.13 \pm 0.02({\rm stat}) 
\pm 0.09({\rm sys}) \pm 0.04({\rm norm}))\%$, which is about $2.3\sigma$ higher than the 
current PDG value~\cite{Hagiwara:fs} for this quantity. The higher E865 branching ratio
translates into a correspondingly higher value of the product $\vert V_{us}  
f_+^{K^-\pi^0}(0)\vert$:
\be
\vert V_{us} \, f_+^{K^-\pi^0}(0) \vert =0.2239 \pm  0.0022({\rm 
rate}) \pm 0.0007(\lambda_t)\,,
\ee
which on using the result from Cirigliano {\it et al}.~\cite{Cirigliano:2001mk}
for $f_+^{K^-\pi^0}(0)$ yields
\bea
 \left \vert V_{us} \right \vert_{\rm E865} = 
0.2272 \pm 0.0023({\rm rate}) 
\pm 0.0007(\lambda_t)\pm 0.0018(f^+)\,.
\label{e865vus}
\eea
This differs from the older $K_{\ell 3}$ result (\ref{oldvus}) by more than 
$2\sigma$.
Interestingly, the E865 value of $\vert V_{us} \vert$, together with the world average 
for $\vert V_{ud} \vert$ given in (\ref{vudwa}), leads to  perfect agreement with 
the CKM-unitarity! Denoting the departure from unity in the first row of $V_{\rm 
CKM}$ by 
$\Delta_1$ (defined earlier), the value obtained by the 
E865 group yields $\Delta_1=0.0001 \pm 0.0016$~\cite{Sher:2003fb}.

The other new addition to this subject is the measurements of 
$\vert V_{us} \vert$ from the production of the $\phi$-meson at DA$\Phi$NE 
and its decays into $K^+K^-$ and $K_LK_S$ pairs with the subsequent 
$K_{\ell 3}$-decays. The KLOE collaboration at DA$\Phi$NE will eventually 
measure $\vert V_{us} \vert$ precisely (to an accuracy of better than 1\%) from
all four channels of the $K_S$ and $K_L$ decays (involving the final states 
$\pi \ell \nu_\ell; \ell=e,\mu$) as well as from the $K^\pm_{\ell 3}$ decays.
Their preliminary results are available in conference 
reports~\cite{Sciascia:2003qp,Moulson:2003zu} on the following three modes:
$K_S \to \pi e \nu_e$, $K_L \to \pi \mu \nu_\mu$ and $K_L \to \pi e \nu_e$.
Of these, the analysis of the $K_S$ decay mode is more advanced in terms 
of the systematics. Concentrating on this decay, its branching ratio has been
measured by KLOE as 
${\rm BR}(K_S \to \pi^- e^+ \nu_e +{\rm c.c.})=(6.81 \pm 0.12 \pm 
0.10)\times 10^{-4}$. The lifetime of the $K_S$-meson has been 
recently measured by the NA48 experiment~\cite{Lai:2002kd}: 
$\tau(K_S)=0.89598(48)(51)\times 10^{-10}$~s, allowing 
to have a new precise measurement of the ratio ${\rm BR}(K_S \to \pi e 
\nu_e)/\tau(K_S)$. Using the theoretical analysis of the $K^0_{e3}$ 
mode discussed earlier, this yields~\cite{Moulson:2003zu}:
\be
\left \vert V_{us} \right \vert 
f_+^{K^0 \pi^-}(0)=0.2109 \pm 0.0026\,,
\label{fpluskloe}
\ee
in excellent agreement with the value given in (\ref{cirivusf}), obtained 
from the earlier results on $K_{\ell 3}$ decays.The corresponding 
(preliminary) values from the two $K_{\ell 3}$ decay modes of the
$K_L$-meson are similar~\cite{Sciascia:2003qp,Moulson:2003zu},
 with $\vert V_{us} \vert f_+^{K^0 \pi^-}(0)=0.2085 \pm 0.0019$ 
(for the $\pi e \nu_e$ mode) and $\vert V_{us} 
\vert f_+^{K^0 \pi^-}(0)=0.2106 \pm 0.0028$ (for the $\pi \mu \nu_\mu$ mode). 
However, as the systematic errors (in particular, for the $K_L$ modes) 
have not yet been finalized, these numbers should not be averaged yet. 
Following the advice of the KLOE collaboration\footnote{Helpful 
communications with Matt Moulson are gratefully acknowledged.}, we take the 
value of $\vert V_{us} \vert$ obtained from the better studied $K_S$ mode, 
as the preliminary value of this quantity from the current DA$\Phi$NE 
measurements:
\be
\vert V_{us} \vert_{{\rm DA\Phi NE;} K_S} = 0.2194 \pm 0.0030\,.
\label{vusdafne}
\ee
This is in comfortable agreement with the earlier determinations of this 
quantity given in (\ref{oldvus}).

While still on the subject of determining $\vert V_{us} \vert$, there 
are two non-$K_{\ell 3}$ estimates of this matrix element available in the 
literature, the first estimate is from the study of the semileptonic 
hyperon decays and the second is from the hadronic decays of the 
$\tau$-lepton. We discuss them in turn.

The value listed in the PDG review for $\vert V_{us} \vert$ from hyperon decays
$\vert V_{us} \vert=0.2176 \pm 0.0026$  is similar in its precision as the 
one in (\ref{oldvus}), obtained from the $K_{\ell 3}$ decays. However, 
based on the observation that the value obtained from hyperon decays is 
illustrative as it depends on the models to incorporate the 
$SU(3)$-symmetry-breaking corrections, and the theoretical dispersion 
(model-dependence) is significant, this value of $\vert V_{us} \vert$ 
is not included in the world average of $\vert V_{us} \vert$ by the PDG. 
This state of affairs was considered more or less as a theoretical
{\it fait accompli} and no significant attempt was undertaken to
reduce this model dependence. Recently, Cabibbo et al.~\cite{Cabibbo:2003ea} 
have taken a somewhat different approach and have reported an 
analysis of the hyperon decays to extract $\vert V_{us} \vert$. Their main 
assumption and results are summarized below.

Denoting a typical hyperon decay by ${\cal N}_1 \to {\cal N}_2 e^- 
\overline{\nu_e}$, the following four decays are reanalyzed by Cabibbo et 
al.~\cite{Cabibbo:2003ea}:
$({\cal N}_1, {\cal N}_2)= (\Lambda, p)\,, (\Sigma^-, n)\,, 
(\Xi^-,\Lambda)\,, (\Xi^0, \Sigma^+)$. The matrix elements for these decays 
can be expressed as follows
\be
{\cal M}=\frac{G_S}{\sqrt{2}} \,  \langle {\cal N}_2 \vert
J_\alpha (0) \vert {\cal N}_1 \rangle \,
\left [\bar{u}_e \gamma^\alpha (1+ \gamma_5) v_{\nu} \right ]\,,
\ee
where 
\bea
\langle {\cal N}_2 \vert J_\alpha (0) \vert {\cal N}_1 
\rangle = f_1(q^2) \,\gamma_\alpha + \frac{f_2(q^2)}{M_{{\cal N}_1}} \,
\sigma_{\alpha\beta}\, q^\beta 
+ g_1(q^2)\,\gamma_\alpha \,\gamma_5  + \frac{g_2(q^2)}{M_{{\cal N}_1}}\,
\sigma_{\alpha\beta}\,q^\beta \,\gamma_5\,,
\label{hypme}
\eea  
$G_S=G_F V_{us}(G_F V_{ud})$ for $\vert \Delta S \vert =1(\Delta S =0)$  
processes, and the contribution proportional to the electron 
mass has been dropped.  
The analysis by Cabibbo~{\it et al}.~\cite{Cabibbo:2003ea} focuses 
on the experimentally measured decay rates and the measured quantity $g_1/f_1$,
which liberates them from estimating this ratio from theory.  
This is then used with the theoretical values of $f_1$, $f_2$, and $g_2$, 
calculated in the SU(3)-symmetry limit to determine $\vert V_{us} \vert$. 
Deviations from the SU(3)-symmetry limits of these quantities are expected 
to be of varying magnitude. Corrections to $f_1$ are of second order, 
due to the Ademollo-Gatto theorem, but the weak magnetism $f_2$ is not 
protected by this theorem. Likewise, $SU(3)$-breaking effects invalidate 
the usual argument based on the absence of the 
second class currents and $SU(3)$ symmetry, which yields $g_2=0$. No 
precise experimental information is available on $g_2$. Expressing $f_2/f_1$ 
in terms of the anomalous magnetic moments of the neutron and the proton, 
and applying the $SU(3)$-symmetry to the ratio $f_2/M_{{\cal N}_1}$, 
where $M_{{\cal N}_1}$ is 
the mass of the parent hyperon, yields~\cite{Cabibbo:2003ea}:
\bea
&& \vert V_{us} \vert\, (\Lambda \to p e^- \bar{\nu})= 0.2224 \pm 0.0034\,, 
\nonumber\\[1.7mm] 
&& \vert V_{us} \vert\, (\Sigma^- \to n e^- \bar{\nu})= 0.2282 \pm 0.0049\,, 
\nonumber\\[-1.5mm] 
\label{vushyper}\\[-1.5mm] 
&& \vert V_{us} \vert\, (\Xi^- \to \Lambda e^- \bar{\nu}) 
= 0.2367 \pm 0.0099\,, 
\nonumber\\[1.7mm] 
&& \vert V_{us} \vert\, (\Xi^0 \to \Sigma^+ e^- \bar{\nu}) 
= 0.209 \pm 0.027\,, 
\nonumber
\eea
giving an average value
\be
\vert V_{us} \vert_{\rm Hyperon} = 0.2250 \pm 0.0027\,.
\label{vushyperav}
\ee
While this analysis is internally consistent, namely that the values of 
$\vert V_{us} \vert$ returned from the four decays are compatible 
with each other, and this observation is used 
by Cabibbo~{\it et al}.~\cite{Cabibbo:2003ea} to argue that the
data are compatible with the assumption that the residual 
SU(3)-breaking corrections are small, this feature is less transparent 
in model-dependent theoretical studies. It is difficult to quantify in a 
model-independent way the effects of $SU(3)$-breaking in $f_1$ and $f_2$ 
(as well as a non-zero value of $g_2$), which are bound to renormalize 
the value of $\vert V_{us} \vert$. Lattice calculations can clarify the 
theoretical issues involved, assuming that they will reach the required 
precision. Interestingly, the combined value of $\vert V_{us} \vert$ from the
hyperon decays in (\ref{vushyperav}) together with the value of 
$\vert V_{ud} \vert$ in (\ref{vudwa}) leads to perfect agreement with the 
CKM unitarity for the elements in the first row.

Finally, we discuss the novel method advocated by 
Gamiz~{\it et al}.~\cite{Gamiz:2002nu} to determine $\vert 
V_{us} \vert$ from the analysis of the hadronic decays of the $\tau$-lepton
using the spectral function sum rules. In this method, 
$\vert V_{us} \vert$ and $m_s$, the $s$-quark mass, are highly correlated 
and it is difficult to determine both. Since $m_s$ is known from other 
methods, one could fix its value in the current 
range, and optimise the analysis to determine $\vert V_{us} \vert$. 
This is what has 
been done by Gamiz~{\it et al}., which we briefly summarize below.

The starting point of this analysis is the moments $R_\tau^{kl}$ of the  
invariant mass distributions of the final state hadrons in the decay $\tau \to 
\nu_\tau +X$:
\be
R_\tau^{kl} \equiv \int_{0}^{m_\tau^2} ds \,
\left (1-\frac{s}{m_\tau^2}\right )^k\,
\left (\frac{s}{m_\tau^2}\right )^l\, \frac{dR_\tau}{ds}\,.
\label{hadmoment}
\ee
Here $R_\tau^{00}=R_\tau$, with $R_\tau$ defined as follows:
\be
R_\tau\equiv\frac{\Gamma(\tau^- \to hadrons + \nu_\tau(\gamma))}{\Gamma(\tau^- \to 
e^- 
\bar{\nu}_e \nu_\tau(\gamma))}=R_{\tau, V} + R_{\tau, A} + R_{\tau, S}\,,
\ee
where the vector $(V)$, axial-vector $(A)$ and scalar $(S)$ contributions
are indicated, with the scalar contribution coming essentially from the $us$
branch of the decay $\tau^- \to \nu_\tau + \bar{u}s$.

The moments can be expressed in a form in which the dependence 
on $\vert V_{ud} \vert^2$
and $\vert V_{us} \vert^2$ becomes explicit:
\be
R_\tau^{kl}= 3(\vert V_{ud}\vert^2 + \vert V_{us} \vert^2)\Delta_R \left ( 1 +
\delta^{kl(0)} + \sum_{D \geq 2}\, \left (\cos^2\theta_C 
\delta_{ud}^{kl(D)} + \sin^2 \theta_C \delta_{us}^{kl(D)}\right ) \right )\,,
\label{rtaukl}
\ee 
where the short-distance radiative correction $\Delta_R$ has been encountered 
earlier, $\delta^{kl(0)}$ is the perturbative dimension-0 contribution, and
$\delta_{ij}^{kl(D)}\equiv (\delta_{ij,V}^{kl(D)}+ \delta_{ij,A}^{kl(D)})/2$
stand for the average of the vector and axial vector contributions 
to the $(kl)$ moments from dimension 
$D\geq 2$ operators in the operator product expansion of the two-point current 
correlation function governing $\tau$-decays.

Theoretical analysis in the determination of $\vert V_{us}\vert$ 
is carried out 
in terms of the SU(3)-breaking differences defined as:

\vspace*{-2mm}

\be
\delta R_\tau^{kl}\equiv \frac{R_{\tau,V+A}^{kl}}{\vert V_{ud} \vert^2} -
 \frac{R_{\tau,S}^{kl}}{\vert V_{us} \vert^2} = 3\Delta_R\sum_{D\geq 2}
\left (\delta_{ud}^{kl(D)}-\delta_{us}^{kl(D)} \right )\,,
\label{masterf}
\ee
which do not involve the perturbative correction $\delta^{kl(0)}$ and vanish 
in the SU(3) limit.
Concentrating on the $(0,0)$ moment for the analysis, for which Gamiz {\it et 
al}.~\cite{Gamiz:2002nu} calculate 
$\delta R_{\tau}=0.229 \pm 0.030$ for the r.h.s. of the above equation,
using the 
experimental input~\cite{Davier:2002mn} $R_\tau=3.642 \pm 0.012$
and $R_{\tau, S}=0.1625 \pm 0.0066$, and invoking 
CKM unitarity to express $\vert V_{ud}\vert$ in terms of $\vert V_{us} \vert$, 
yields~~\cite{Gamiz:2002nu}
\be
\vert V_{us} \vert_{\tau-{\rm decays}} = 0.2179 \pm 0.0044({\rm exp}) \pm 
0.0009({\rm th})
=0.2179 \pm 0.0045\,.
\label{vusgamiz}
\ee 
The first error is the experimental uncertainty due to the measured values of
$R_{\tau}$ and $R_{\tau, S}$, which is the dominant error at present but can be
greatly reduced if the $B$-factory data on $\tau$ decays is brought to bear on this 
problem, and the second error stems from the theoretical error in the calculation 
of $\delta R_\tau$, which is dominated by the assumed value for the $s$-quark mass: 
$m_s (2~{\rm GeV})=105 \pm 20$ MeV, and should also decrease in future as the
$s$-quark mass gets determined more precisely. While the current error on $\vert 
V_{us} \vert$ from $\tau$-decays is approximately a factor 2 larger at present than 
the corresponding error on this quantity  
from the $K_{\ell 3}$ analysis, potentially $\tau$-decays may provide a very 
competitive measurement of $\vert V_{us} \vert$. Of course, we also expect 
substantial progress on the $K_{\ell 3}$ front from the ongoing experiments.
 
The present status of $\vert V_{us} \vert$ is summarized 
in~Fig.~\ref{figvusall},
and is based on the following five measurements: 
(i)  From the old $K_{\ell 3}$ data,
(ii) from the $K^+_{e3}$ measurements by the BNL-E865 collaboration,
(iii) from the KLOE data on $K_S$ decays (still preliminary),
(iv) from hyperon semileptonic decays, and
(v) from $\tau$-decays. The current world average based on
these measurements
\be
 \vert V_{us} \vert_{\rm WA}  =0.2224 \pm 0.0017\,,
\label{vuswa}
\ee
is also shown in this figure. We have added the statistical and 
systematic errors 
in quadrature. As not all the measurements are compatible with each 
other, we have used a scale factor of 1.3 in quoting the error. The resulting 
value of $\vert V_{us} \vert$ is somewhat larger but compatible with the 
corresponding PDG value, $\vert V_{us} \vert_{\rm PDG}=0.2196 \pm 0.0026$.
%
%
\begin{figure}[htbp]
\vspace{10pt}
\centerline{\psfig{width=7.5cm,file=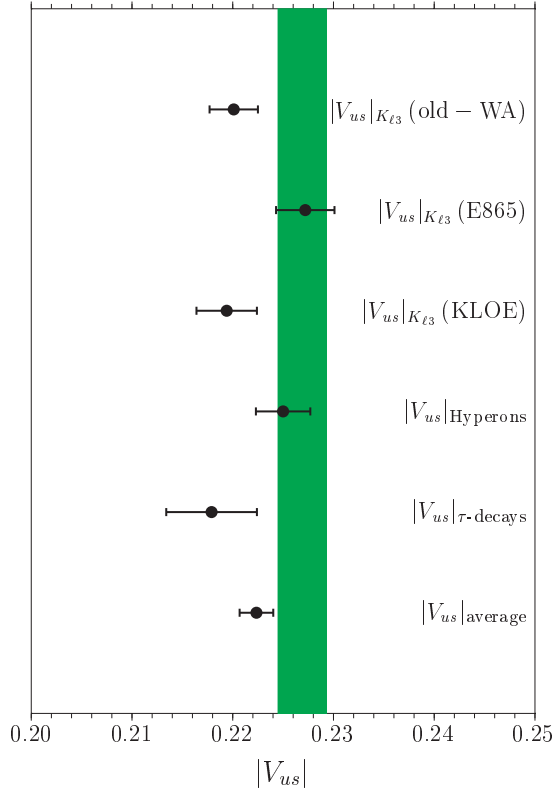}}
\caption{Present status of $\vert V_{us} \vert$ measured in various 
experiments. Note that the KLOE entry is based on $K_S$ decays.
The band results from the current measurements of 
$\vert V_{ud} \vert$ and $\vert V_{ub} \vert$, and imposing the 
CKM unitarity constraint.}
\label{figvusall}
\vskip -0.2cm
\end{figure}

\subsection{Unitarity constraint for the first row in $V_{\rm CKM}$}
In discussing the test of the CKM unitarity in the first row, we use the 
current world average of the matrix elements $\vert V_{ud} \vert~=0.9739 
\pm 0.0005$ to determine from the unitarity constraint a value for $\vert V_{us} 
\vert$:
\be
\vert V_{us} \vert_{\rm unit}=0.2269 \pm 0.0024\,.
\label{vuswall}
\ee
This is shown as a vertical band in Fig.~\ref{figvusall}. The  current 
world average $\vert V_{us} \vert_{\rm WA}$ from direct measurements 
(\ref{vuswa}) differs from its value $\vert V_{us} \vert_{\rm unit}$
by 1.5$\sigma$. 
In this mismatch, $\vert V_{ub}\vert$ plays no role, as its current 
value $\vert V_{ub}\vert=(3.80 ^{+0.24}_{-0.33} \pm 0.45)
\times 10^{-3}$~\cite{Schubert:2003} is too small.

 However, in averaging the
value of $\vert V_{us} \vert$, if one leaves out the entries from the 
BNL-E865 
and Hyperon data, the former on the grounds of being at variance with the PDG 
value for BR$(K^+_{\ell 3})$, and the latter due 
to the neglect of the SU(3)-breaking corrections in some of the form factors, 
which can only be estimated in model-dependent ways, then the resulting world 
average goes down, yielding  $\vert V_{us} \vert=0.2195 \pm 
0.0017$. The error is almost the same as the one shown in (\ref{vuswa}), as the 
remaining measurements are compatible with each other, and hence do not require a 
scale factor $(S=1)$. This value of $\vert V_{us} \vert$, together with  
$\vert V_{ud} \vert$ given above, yields $\Delta_1 =(3.3 \pm 1.3) \times 
10^{-3}$, corresponding to a 2.5$\sigma$ violation of unitarity in the first row 
of the CKM matrix.  Tentatively, we conclude that the 
deviation of $\Delta_1$ from zero is currently not established at a significant 
level. We look forward to the forthcoming 
results from the KLOE collaboration (as well as from NA48), on $\tau$-decays from 
the B-factories, and also
improved measurements of $g_A/g_V$ in polarized neutron beta-decays, 
which will yield more precise measurements of $\vert V_{us} \vert$ and $\vert 
V_{ud}\vert$, enabling us to undertake a definitive test of the unitarity 
involving the first row of the CKM matrix.

\section{Current Status of $\vert V_{cd}\vert$ and $\vert V_{cs}\vert$} 
Concerning the determination of $\vert V_{cd} \vert$, nothing much has 
happened during the last decade! Current value of this matrix 
element is deduced from neutrino and antineutrino production of charm off 
valence $d$ quarks in a nucleon, with the basic process being $\nu_\mu d 
\to \mu^- c$,
followed by the semileptonic charm quark decay $c \to s \mu^+ \nu_\mu$,
and the charge conjugated processes involving an initial $\bar{\nu_\mu}$ 
beam. Then, using the~relation
\be
\frac{ \sigma(\nu_\mu \to \mu^+ \mu^-) - \sigma(\bar{\nu}_\mu \to \mu^+ 
\mu^-)}{\sigma(\nu_\mu \to \mu^-) - \sigma(\bar{\nu}_\mu \to \mu^+)} 
=\frac{3}{2}\, {\cal B}(c \to \mu^+X)\, \vert V_{cd} \vert^2\,,
\ee
one obtains $\vert V_{cd} \vert$ from the current average of
the l.h.s., quoted by the PDG 
~\cite{Hagiwara:fs}  as $(0.49 \pm 0.05) \times 10^{-2}$, 
and ${\cal B}(c \to \mu^+X)$, for which the PDG average is ${\cal B}(c 
\to \mu^+X)=0.099 \pm 0.012$, yielding
\be
\vert V_{cd} \vert = 0.224 \pm 0.016\,.
\label{pdgvcd}
\ee
Compared to $\vert V_{us} \vert$, the precision on $\vert V_{cd} \vert$ is
not very impressive, with $\delta \vert V_{cd} \vert/\vert V_{cd} \vert 
=7\%$.

\vspace*{1mm}

Concerning $\vert V_{cs} \vert$, three methods have been used in its 
determination:
\begin{enumerate}
\item Semileptonic decays $D \to K \ell^+ \nu_\ell$,
\item Decays of real $W^\pm$ at LEP: $W^+ \to c \bar{s} (g)$ and $W^- \to \bar{c} s 
(g)$,
\item Measurement of the ratio $\Gamma(W^\pm \to hadrons)/\Gamma(W^\pm \to 
\ell^\pm \nu_\ell)$.
\end{enumerate}
We briefly discuss them in turn.\\

\noindent

\underline{\sl $\vert V_{cs} \vert$ from $D \to K \ell^+ \nu_\ell$:} \\

 This makes use of the following relation:
\be
\Gamma(D \to K e^+ \nu_e) =\frac{{\cal B}(D \to K e^+ 
\nu_e)}{\tau_D} =\frac{G_F^2 \vert V_{cs} \vert^2}{192 \pi^3}\, \Phi 
\, \vert f_+(0)\vert^2\, ( 1 + \delta_R)\,,
\ee
where $\Phi$ is the phase space factor, $f_+(0)$ is the dominant form factor 
in the $D_{\ell 3}$ decay evaluated at $q^2=0$ (requiring extrapolation of 
data to $q^2=0$), and $\delta_R$ arises from the $q^2$-dependence of 
this form factor. Using~\cite{Aliev:1989mr} $f_+(0)=0.7\pm 0.1$, coming from the early 
epoch of the QCD sum rules, the earlier version of the
PDG CKM review~\cite{Gilman:sb} quotes a value $\vert V_{cs} \vert=1.04\pm 
0.16$. The decays $D \to K \ell \nu_\ell$ and
$D \to K^* \ell \nu_\ell$ have also received quite a lot of attention by the
lattice groups in the past. There is almost a decade old result from the UKQCD
collaboration~\cite{Bowler:1994zr}, $f_+(0)=0.67^{+0.07}_{-0.08}$, and a relatively
recent result by the Rome Lattice group~\cite{Abada:2000ty},
$f_+(0)=0.77\pm 0.04^{+0.01}_{-0.0}$. In fact, the lattice technology has advanced
to a point where precision calculations of the relevant form factors in
$D \to K$ and $D \to K^*$ can be undertaken. On the experimental side, one already
has a measurement of the form factor in $D \to K$, and ratios of the form factors in $D
\to K^* \ell \nu_\ell$ have now been measured with good accuracy, most recently by the 
FOCUS photoproduction experiment at Fermilab~\cite{Link:2002wg}. However,
in the current version of the PDG review~\cite{Hagiwara:fs}, the determination of 
$\vert V_{cs} \vert$ from $D \to K$ transitions has been dropped. We hope that in
future, with improved theory and experiments, this value judgement on the part of the 
PDG will be revised.\\

\noindent
\underline{\sl $\vert V_{cs} \vert$ from $W^+ \to c \bar{s} (g)$ 
and $W^- \to \bar{c} s 
(g)$}:\\

  This method involves the process $e^+ e^- \to W^+ 
W^-$, well measured at CERN, and subsequent charmed-tagged $W^\pm$-decays. 
The ratio
\be
r^{(cs)}= \Gamma(W^+ \to c \bar{s})/\Gamma(W^+ \to {\rm 
hadrons})\,,
\ee
then allows to determine $\vert V_{cs}\vert$. The weighted average of the
ALEPH~\cite{Abreu:1998ap} and DELPHI~\cite{Barate:1999ud} measurements 
yields~\cite{Hagiwara:fs}
\be
\vert V_{cs} \vert =0.97 \pm 0.09\,({\rm stat.)}
\pm 0.07\, ({\rm syst.})\,.
\ee
The current precision on $\vert V_{cs} \vert$
from direct $W^\pm$-measurements is $\delta \vert V_{cs} \vert/\vert 
V_{cs} \vert =11\%$.\\ 

\noindent
\underline{\sl Measurement of the ratio $\Gamma(W^\pm \to 
hadrons)/\Gamma(W^\pm \to \ell^+ \nu_\ell)$}:\\

 A tighter determination of $\vert V_{cs} \vert$ follows 
from the ratio of the hadronic $W$ decays to leptonic decays, which has 
been measured at LEP. Using the relation
\be
\frac{1}{{\cal B}(W \to \ell \nu_\ell)}= 3 \left ( 1 + 
\frac{\alpha_s(M_W^2)}{\pi}\, \sum_{i=u,c; j=(d,s,b)} \vert V_{ij} 
\vert^2 \right ) \,,
\ee
yields~\cite{Abbaneo:2001ix}
\be
\sum \vert V_{ij}\vert^2=2.039 \pm 0.025\, ({\cal B}(W \to \ell \nu_\ell))
\pm 0.001\, (\alpha_s)\,,
\label{wunit}
\ee  
and gives (on using the known values of the other matrix elements)
\be
\vert V_{cs} \vert=0.996 \pm 0.013\,.
\ee
The measurement (\ref{wunit}) provides a quantitative test of the CKM 
unitarity involving the first two rows of the CKM matrix. This amounts
to a violation of unitarity by $1.6\sigma$, 
and hence is statistically not~significant. 

The matrix elements $\vert V_{cd} \vert$ and $\vert V_{cs} \vert$ will be 
measured very precisely in the decays $D \to K\ell \nu_\ell$ and $D \to 
\pi \ell \nu_\ell$ by the CLEO-C and BES-III experiments,
with anticipated integrated luminosity of $3~{\rm fb}^{-1}$ and
$30~{\rm fb}^{-1}$ at $\psi(3770)$, respectively. These experiments will 
also allow, for the first time, a complete set of measurements in $D \to 
(K,K^*) \ell \nu_\ell$ and $D \to (\pi, \rho) \ell \nu_\ell$ of the 
magnitude and slopes of the form factors to a few per cent level. 
From a theoretical 
point of view, this is an area where the Lattice-QCD techniques can be
reliably applied to enable a very precise determinations of the matrix 
elements $\vert V_{cs} \vert$ and $\vert V_{cd} \vert$. Typical
projections~\cite{Stoeck:2003} 
at the CLEO-C are: $\delta \vert V_{cs}\vert/\vert V_{cs} \vert =1.6\%$
and a similar precision on $\delta \vert V_{cd}\vert/\vert V_{cd} \vert$. 

\section{Present Status of $\vert V_{cb}\vert$ and $\vert V_{ub} \vert$}
 The matrix elements $\vert 
V_{cb} \vert$ and $\vert V_{ub} \vert$  
play a central role in the quantitative tests of the CKM theory in 
current experiments. In particular, these matrix elements enter in the 
following unitarity relation

\beq
V_{ud} V_{ub}^* + V_{cd} V_{cb}^* + V_{td} V_{tb}^* = 0\,.
\label{vudtriangle}
\eeq
This is a triangle relation in the complex plane 
(i.e.\ $\bar{\rho}$--$\bar{\eta}$ space). The three angles of this 
triangle are defined as:
\be
\alpha \equiv \arg\left(-\frac{V_{tb}^* V_{td}}{V_{ub}^*V_{ud}}\right)\,,
\qquad
\beta \equiv \arg\left(-\frac{V_{cb}^* V_{cd}}{V_{tb}^*V_{td}}\right)\,,
\qquad
\gamma \equiv \arg\left(-\frac{V_{ub}^* V_{ud}}{V_{cb}^*V_{cd}}\right)\,.
\label{abgamma}
\ee
The BELLE convention for these phases is: $\phi_2=\alpha$, $\phi_1=\beta$ and 
$\phi_3=\gamma$. In the Wolfenstein parametrization given above, the matrix 
elements $V_{ud}$, $V_{cd}$, $V_{cb}$ and $V_{tb}$ entering in the above 
relations are real, to $O(\lambda^3)$. Hence, the angles 
$\beta$ and $\gamma$ have a simple interpretation: They are the phases 
of the matrix elements $V_{td}$ and $V_{ub}$, respectively:
\be
V_{td}=\vert V_{td}\vert e^{-i\beta}\,, \qquad 
V_{ub}=\vert V_{ub} \vert e^{-i \gamma}\,;
\label{betagamma}
\ee
and the phase $\alpha$ defined by the triangle relation: $\alpha=\pi-\beta 
-\gamma$. The unitarity relation (\ref{vudtriangle}) can be written as
\be
R_b {\rm e}^{i\gamma} + R_t {\rm e}^{-i \beta} =1\,,
\label{trianglerel} 
\ee
where
\bea
R_b & \equiv & \frac{\vert V_{ub}^* V_{ud} \vert}{\vert V_{cb}^*V_{cd}\vert} 
=\sqrt{\bar{\rho}^2 + \bar{\eta}^2} =\left (1-
\frac{\lambda^2}{2}\right )\, \frac{1}{\lambda} \,\left \vert 
\frac{V_{ub}}{V_{cb}} \right \vert\,,
\nonumber\\[-1.5mm] 
\label{rbrtdef}\\[-1.5mm] 
R_t &\equiv& \frac{\vert V_{tb}^* V_{td} \vert}{\vert V_{cb}^*V_{cd}\vert} 
=\sqrt{(1-\bar{\rho})^2 + \bar{\eta}^2}=\frac{1}{\lambda} \,
\left \vert 
\frac{V_{td}}{V_{cb}} \right \vert\,.
\nonumber
\eea
The unitarity triangle with unit base, and the other two sides 
given by $R_b$ and 
$R_t$, and the apex defined by the coordinates $(\bar{\rho}, \bar{\eta})$  
is shown in Fig.~\ref{triangle}.
Quantitative tests of the CKM unitarity, being carried out at the 
$B$ factories, consists of determining the sides of this 
triangle through the measurements of 
$\vert V_{ub}\vert$, $\vert V_{cb}\vert$ and 
$\vert V_{td}\vert$ as precisely as possible, which allows to
determine {\it indirectly} the three inner angles $\alpha$, $\beta$,
$\gamma$, and  confronting this information with the {\it direct}
 measurements of the three angles $(\alpha, \beta, \gamma)$ 
(or $\phi_1, \phi_2, \phi_3$) through the CP-violating 
asymmetries. We will return to a quantitative discussion of these tests 
in Section~6.
%
\begin{figure}[htbp]
\centerline{\psfig{width=0.60\textwidth,file=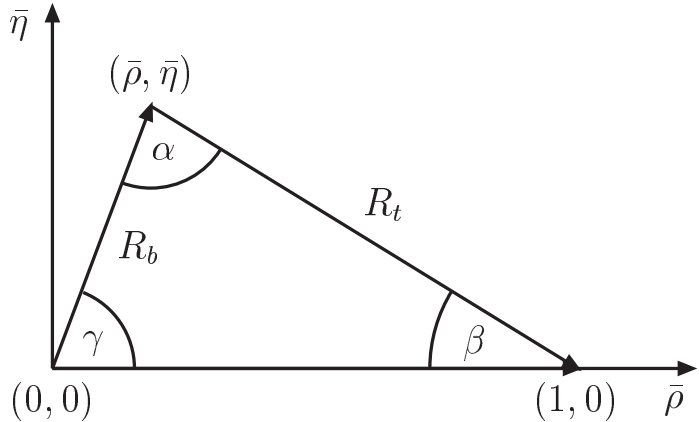}}
\caption{The unitarity triangle with unit base 
in the $\bar{\rho}$ - $\bar{\eta}$ plane. The angles 
$\alpha$, $\beta$ and $\gamma$ are defined in (\ref{abgamma}) 
and the two sides $R_b$ and 
$R_t$ are defined in (\ref{rbrtdef}).}
\label{triangle}
\end{figure}

Current status of the matrix elements $\vert V_{cb}\vert$ and $\vert V_{ub}
\vert$ has been discussed in the literature in great detail. These include 
proceedings of the research workshops and conferences on flavour physics held
recently~\cite{Battaglia:2003in,ckm2003,fpcp2003}, in particular, the
experimental reviews by Gibbons~\cite{Gibbons:2003gq}, 
Thorndike~\cite{Thorndike:2003}, and Calvi~\cite{Calvi:2003-Talk}, 
and theoretical developments reviewed by Luke~\cite{Luke:2003nu}, 
Ligeti~\cite{Ligeti:2003hp}, Lellouch~\cite{Lellouch:2003},
and Uraltsev~\cite{Uraltsev:2003uq}. Updated data are available on
the webcites of the working groups established to perform the averages of the
experimental results in flavour physics~\cite{hfag03} and the CKM
matrix elements~\cite{Hocker:2001xe}. We shall make extensive use of these 
resources, focusing on some of the principal results achieved, 
and discuss their theoretical underpinnings.

\subsection{Present status of $\vert V_{cb}\vert$}
Measurements of $\vert V_{cb} \vert$ are based essentially on the 
semileptonic decays of the $b$-quark $b \to c \ell \nu_\ell$. In the 
experiments, one measures hadrons, and hence the
inclusive hadronic states $B \to X_c \ell \nu_\ell$ and some 
selected exclusive states, such as $B \to (D,D^*) \ell \nu_\ell$,
is as close as one gets to the underlying partonic weak transition. In the 
interpretation of data, QCD is intimately involved. 
In fact, quantitative studies of heavy mesons, in particular $B$-mesons, 
have led to novel applications  of QCD, of which HQET~\cite{Manohar:dt} in 
its various incarnations is at the forefront. Semileptonic 
$B$-decays  have also received a great deal of theoretical attention in 
methods which involve non-perturbative techniques, foremost among them 
are the QCD sum rules~\cite{Colangelo:2000dp} 
and Lattice QCD~\cite{Lattice:2002}. We shall restrict the
theoretical discussion to these frameworks. 

\subsection{Determination of $\vert V_{cb} \vert$ 
from inclusive decays $B \to X_c \ell \nu_\ell$}
The theoretical framework to study inclusive decays is based on the operator 
product expansion (OPE), which allows to calculate the decay rates in terms 
of a perturbation series in $\alpha_s$ and power corrections in $\Lambda_{\rm 
QCD}/m_b$ (and $\Lambda_{\rm QCD}/m_c$). This tacitly assumes quark-hadron 
duality, which is supposed to hold for inclusive decays and also for partial 
decay rates and distributions, if summed over sufficiently large 
intervals ($\gg \Lambda_{\rm QCD}$), and weighted distributions (moments). 
Deviations from this duality are, however, hard to quantify, and they will be 
the limiting factor in theoretical precision ultimately.
The first term of this QCD corrected series is the parton model result for the 
decay $b \to c \ell nu_\ell$. Leading $O(\alpha_s)$ corrections 
were obtained some time ago for inclusive decay rates~\cite{Cabibbo:1978sw} 
and lepton energy 
distribution~\cite{Ali:1979is,Altarelli:1982kh,Jezabek:1988iv}. In 
the meanwhile, the QCD perturbative corrections to the decay rates are 
known up to order $\alpha_s^2\beta_0$~\cite{Luke:1994yc}, 
where $\beta_0=11-2n_f/3$, with $n_f$ being the number of active quarks. 
This term usually dominates the $O(\alpha_s^2)$ corrections, though there 
exists at least one counter example, namely the inclusive decay width 
$\Gamma(B \to X_s \gamma)$, where the contribution to the width in
$\alpha_s^2\beta_0$ is small~\cite{Bieri:2003ue}.  
The result in the $\overline{\rm MS}$ scheme is~\cite{Luke:1994yc}
\be
\Gamma (b \to ce \bar{\nu}_e (+ g)) = \vert V_{cb}\vert^2 \,
\frac{G_F^2m_b^5}{192 \pi^3} \, 
0.52 \, \bigg[1- 1.67 \, (\frac{\bar{\alpha}_s(m_b)}{\pi})
- 15.1 \, (\frac{\bar{\alpha}_s(m_b)}{\pi})^2\bigg]\,,
\ee
where the numerical coefficients correspond to the choice $m_c/m_b=0.3$.

The leptonic and hadronic distributions in $B$ decays 
are now calculated using techniques based on the OPE.
While the distributions themselves are not calculable from first principles and
invariably involve models, called the shape functions, inclusive decay rates, 
partially integrated spectra, and moments are calculable in the OPE approach in terms of 
the matrix elements of higher (than four) dimension operators.

The book-keeping of the power corrections is as follows. The leading $\Lambda/m_b$
correction to the decay rates vanishes in the heavy quark limit~\cite{CGG}, and the
$O(\Lambda^2/m_b^2)$ effects can be  parametrized in terms of two parameters
$\lambda_1$ and $\lambda_2$, defined as~\cite{incl,MaWi,Blok:1993va,Mannel:1993su}:
\be
\lambda_1=\frac{1}{2 m_B} \langle B(v)|\bar{b}_v (iD)^2 b_v|B(v)\rangle\,,
\qquad
\lambda_2=\frac{1}{6m_B} \langle B(v)|\bar{b}_v \frac{g_s}{2}
\sigma_{\mu\nu}G^{\mu\nu} b_v|B(v)\rangle\,,
\label{lamda12}
\ee
where $b_v$ denotes the $b$ quark field in HQET, with $D_\mu$ and $G_{\mu\nu}$
being the covariant derivative and the QCD field strength tensor, respectively.
While $\lambda_1$ is not known precisely, having a value typically in the range
$\lambda_1= [-0.1, -0.5]~{\rm GeV}^2$, but $\lambda_2$ is known from the 
$B^* -B$ mass difference to be $\lambda_2(m_b)\simeq 0.12~{\rm GeV}^2$.
Corrections of order $\alpha_s/m_b^2$ are still not available, 
but $O(1/m_b^3)$ corrections to the decay widths have been 
calculated~\cite{Gremm:1996df}. They are expressed in 
terms of six additional non-perturbative parameters, $\rho_1$, $\rho_2$,
and ${\cal T}_i$, $i=1,...,4$. There are two constraints on these six 
parameters, which reduces the number of free parameters relevant for 
$B$-decays in this order to four. So, including the $O(1/m_b^3)$ terms, 
there are in all six non-perturbative parameters which have to be determined 
from experimental analysis. They can be determined, or at least bounded, from
the already measured lepton- and hadron-energy moments in 
$B \to X_c \ell \nu_\ell$ and 
photon-energy moments in the inclusive decay $B \to X_s \gamma$.
 
Theoretical results for the inclusive rate and moments depend on the scheme 
for defining the $b$-quark mass. They influence the decay rates more, 
in particular the branching ratio ${\cal B}(B \to X_s \gamma)$, where 
the scheme-dependence of $m_b$ and $m_c$ is currently the largest 
theoretical uncertainty, typically of order $10\%$~\cite{Gambino:2001ew}, 
but the moments are less sensitive. We shall 
discuss here the results in the so-called $\Upsilon(1S)$ 
scheme~\cite{upsexp,Hoang:1999zc} ($m_b^{1S}$ denotes the $b$-quark mass 
in this scheme and $m_\Upsilon$ is the $\Upsilon$-meson mass),
as this scheme is {en vogue} in the analyses of the moments  
by experimental groups. Taking into account the perturbative corrections 
to order $\alpha_s^2 \beta_0$ and power corrections to order $1/m_b^3$, 
the result for the semileptonic decay width $\Gamma(B \to 
X_c \ell \nu_\ell)$ in $\Upsilon(1S)$ mass scheme reads as 
follows~\cite{Bauer:2002sh} 
\bea
& & \Gamma(B \to X_c \ell \nu_\ell)=\frac{G_F^2 \vert V_{cb} \vert^2}
{192\pi^3} \left (\frac{m_\Upsilon}{2}\right )^5
\bigg[0.534 -0.232\, \Lambda -0.023\, \Lambda^2 
-0.11\, \lambda_1 
-0.15\, \lambda_2 - 0.02 \, \lambda_1\Lambda + 
\label{bauersldecay} \\
&+& 0.05\, \lambda_2\Lambda
- 0.02\, \rho_1 + 0.03 \, \rho_2 - 0.05 \,{\cal T}_1 + 0.01 \,{\cal T}_2
-0.07\,  {\cal T}_3 
- 0.03\, {\cal T}_4 
-0.051\, \epsilon -0.016\, \epsilon_{\rm BLM}^2 
+0.016\, \epsilon \Lambda\bigg]\,,
\nonumber
\eea
where $\epsilon$ and $\epsilon_{\rm BLM}$ are parameters in the perturbative part,
entering through the $\alpha_s^2 \beta_0$ term~\cite{Luke:1994yc};
the parameter denoted by $\Lambda$ is called $\bar{\Lambda}=m_B-m_b$
in the HQET jargon and is defined above in terms of the $\Upsilon (1S)$ mass 
by $\Lambda=m_\Upsilon/2-m_b^{1S}$. The corresponding
expressions for the decay widths in other schemes for the $b$-quark mass
can be seen in the paper by Bauer et al.~\cite{Bauer:2002sh}.  

Determinations of $\vert V_{cb} \vert$ from this method are based on the 
analysis of the following measures. For the lepton energy spectrum, partial 
rates and  
moments are defined by cuts on the lepton energy ($E_\ell > E_0,E_1$):
\be
R_0(E_0,E_1)=\frac{\int_{E_1}\frac{d\Gamma}{dE_\ell} dE_\ell}
{\int_{E_0} \frac{d\Gamma}{dE_\ell}dE_\ell}\,, \qquad 
R_n(E_0) = \frac{\int_{E_0} E_\ell^n\frac{d\Gamma}{dE_\ell} dE_\ell}
 {\int_{E_0} \frac{d\Gamma}{dE_\ell}dE_\ell}\,,
\ee
where $d\Gamma/dE_\ell$ is the charged lepton spectrum in the $B$ rest frame. 
The moments $R_n$ are known to order $\alpha_s^2\beta_0$~\cite{Gremm:1996gg} 
and $\Lambda_{\rm QCD}^3/m_b^3$~\cite{Gremm:1996df}. For the hadronic moments, 
the quantities analyzed are the mean hadron invariant mass and its variance,
both with lepton energy cuts 
$E_0$: $S_1(E_0) = \langle m_X^2-\bar{m}_D^2\rangle |_{E_\ell > E_0}$ and 
$S_2(E_0)=\langle (m_X^2 - \langle m_X^2\rangle)^2\rangle|_{E_\ell > 
E_0}$, where $\bar{m}_D=(m_D + 3 m_{D^*})/4$. $S_n$ are known to order 
$\alpha_s^2\beta_0$~\cite{Falk:1997jq} and $\Lambda_{\rm 
QCD}^3/m_b^3$~\cite{Gremm:1996df}. For the decay $B \to X_s \gamma$, the mean 
photon energy and variance with a photon energy cut, $E_\gamma > E_0$,
calculated in the $B$ rest frame, have been 
used: $T_1(E_0) =\langle E_\gamma \rangle|_{E_\gamma > E_0}$ and
$T_2(E_0) =\langle (E_\gamma -\langle E_\gamma \rangle)^2\rangle|_{E_\gamma > 
E_0}$. Also, $T_{1,2}$ are known to order 
$\alpha_s^2\beta_0$~\cite{Ligeti:1999ea}
and $\Lambda_{\rm QCD}^3/m_b^3$~\cite{Bauer:1997fe}.

This theoretical framework has been used by the CLEO~\cite{Mahmood:2002tt},
BABAR~\cite{Aubert:2003dr} and DELPHI~\cite{delphivcb} collaborations, 
and their 
results for $ \vert V_{cb} \vert$ using the moment analysis are as follows:

\vspace*{-3mm}

\bea
\vert V_{cb} \vert &=& (40.8 \pm 0.6 \pm 0.9) \times 10^{-3} \qquad [{\rm 
CLEO}]\,,
\nonumber\\[0.5mm]
\vert V_{cb} \vert &=& (42.1 \pm 1.0 \pm 0.7) \times 10^{-3} \qquad [{\rm
BABAR}]\,,\
\label{vcbinclexp} \\[0.5mm]
\vert V_{cb} \vert &=& (42.4 \pm 0.6 \pm 0.9) \times 10^{-3} \qquad [{\rm
DELPHI}]\,.
\nonumber
\eea

\noindent
The moments (and hence the values for $\vert V_{cb} \vert$) are strongly correlated with 
$m_b$, and are scheme-dependent. This aspect should not be missed in comparing or
combining various determinations of $\vert V_{cb} \vert$. They can be averaged 
to get  $\vert V_{cb} \vert$ from inclusive decays
\be
\vert V_{cb} \vert_{\rm incl}= (42.1 \pm 0.7_{\rm exp} \pm 0.9_{\rm theo}) \times 
10^{-3}\,,
\label{vcbinclsch}
\ee 
where we have kept the theoretical error as $0.9$, which is an 
underestimate as no account is taken of the duality error, probably not 
negligible at this precision. 
A value very similar to the above results was obtained by
Bauer et al.~\cite{Bauer:2002sh}, using the CLEO 
data~\cite{Cronin-Hennessy:2001fk,Briere:2002hw,Chen:2001fj}, the earlier BABAR 
data~\cite{Aubert:2002pm}, and data from the DELPHI 
Collaboration~\cite{delphi2002}:
\be
\vert V_{cb} \vert_{\rm incl} = (40.8 \pm 0.9) \times 10^{-3}~~[{\rm Bauer}~et 
~al.]\,.
\ee

A mismatch 
between the earlier BABAR measurements~\cite{Aubert:2002pm} of the hadron 
invariant mass spectrum presented as a function of the lepton energy cut $(E_0)$ and 
the corresponding OPE-based theoretical analysis of Bauer et al.~\cite{Bauer:2002sh} is  
now largely gone. The updated BABAR~\cite{Aubert:2003dr} and CLEO~\cite{Mahmood:2002tt}
data are in agreement with each other and with the OPE-based
theory. This is depicted in Fig.~\ref{CLEO-momvsth} showing the hadron moment $\langle 
M_X^2-\bar{M}_D^2 \rangle$ vs. lepton energy cut. Theory bands taking into account 
the variations in the input parameters are also given and the details
of the analysis can be seen in the CLEO paper~\cite{Huang:2003ay}.
 
Mutual consistency of the experiments in terms of $\vert V_{cb} \vert$ and $m_b^{1S}$  
is shown in Fig.~\ref{babarlephad}. The contours represent  
the best fits ($\Delta \chi^2=1$) for the hadron moments (from the
BABAR~\cite{Aubert:2003dr}, CLEO~\cite{Briere:2002hw} and DELPHI~\cite{Calvi:2002wc} 
data) and lepton moments (from 
the CLEO~\cite{Mahmood:2002tt} and DELPHI~\cite{Calvi:2002wc} data).
One observes from these correlations that there is still some residual difference 
between the best fit contours resulting from the analysis of the lepton- and 
hadron-energy moments. The current mismatch remains to be clarified in future 
experimental and theoretical analyses. Once the experimental issues are resolved,
the $(\vert V_{cb}\vert$ - $m_b^{1S})$ correlation from the hadron and lepton moments can 
be used to quantify the quark-hadron duality violation.

\begin{figure}
\begin{center}
\vspace{10pt}
\includegraphics[width=9.5cm]{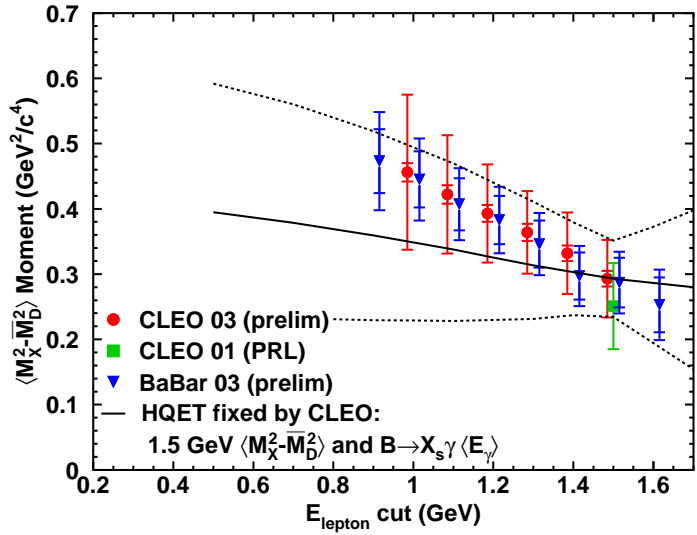}
\vskip 0.2cm
\caption{Comparison of BABAR '03~\protect\cite{Aubert:2003dr},
 CLEO '03~\protect\cite{Huang:2003ay} and 
CLEO '01 ~\protect\cite{Cronin-Hennessy:2001fk} measurements of the hadron 
invariant mass moment
$\langle M_X^2 -\bar{M}_D^2 \rangle$ vs.lepton energy cut. The bands show the
parametric uncertainties from theory~\protect\cite{Bauer:2002sh}.
(Figure taken from the CLEO paper~\protect\cite{Huang:2003ay}\,.)}
 \label{CLEO-momvsth}
\end{center}
\end{figure}
%

\begin{figure}
\begin{center}
\vspace{10pt}
\includegraphics[width=9.5cm]{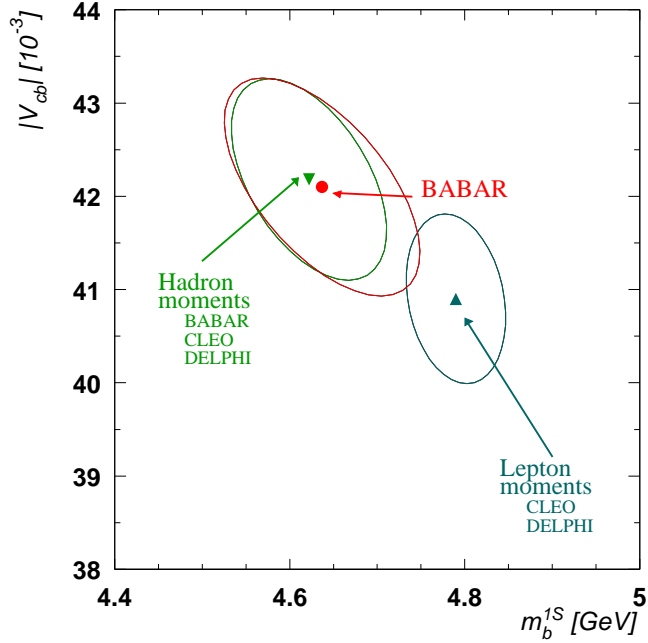}
\vskip 0.2cm
\caption{Constraints on the HQET parameters obtained from hadronic moments as measured by 
BABAR~\protect\cite{Aubert:2003dr}, hadronic moments combined 
(BABAR~\protect\cite{Aubert:2003dr}, CLEO~\protect\cite{Briere:2002hw}, and
~DELPHI~\protect\cite{Calvi:2002wc}) and the combined lepton energy 
moments (CLEO~\protect\cite{Mahmood:2002tt} and DELPHI~\protect\cite{Calvi:2002wc}) in 
the $(\vert V_{cb} \vert, m_b^{1S})$ plane. (Figure taken from 
the BABAR paper~\protect\cite{Aubert:2003dr}\,.)}
 \label{babarlephad}
\end{center}
\end{figure}
\subsection{Determination of $\vert V_{cb} \vert$ from exclusive decays $B \to 
(D,D^*) \ell \nu_\ell$}
In exclusive decays, $ B \to (D,D^*) \ell \nu_\ell$, one needs to know the 
hadronic matrix elements of the charged
weak current, $\langle D| J_\mu | B \rangle$ and $\langle D^*| J_\mu | B \rangle$.
The former involves two form factors, called $F_{+}(q^2)$ and $F_{0}(q^2)$,
and for the transition to the $D^*$-meson, one has four such form factors, called in
the literature by the symbols $V(q^2)$, $A_i(q^2)$, $i=1,2,3$. If these form 
factors can be measured over a large-enough range of $q^2$ and the same can be
obtained from a first principle calculation, such as lattice QCD, then 
exclusive decays would provide the best determination of $\vert V_{cb} \vert$. 
In the absence of a first principle calculation of these form factors, HQET 
provides a big help in that the heavy quark symmetries in HQET allow to 
reduce the number of independent form factors from six in the decays 
at hand to just one,
called the Isgur-Wise (IW) function~\cite{Isgur:vq}  
${\cal F}(\omega)$, where $\omega =v.v^\prime$, with $v$ and $v^\prime$ 
being the four-velocities of the $B$ and $D(D^*)$ meson, respectively. 
Moreover, HQET provides a normalization of the IW function at the symmetry 
point, $\omega =1$. A lot of attention has been paid to the decay 
$B \to D^* \ell \nu_\ell$ due to Luke's theorem~\cite{Luke:1990eg}, which
states that symmetry-breaking corrections to ${\cal F}(\omega=1)$ are of 
second order, a situation very much akin to the Ademollo-Gatto theorem for 
the $K_{\ell 3}$ form factor $f_{+}(0)$ discussed earlier.

The differential decay rate for $B \to D^* \ell \nu_\ell$ can be written as
\be
\frac{d\Gamma}{d\omega} =\frac{G_F^2}{4 \pi^3} 
\left (\omega^2 -1 \right )^{1/2} \, m_{D^*}^3 \, 
(m_B - m_{D^*})^2 \,
{\cal G}(\omega) \, \vert V_{cb} \vert^2 \, \vert {\cal F} (\omega) \vert^2\,,
\label{dgammaomeg}
\ee
where ${\cal G}(\omega)$ is a phase space factor with ${\cal G}(1)=1$. 
Theoretical issues are then confined to a precise determination of the 
second order corrections to ${\cal F}(\omega=1)$,
the slope of this function, $\rho^2$, and its curvature~$c$,
\be
{\cal F}(\omega) ={\cal F}(1)\, \left [1 + \rho^2(\omega -1) + 
c(\omega -1)^2 + ...\right ]\,.
\label{iwexpan}
\ee
In terms of the perturbative (QED and QCD) and the non-perturbative
(leading $\delta_{1/m^2}$ and subleading $\delta_{1/m^3}$) corrections,
the normalization of the Isgur-Wise function ${\cal F}(1)$ can be expressed 
as follows:
\be
{\cal F}(1)=\eta \, \left[ 1 + \delta_{1/m^2} + \delta_{1/m^3}\right ]\,,
\ee
where $\eta$ is the perturbative renormalization of the Isgur-Wise function, known to two 
loops, $\eta=0.960 \pm 0.007$~\cite{Czarnecki:1996gu,Czarnecki:1997cf}. The formalism for 
calculating $\delta_{1/m^2}$ corrections in HQET has been developed by
Falk and Neubert~\cite{Falk:1992wt} and Mannel~\cite{Mannel:1994kv}. 
The various non-perturbative parameters entering in $\delta_{1/m^2}$ and the slope 
$\rho^2$ have been studied in the context of
quark models~\cite{Neubert:1994vy}, sum
rules~\cite{Shifman:jh,Bigi:ga,Kapustin:1996dy,Boyd:1996hy,Uraltsev:2000ce}
and quenched Lattice QCD~\cite{Kronfeld:2002cc,Hashimoto:2001nb}.
The default value used in the BABAR {\rm Physics Book}~\cite{BABARbook},  ${\cal 
F}(1)=0.91 \pm  0.04$, has recently been confirmed in the quenched lattice QCD 
calculations, including also
$\delta_{1/m^3}$ corrections~\cite{Kronfeld:2002cc}. The slope $\rho^2$ is
obtained by a simultaneous fit of the data for ${\cal F}(1) \vert V_{cb} \vert$ and 
$\rho^2$, and experiments use a form for ${\cal F}(\omega)$ given by Caprini {\it et 
al}.~\cite{Caprini:1997mu}. The resulting values of $ {\cal F}(1) \vert V_{cb} \vert$ and 
the
${\cal F}(1) \vert V_{cb} \vert$ - $\rho^2$ correlation from a large number of 
measurements 
from the LEP experiments, CLEO, BELLE and BABAR are summarized in 
Fig.~\ref{vcbexcl}.  They lead to the following world averages~\cite{hfag03}:
\be
{\cal F}(1) \vert V_{cb} \vert =(36.7 \pm 0.8) \times 10^{-3}\,, \qquad
\rho^2=1.44 \pm 
0.14\,\,\,\,\,[\chi^2=30.3/14]\,,
\label{hfagwavcbrho}
\ee
which with the value ${\cal F}(1)=0.91 \pm 0.04$ leads to
\be
\vert V_{cb} \vert_{B \to D^* \ell \nu_\ell} = (40.2 \pm 0.9_{\rm exp} \pm 
1.8_{\rm theo})\times 10^{-3}\,.  
\label{hfagwavcb}
\ee
This value is in very good accord with the determinations of $\vert V_{cb} \vert$ 
from the inclusive decays $B \to X_c \ell \nu_\ell$ given in (\ref{vcbinclexp}) for
the $\Upsilon(1S)$-scheme. 
While this consistency is striking, the agreement among the various experiments in 
the 
${\cal F}(1) \vert V_{cb} \vert$ - $\rho^2$ correlation from the decay $B \to D^* \ell 
\nu_\ell$ is less so, having a rather high $\chi^2$,  
$\chi^2/d.o.f. =2.16$. A robust average for $\vert V_{cb} \vert$ from both the inclusive 
and exclusive decays is not yet available from the Heavy Flavor Averaging Group
HFAG~\cite{hfag03}. 
It is also not clear to me how 
to do this as $\vert V_{cb} \vert$ from the inclusive measurements is obtained in 
the $\Upsilon(1S)$ scheme, as the quark masses have been defined in this scheme in 
the analysis of data, whereas $\vert V_{cb} \vert$ from the exclusive decays does not
have this dependence. An average can be given by weighting the two measurements
with their experimental errors only, leaving the theoretical errors as they are,
or one could add them in quadrature assuming they are independent. This gives
\be
\vert V_{cb} \vert =(41.2 \pm 0.8_{\rm exp} \pm 2.0_{\rm theo}) \times 10^{-3}
= (41.2 \pm 2.1) \times 10^{-3}\,,
\ee
yielding a precision $\delta \vert V_{cb}\vert/\vert V_{cb} \vert\simeq 5\%$.
In view of the still open issues (such as duality-related error, quark mass 
scheme-dependence in the inclusive decay, large $\chi^2/{\rm d.o.f.}$ in 
$B \to D^* \ell \nu_\ell$ decay, which makes the various experiments  
compatible with each other only at the expense of an increased error, etc.), a more 
precise value of $\vert 
V_{cb}\vert$, in my opinion, is not admissible at present. However, despite all the
caveats, the achieved precision in $\vert V_{cb}\vert$ is indeed remarkable.   
 
%
%
\begin{figure}[tbp]
\vspace{10pt}
\includegraphics[width=.5\textwidth]{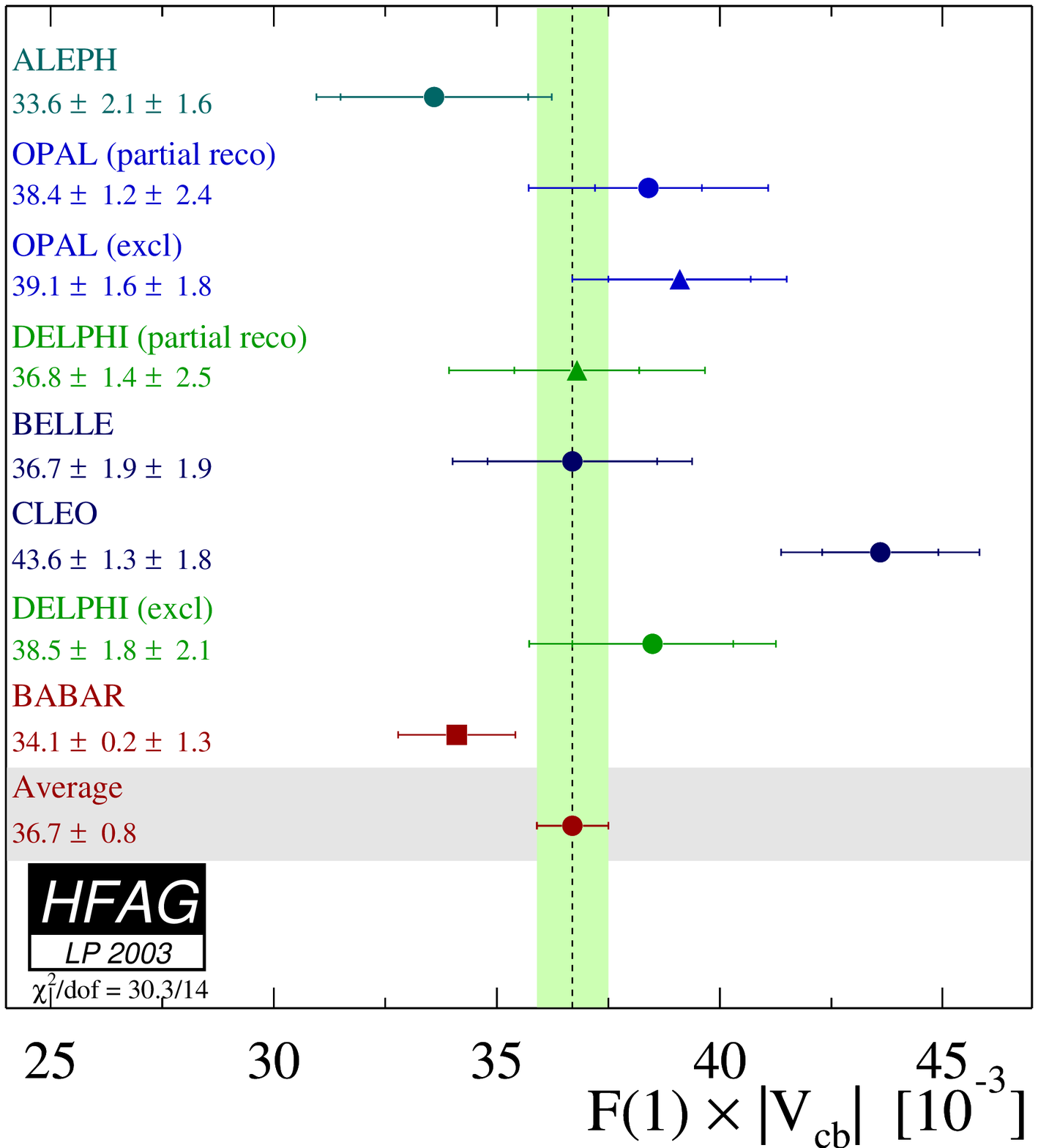}
\hspace*{0.5mm}
\includegraphics[width=.52\textwidth]{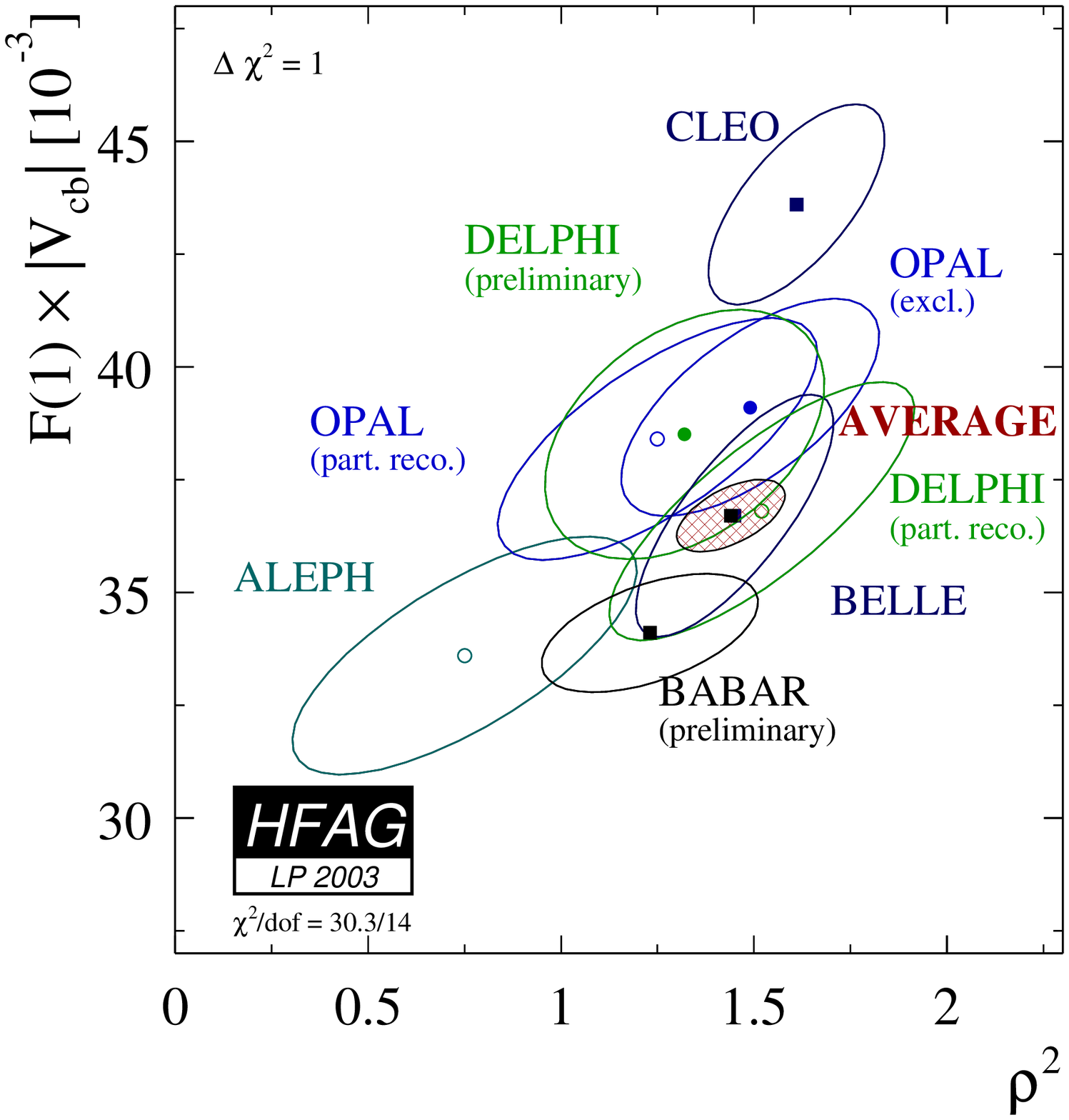}
\vskip 0.2cm
\caption{ Present status of $ F(1) \vert V_{cb} \vert$ (left frame)
and the  $F(1) \vert V_{cb} \vert$ - $\rho^2$ correlation (right frame)
from $B \to D^* \ell \nu_\ell$ decays. Note that $F(1)$ is called ${\cal F}(1)$ in 
the text. (Figures taken from
 the Heavy Flavor Averaging Group [{\rm HFAG
LP2003}]~\protect\cite{hfag03}\,.)}
%
\label{vcbexcl}
\vskip -0.2cm
\end{figure}

\subsection{Present status of $\vert V_{ub}\vert$}
Essentially, there are two methods to measure $\vert V_{ub} \vert$. The first 
one analyses the inclusive decays $B \to X_u \ell \nu_\ell$,
for which the branching ratio lies about a factor 60 below the dominant process
$ B \to X_c \ell \nu_\ell$. This circumstance makes it mandatory 
to apply harsh cuts to 
tag well the $b \to u$ events, invariably bringing 
in its wake theoretical problems involving enhanced non-perturbative and 
perturbative effects. The second method involves  
the exclusive decays, such as $B \to (\pi,\rho,\omega) \ell \nu_\ell$, which
require good knowledge of the form factors, not yet completely under 
theoretical control. We briefly summarize the present status 
for both the inclusive 
and exclusive determinations of $\vert V_{ub} \vert$.
\subsection{$\vert V_{ub} \vert$ from inclusive measurements}
The theoretical framework to study the inclusive decays $B \to X_u \ell 
\nu_\ell$ is also based on the operator product 
expansion. Up to $O(1/m_b^2)$ order,
the result for $\Gamma(B \to X_u \ell \nu_\ell)$ can be expressed as
follows~\cite{Czarnecki:1997hc}:
\be
\Gamma(B \to X_u \ell \nu_\ell) = \frac{G_F^2 m_b^5 \vert V_{ub} \vert^2}{192
\pi^3}\, \left (1 -2.41 \, \frac{\alpha_s(m_b)}{\pi} - 21.3 \,
\left (\frac{\alpha_s}{\pi} \right )^2
+ \frac{\lambda_1 - 9\lambda_2}{2 m_b^2} +...\right )\,,
\ee
of which the first three terms are coming from the perturbative-QCD improved
parton decay $b \to u \ell \nu_\ell$. The largest uncertainty 
in the decay rate is due to the $b$-quark mass, $m_b$, which is also scheme 
dependent, as already discussed.
In the $\Upsilon(1S)$-scheme, the result for $\vert V_{ub} \vert$
is numerically expressed  as follows~\cite{upsexp}:
\be
\label{Vub}
\vert V_{ub}\vert = (3.04 \pm 0.06_{(\rm pert)} \pm 0.08_{(m_b)}) \times 10^{-3}\,
  \bigg( {{\cal B}(B\to X_u \ell\bar\nu)\over 0.001}
  {1.6\,{\rm ps}\over\tau_B} \bigg)^{1/2}\,,
\label{Vubinclig}
\end{equation}
where the first error has a perturbative origin and the second is from $\Delta 
m_b$ in the $\Upsilon(1S)$-scheme, $m_b^{1S}=(4.73\pm 0.05)$ GeV. 
If this rate can be measured without significant cuts, then 
$\vert V_{ub}\vert$ can be measured with an
accuracy of $O(5\%)$. We shall take this as an ideal case and discuss now the 
realistic cases when kinematic cuts are imposed on some of the variables to 
measure the $b \to u$ semileptonic decays.

As already mentioned, the dominant background is from the decays 
$B \to X_c \ell \nu_\ell$. Noting that the lowest mass hadronic state in
$B \to X_c \ell \nu_\ell$ is the $D$-meson, thus its mass $m_D$ is used to 
define the cut region to suppress $b \to c$ transitions. So, the kinematic 
cut is either (i) on the upper end of the lepton energy spectrum, 
with $E_\ell > (m_B^2 - m_D^2)/2m_B$, or (ii) on the momentum transfer 
squared of the $\ell \nu_\ell$ pair, $q^2 > (m_B - m_D)^2$,
or (iii) on the hadron invariant mass, $m_{X} < m_D$, or (iv) an optimized 
combination of some or all of them. These cuts reduce the experimental 
rates for $B \to X_u \ell \nu_\ell$, a handicap which will be
overcome at B-factories with $O(10^8)$ $B\bar{B}$ 
mesons already at hand. However, the cuts also make the theoretical rates 
less rapidly convergent in terms of the perturbation series in 
$\Lambda_{\rm QCD}/m_b$ and $\alpha_s(m_b)$. 
The other disadvantage is that the theoretical rate with a 
cut depends sensitively (except for a cut on $q^2$) on the 
details of the $B$-meson wave function, or the shape function 
$f(k_+)$~\cite{shape}. Here $k_+$ is the $+$ component 
(using light cone variables) of a 
residual momentum $k_\mu$ of order $\Lambda_{\rm QCD}$, entering
through the relation $p_b^\mu=m_b v^\mu + k^\mu$, where $p_b^\mu$ is the
momentum of the $b$-quark in the $B$ meson, and $v^\mu$ is the 
four-velocity of
the quark. This can be seen as follows.
With either of the two cuts, $E_\ell > (m_B^2 - m_D^2)/2 m_B$,
or for the small hadronic invariant mass region, $m_X < m_D$, we have
\be
E_X \sim m_b\,; \quad m_X^2=(m_b v -q)^2 + 2E_X k_+ +...,
\ee
bringing in the dependence on $k_+$. This dependence is rather mild using the 
cuts on $q^2$. The effects of the kinematic cuts on the decay 
distributions and 
rates have been studied at great length in the literature. 
In fact, this enterprise
has led to a flourishing industry - the (kinematic) cutting
technology using 
HQET~\cite{mass,FLW,BDU,DeFazio:1999sv,llrhadron,energy,BLL1,BLL2,KoMe,ugo}!
Some applications are discussed here.

In the leading order in $\Lambda_{\rm QCD}/m_b$,
there is a universal function which governs the shape of the charged lepton 
energy spectrum, the hadronic invariant mass spectrum 
in $B \to X_u \ell \nu_\ell$
and the photon energy spectrum in $B \to X_s \gamma$, defined 
as follows~\cite{Neubert:1993um}
\be
f(k_+)=\frac{1}{2 m_B} \langle B(v)| \bar{b}_v \delta(k_+ + iD.n)b_v|B(v)\rangle\,,
\label{fkneubert}
\ee
where $n$ is a light-like vector satisfying $n.v=1$ and $n^2=0$.
The physical spectra are obtained by convoluting the universal shape function
with the perturbative-QCD expressions. Ignoring the perturbative and
subleading power corrections, a measurement of the photon energy spectrum is a 
measurement of the shape function $f(k_+)$:
\begin{eqnarray}
\frac{m_b}{2 \Gamma^{(0)}_{\rm sl}} \, \frac{d \Gamma}{d E_l}
(B \to X_u l \nu_l) & = & \int d\omega \,
\theta \left (m_b - 2 E_l - \omega \right ) \, f (\omega) + \ldots , \\
\frac{1}{\Gamma^{(0)}_\gamma} \, \frac{d \Gamma}{d E_\gamma}
(B \to X_s \gamma) & = & \int d\omega \,
\delta \left (m_b - 2 E_\gamma - \omega \right ) \, f (\omega) + \ldots  =
f (m_b - 2 E_\gamma) + \ldots ,
\nonumber
\label{shapespec}
\end{eqnarray}
where the normalization constants for the semileptonic and radiative
$B$-meson decays are:
\begin{eqnarray}
\Gamma^{(0)}_{\rm sl} & = & \frac{G_F^2 |V_{ub}|^2 m_b^5}{192 \pi^3}
\equiv C_{\rm sl} \, |V_{ub}|^2 ,
\nonumber \\
\Gamma^{(0)}_\gamma & = &
\frac{G_F^2 \alpha |V_{tb} V_{ts}^*|^2 m_b^5}{32 \pi^4} \,\,
|C_7^{\rm eff}|^2 \equiv C_\gamma \, |V_{tb} V_{ts}^*|^2 ,
\end{eqnarray}
and $C_7^{\rm eff}$ is the effective
Wilson coefficient governing the decay $B \to X_s \gamma$. 
Thus, combining the data on $B \to X_s \gamma$ 
and the lepton energy spectrum from $B \to X_u \ell \nu_\ell$, one can
determine in the SM the following~ratio:
\be
\left \vert \frac{V_{ub}}{V_{tb} V_{ts}^*}\right \vert^2 =
3 \, \frac{\alpha}{\pi}  \, \left \vert C_7^{\rm 
eff}\right \vert^2 \frac{\Gamma_u(E_c)}{\Gamma_s(E_c)} + 
O\left (\alpha_s \right ) + 
O\left (\frac{\Lambda_{\rm QCD}}{m_b} \right )\,,
\label{leadingshape}
\ee
where $\Gamma_u(E_c)$ and $\Gamma_s(E_c)$ are the cut-off dependent  
decay width in $\Gamma(B \to X_u \ell \nu_\ell;~E_\ell > E_c)$ and the cut-off
dependent first moment of the photon energy spectrum, respectively
\bea
\Gamma_u(E_c) &\equiv& \int_{E_c}^{m_B/2} dE_\ell \,
\frac{d \Gamma_u}{dE_\ell}\,,\nonumber \\ 
\Gamma_s(E_c) &\equiv& \frac{2}{m_b}\int_{E_c}^{m_B/2} dE_\gamma 
\left (E_\gamma -E_c \right ) \frac{d \Gamma_s}{dE_\gamma}\,.
\label{gammaus}
\eea
It should be stressed that the ratio (\ref{leadingshape}) 
holds not only in the SM, but also in models where the flavour changing (FC) 
transition  $b \to s$ is enacted solely in terms of $V_{\rm CKM}$, such as the 
minimal flavour violating supersymmetric models. An example where this relation 
does not hold is a general supersymmetric model in which the couplings $d_i 
\tilde{s}_j\tilde{g}$,  involving a down-type quark, a squark and gluino, are not diagonal 
in the flavour $(ij)$ space. In that case, the decay width for $B \to X_s \gamma$
does not factorize in $\vert V_{tb}V_{ts}^* \vert^2$ and depends on additional 
FC parameters. 

The relation (\ref{leadingshape}) has been put to good use, in the context of the SM,  
by the CLEO collaboration~\cite{Bornheim:2002du} through the measurement of the 
photon-energy spectrum in $B \to X_s \gamma$~\cite{Chen:2001fj} and the lepton 
energy spectrum in $B  \to X_u \ell \nu_\ell$, with 2.2 GeV $< E_\ell < 2.6$ GeV,
as the photon energy spectrum has been well measured in the overlapping range for 
$E_\gamma$, yielding

\vspace*{-2mm}

\be
\vert V_{ub} \vert =(4.08 \pm 0.34 \pm 0.44 \pm 0.16 \pm 0.24)\times 10^{-3}\,,
\label{vubcleobsg}
\ee

\vspace*{1mm}

\noindent
where the first two uncertainties are of experimental origin and the last two
are theoretical, of which $\delta \vert V_{ub} \vert=\pm 0.24 \times 10^{-3}$ is
an assumed uncertainty in the relation (\ref{leadingshape}). Combining all the 
errors in quadrature leads to 
$\vert V_{ub} \vert=(4.08 \pm 0.63)\times 10^{-3}$,
which is a $\pm 15\%$ measurement of this matrix element.

This method of determining $\vert V_{ub} \vert$
pioneered by the CLEO collaboration  has received a lot of theoretical 
attention lately. In particular,  
the subleading-twist contributions to the lepton and photon energy spectra in 
the decays $B \to X_u \ell \nu_\ell$ and $B \to X_s \gamma$, respectively, 
have been calculated in a number of 
papers~\cite{Bauer:2001mh,Bauer:2002yu,Leibovich:2002ys,Neubert:2002yx}, 
providing  estimates of the  subleading  correction in (\ref{leadingshape})
indicated as $O(\frac{\Lambda_{\rm QCD}}{m_b})$. An important feature 
emerging from these studies is that the various spectra are no longer 
governed by a universal shape function $f(\omega)$, a feature which is 
valid only in the leading twist. In subleading twist, HQET shows its rich 
underlying structure leading to a number of additional subleading shape 
functions, which are no longer universal. The operators ${\cal O}_i$ needed to 
calculate the subleading twist contributions and 
the corresponding matrix elements (shape functions) of these 
operators $\langle B| {\cal O}_i |B \rangle$ can be found, for example, 
in the papers by Bauer, Luke and Mannel~\cite{Bauer:2001mh,Bauer:2002yu}. 
The photon energy spectrum in $B \to X_s \gamma$ can now be expressed 
in terms of three structure functions $F(\omega)$, $h_1(\omega)$ 
and $H_2(\omega)$ as follows~\cite{Bauer:2001mh}
\begin{eqnarray}
\frac{m_b}{C_\gamma}\frac{d \Gamma}{d E_\gamma} = 
\left \vert V_{tb}V_{ts}^* \right \vert^2 \left [ 
(4 E_\gamma -m_b) F(m_b -2E_\gamma) 
+ \frac{1}{m_b}\bigg(h_1(m_b-2E_\gamma)+H_2(m_b-2E_\gamma)\bigg)
+ {\cal O}\left (\frac{\Lambda_{\rm QCD}^2}{m_b^2}\right )\right ],
\nonumber\\[0.5mm] 
\label{cleobsg}
\end{eqnarray}
where $C_\gamma$ has been defined earlier. Here, $F(\omega)$
contains both the leading and sub-leading parts
with $F(\omega) =f(\omega) + {\cal O}(\Lambda_{\rm QCD}/m_b)$, and
$h_1(\omega)$ and $H_2(\omega)$ are the subleading shape functions.

The corresponding lepton energy spectrum in the decay $B \to X_u \ell \nu_\ell$
now has the following form~\cite{Bauer:2002yu}:
\begin{eqnarray}
\frac{m_b}{2C_{\rm sl}}\frac{d \Gamma}{d E_\ell}= \vert V_{ub} \vert^2
\left [\int d \omega\,\theta\,(m_b -2 E_\ell -\omega) 
\left ( F(\omega)\left (1-\frac{\omega}{m_b}\right )
-\frac{1}{m_b}h_1(\omega) + \frac{3}{m_b}H_2(\omega)\right )
+{\cal O} \left (\frac{\Lambda_{\rm QCD}^2}{m_b^2}\right ) \right ],
\nonumber\\
\end{eqnarray}
where $C_{\rm sl}$ has also been defined earlier. It is obvious that the
measurement of
either the photon energy spectrum or the lepton energy spectrum does not allow
to determine all three shape functions. Hence, they will have to be modeled.
These subleading corrections modify the relation (\ref{leadingshape}) used in
extracting $\vert V_{ub} \vert$, which can be written as\cite{Bauer:2002yu}:   
\be
\left \vert \frac{V_{ub}}{V_{tb} V_{ts}^*}\right \vert =
\bigg(3 \, \frac{\alpha}{\pi} \, \vert C_7^{\rm eff}\vert^2 \,
\frac{\Gamma_u(E_c)}{\Gamma_s(E_c)}\bigg)^{1/2}
\left (1 + \delta(E_c) \right )\,.  
\label{subleadingshape}
\ee
Again, $\delta(E_c)$ can only be estimated in a model-dependent way. Typical 
estimates are $\delta(E_c=2.2~{\rm GeV}) \simeq 0.15$, with $\delta(E_c)$ decreasing
as the lepton-energy cut $E_c$ decreases, estimated as $O(10\%)$ for $E_c=2.0~{\rm 
GeV}$. 
One should use the order of magnitude estimate of the subleading twist contribution 
$\delta(E_c)$ to  set the theoretical uncertainty on $\vert V_{ub} \vert$ from this 
method, which typically is $15\%$.

These uncertainties can be reduced if one considers more 
complicated kinematic cuts, such as a simultaneous cut on $m_X$ and 
$q^2$~\cite{BLL2,Luke:2003nu}, whose effect has  been studied using a model  
for the leading-twist shape function $f(k_+)$. The
sensitivity of the partial decay width $\Gamma(q^2> q^2_{\rm cut}, m_X < m_{\rm 
cut})$ on $f(k_+)$ is found to be small, and this is likely to hold also if the 
subleading shape functions are included.  This method of determining $\vert V_{ub} \vert$ 
has been applied by BELLE using two techniques. The first uses the decays $B \to 
D^{(*)} \ell \nu_\ell$ as a tag, and the other uses 
the neutrino reconstruction technique, as in exclusive semileptonic decays,
combined with a sorting algorithm (called ''annealing'') to separate the event in
a tag and a $b \to u \ell \nu_\ell$ side. The method based on $D^{(*)}$-tagging 
yields\cite{schwanda:2003}
$\vert V_{ub} \vert =(5.0 \pm 0.64 \pm 0.53) \times 10^{-3}$. 
The result using the annealing method with the cuts $m_X < 1.7$ GeV, $q^2 > 8$ 
GeV$^2$ 
is~\cite{Kakuno:2003fk}
\be
\vert V_{ub} \vert = (4.66 \pm 0.28 \pm 0.35 \pm 0.17 \pm 0.08 \pm 0.58) 
\times 
10^{-3}\,,
\label{bellevubanneal}
\ee
where the errors are statistical, detector systematics, modeling $b \to c$, modeling 
$b \to u$, and theoretical, respectively. Combining all the errors yields
$\vert V_{ub} \vert =(4.66 \pm 0.76) \times 10^{-3}$.

A similar analysis by the BABAR collaboration, in which one of the two $B$-mesons is
constructed through the hadronic decays $B \to D^{(*)}h$, and the inclusive 
semileptonic decay of the other $\bar{B}$-meson is measured with the cuts $E_\ell > 1$ 
GeV and $m_X < 1.55$ GeV~\cite{Sarti:2003}, yields
\be
\vert V_{ub} \vert = (4.52 \pm 0.31 \pm 0.27 \pm 0.40 \pm 0.25) \times
10^{-3}\,,
\label{babarvubinc}
\ee
where the errors are statistical, systematics, due to extrapolations to the full phase 
space, and from the HQET parameters, respectively.

Finally, the effects of the so-called weak annihilation (WA)~\cite{WA}, which 
are formally  of $O(\Lambda_{\rm QCD}^3/m_b^3)$ but are enhanced by the 
phase space factor $16 \pi^2$ (compared to that of $b \to u \ell \nu_\ell$),
introduce an additional theoretical uncertainty~\cite{Voloshin:2001xi}. They 
stem from the dimension-6 four-quark operators in the OPE,
\be
O_{V-A}=(\bar{b}_v \gamma_\mu P_Lu)(\bar{u} \gamma^\mu P_L b_v)\,,
\qquad 
O_{S-P}=(\bar{b}_v P_L u)(\bar{u} P_R b_v)\,,
\ee
where $P_{L,R}=(1\pm \gamma_5)/2$. In the lepton energy 
spectrum from $B\to X_u \ell \nu_\ell$, they enter as delta functions near the
end-point~\cite{Voloshin:2001xi}:
\be
\frac{d\Gamma_{(6)}}{dy}=-\frac{G_F^2 m_b^2 \vert V_{ub} \vert^2}{12\pi} \,
f_B^2 \, m_B \, (B_1 - B_2)\, \delta(1-y)\,,
\ee
where $y=2E_\ell/m_b$ and $f_B \simeq 200$ MeV is the $B$ meson decay 
constant; $B_{1}$ and $B_{2}$ parameterize the matrix elements of the 
operators $O_{V-A}$ and $O_{S-P}$, respectively:
\be
\frac{1}{2 m_B} \langle B|O_{V-A} |B\rangle =\frac{f_B^2 m_B}{8}\, B_1\,,
\qquad
\frac{1}{2 m_B} \langle B|O_{S-P} |B\rangle =\frac{f_B^2 m_B}{8}\, B_2\,.
\ee
In the vacuum insertion approximation, i.e., assuming factorization, their
effect in the spectrum vanishes, as in this approximation, $B_1=B_2=1(0)$ for 
the charged (neutral) $B$ mesons. Hence, they are generated by non-factorizing 
contributions and are not yet quantified. These matrix elements are also 
encountered in calculating the differences in the $B^\pm$ and $B^0$  
lifetimes~\cite{Neubert:1996we}, and we refer to a recent discussion 
in the context of lattice QCD~\cite{Becirevic:2001fy}. However, comparing the
extraction of $\vert V_{ub} \vert$ from $B^\pm \to X_u \ell \nu_\ell$ and
 $B^0 \to X_u \ell \nu_\ell$ near the end-point of the lepton energy 
spectrum, one can determine the size of the WA effects. For the $B^\pm$ 
decays, they can also be estimated from a related process  
$B^\pm \to \ell^\pm \nu_\ell 
\gamma$~\cite{Grinstein:2000pc}.

The current results on $\vert V_{ub} \vert$ from various inclusive 
measurements by the LEP, CLEO, BABAR and BELLE experiments are summarized
by HFAG~\cite{hfag03}. No averaging for $\vert V_{ub} \vert$ has been 
undertaken so far by this working group. However, the results 
in (\ref{vubcleobsg}), (\ref{bellevubanneal}) and (\ref{babarvubinc}) 
from the CLEO, BELLE and
BABAR collaborations, respectively, have been averaged by 
Muheim~\cite{Muheim:2003} to get $\vert V_{ub} \vert$ from the 
$\Upsilon(4S)$ data, yielding
\be
\vert V_{ub} \vert_{\rm incl}= (4.32 \pm 0.57) \times 10^{-3}\,,
\label{vubmuheim}
\ee
in agreement with the LEP average~\cite{Abbaneo:2000ej} $\vert 
V_{ub}\vert=(4.09\pm 0.70) \times 10^{-3}$.

\subsection{$\vert V_{ub} \vert$ from exclusive measurements}
First measurements of the exclusive decays $B \to \pi \ell \nu_\ell$
and $B \to \rho \ell \nu_\ell$ were reported by the CLEO collaboration
in 1996~\cite{Alexander:1996qu}. Improved measurements of the rates
for $B \to \rho \ell \nu_\ell$  were published 
subsequently~\cite{Behrens:1999vv}. This year, results based on the entire 
CLEO data ($9.7\times 10^{6}~B\overline{B}$ pairs) were reported
~\cite{Athar:2003yg}, including the  measurement of the branching ratio 
for  $B^+ \to \eta \ell^+ \nu_\ell$ (charge conjugation average is implied):
\begin{eqnarray}
{\cal B}(B^0 \to \pi^- \ell^+ \nu_\ell)&=& (1.33 \pm 0.18 \pm 
0.11  \pm 0.01 \pm 0.07) \times 10^{-4}\,,
\nonumber\\  
{\cal B}(B^0 \to \rho^- \ell^+ \nu_\ell)&=& (2.17 \pm 0.34 
^{+0.47}_{-0.54} \pm 0.41  \pm 0.01) \times 10^{-4}\,,
\label{cleoexbrs} \\
{\cal B}(B^+ \to \eta \ell^+ \nu_\ell)&=& (0.84 \pm 0.31 \pm
0.16 \pm 0.09) \times 10^{-4}\,,
\nonumber
\end{eqnarray}
where the errors are statistical, experimental systematic, form factor 
uncertainties in the signal, and form factor uncertainties in the 
cross-feed modes, respectively. Rough measurement of the  $q^2$ 
distributions $d\Gamma(q^2)/dq^2$
for the $(\pi \ell \nu_\ell)$ and $(\rho \ell \nu_\ell)$ modes by
splitting the data in three $q^2$-bins were also reported.
BABAR has also measured the decay $B \to \rho \ell 
\nu_\ell$~\cite{Aubert:2003zd}:
\be
{\cal B}(B^0 \to \rho^- \ell^+ \nu_\ell)=(3.29 \pm 0.42 \pm 0.47 \pm
0.60) \times 10^{-4}\,,
\label{babarrho}
\ee
where the errors are statistical, systematic and theoretical, 
respectively.

Extracting $\vert V_{ub} \vert$ from these measurements is done by using
quark models, QCD sum rules, and quenched Lattice-QCD calculations for 
the form factors. We discuss the two main contenders, Lattice QCD and QCD 
sum rules, for the semileptonic decays $B 
\to \pi \ell \nu_\ell$, as this involves (neglecting the lepton mass) only 
one form factor, $F_+(q^2)$, defined as follows:
\be
\langle \pi(p_\pi)|\bar{b} \gamma_\mu q|B(p_B)\rangle = 
\left ( (p_B+p_\pi)_\mu
-\frac{m_B^2 - m_\pi^2}{q^2}q_\mu\right ) \,F_+(q^2)
+ \frac{m_B^2 - m_\pi^2}{q^2} \,F_0(q^2) \, q_\mu \,,
\label{fpiBdef}
\ee
with $F_+(0)=F_0(0)$.

QCD sum rules~\cite{Shifman:bx}, in particular Light cone QCD sum 
rules (LCSRs)~\cite{Balitsky:ry,Chernyak:ag}, have been used extensively 
to study the form factors in $B \to \pi$ transitions (and other related 
processes)~\cite{Colangelo:2000dp}. In the LCSR, one calculates a 
correlation function
involving the weak current and an interpolating  current with the
quantum numbers of the $B$ meson, sandwiched between the vacuum ($\langle 
0|$) and a pion state ($|\pi \rangle)$:
\be
F_\mu(p,q)=i \int dx {\rm e}^{iq.x} \langle \pi (p)| T\{\bar{u}(x) 
\gamma_\mu b(x), m_b\bar{b}(0) i\gamma_5 d(0)\}|0\rangle\,.
\label{srcf}
\ee
For large negative virtualities $(q^2, p.q)$ of these 
currents, the correlation function (CF) in the coordinate space is 
dominated by the dynamics at distances near the light cone, allowing a light-cone
expansion of the CF (hence, the name).
In essence, LCSRs are based on the factorization property of
the CF into non-perturbative light-cone
distribution amplitudes (LCDAs) of the pion, called 
$\phi_\pi^n(u, \mu)$, where $u$ is the fractional momentum of 
the quark in the pion and $\mu$ is a factorization scale, and
process-dependent hard (perturbative QCD) amplitudes $T_H^n 
(\mu,m_b,u,q^2)$, where $q^2$ is the virtuality of the weak current.
 Schematically, the coefficients in front of the Lorentz structures in the
decomposition of the CF (\ref{srcf}) can be written as: 
\be
C(\mu, q^2, m_b) \sim \sum_n T_H^n(\mu, m_b, u,q^2)\otimes \phi^n(u,\mu)\,,
\ee  
where the sum runs over contributions with increasing twist, with twist-2 
being the lowest, and the symbol $\otimes$ implies an integration over the variable 
$u$. The amplitudes $T_H^n$ have an expansion in perturbative QCD (i.e., 
$\alpha_s$).
 The same correlation function can also be written as a 
dispersion relation in the virtuality of the current coupled to the 
$B$-meson. Equating the two, using quark-hadron duality, and
separating the $B$-meson contribution from higher excited and continuum 
states, results in the LCSR. As an illustration, the LCSR for the form 
factor $F_+(q^2)$ in the lowest order in $\alpha_s$ and leading-twist has 
the form~\cite{Colangelo:2000dp}
\be
F_+(q^2) = \frac{1}{2 m_B^2f_B} {\rm e}^{~-\frac{m_B^2}{M^2}}
m_b^2f_\pi \int_{\Delta}^{1}\frac{du}{u} {\rm 
e}^{-\frac{m_b^2-p^2(1-u)}{uM^2}} \phi_\pi(u, \mu_b)\,,
\ee
where $f_\pi=132$ MeV, M is a Borel parameter, characterizing the 
off-shellness of the $b$-quark, and $\phi_\pi(u,\mu_b)$ is the twist-2
LCDA of the pion
\be
\phi_\p(u,\mu) = 6 u (1-u) \left ( 1+ a_2^\pi(\mu) C_2^{3/2}(2u-1)+...
\right )\,.
\ee
Here, $C_2^{3/2}(x)$ is a Gegenbauer polynomial and $a_2^\pi(\mu)$ is a 
non-perturbative coefficient (the second Gegenbauer moment) to be determined, 
for example, from the data on the electromagnetic form factor of the pion. 
The lower integration limit denoted by $\Delta$ is defined through 
$\Delta=(m_b^2-q^2)/(s_0^B -q^2)$, where $s_0^B$ is determined
by the subtraction point of the excited resonances and continuum states 
contributing to the dispersion integral in the $B$ channel. 
Assuming quark-hadron duality, this subtraction is performed 
at $(p+q)^2 \geq s_0^B$.

In principle, given the assumption of quark-hadron duality, the LCSRs 
can be made arbitrarily accurate, by calculating enough 
perturbative and non-leading twist contributions to the CF. In practice, this
framework has a number of parameters (such as $\mu$, $M$, 
$a_i^\pi(\mu)$, $s_0^B$, $m_b$), whose imprecise knowledge restricts 
the precision on the CF. Typically, the state-of-the-art LCSRs for 
$F_+(q^2)$~\cite{Khodjamirian:2000ds,Ball:1998tj,Ball:2001fp},
which include $\alpha_s$ corrections to the leading-twist and part of the 
twist-three  contributions, and tree level for rest of the twist-three and 
twist-four, have an  uncertainty of $\sim \pm  
20\%$~\cite{Colangelo:2000dp,Ball:2003rd}, and it is 
probably difficult to make these estimates more precise.

In Lattice QCD, one
calculates a three-point correlation function involving interpolating 
operators for the $B$ and $\pi$ mesons and the vector current 
$V_\mu \sim \bar{\Psi}_b \gamma_\mu \Psi_q$. In the limit of a large time 
separation, the correlation function has the following behaviour
\be
C_\mu(p,k,t_f,t_s,t_i)={\cal Z}_B^{1/2} \, {\cal Z}_\pi^{1/2} \, 
\frac{\langle B(k)|V_\mu|\pi(p)\rangle}{\sqrt{2E_B}\, \, \sqrt{2 E_\pi}} \, 
{\rm e}^{-E_\pi(t_s-t_i)}\,{\rm e}^{-E_B(t_f-t_s)}\,,
\label{correlator}
\ee
where $E_B(E_\pi)$ is the energy of a $B (\pi)$ meson with the 
three-momentum $k(p)$ and ${\cal Z}_B^{1/2}({\cal Z}_\pi^{1/2})$
is the external line factor calculated from the two-point correlation
functions involving the interpolating $B(\pi)$ fields. Since, one has to 
go to large time separations to suppress the continuum contribution, one is forced to 
restrict the pion momentum in $B \to \pi \ell \nu_\ell$ decay  to low values. Hence, in 
lattice calculations,  there is an upper limit on this momentum,
$\vert p^{\rm max} \vert$, prescribed by the requirement to keep the 
statistical  and discretization errors small. There is also a lower limit 
on $\vert p 
\vert$ dictated by the difficulty in extrapolations in $\vert p\vert$
and light quark masses. Thus, for example, the FNAL Lattice QCD 
calculations for the $B \to \pi$ form factors~\cite{El-Khadra:2001rv} have as 
cut-offs $\vert 
p^{\rm max} \vert=1.0$ GeV and $\vert p^{\rm min} \vert=0.4$ GeV.
This translates into a limited $q^2$ range $q^2_{\rm min} < q^2 < 
q^2_{\rm max}$ close to the zero-recoil point.

 Recalling that the differential decay 
rate for $B \to \pi \ell \nu_\ell$ is given by
\be
\frac{d\Gamma}{d p}(B \to \pi \ell \nu_\ell) = 
\frac{G_F^2 \vert V_{ub} \vert^2}{24 \pi^3}\, 
\frac{2 m_B p^4 \vert F_{+}(E) \vert^2}{E}\,,
\label{pildiff}
\ee
where $E=p_\pi . p_B/m_B$ is the pion energy in the $B$-meson rest frame,
one calculates the dynamical part on the lattice over a limited region of $\vert p
\vert$~\cite{El-Khadra:2001rv}. Defining
\be
T_B(\vert p^{\rm min} \vert, \vert p^{\rm max} \vert)
\equiv \int_{\vert p^{\rm min}\vert}^{\vert p^{\rm max} \vert} dp\, \frac{p^4 
\vert F_+(E) \vert^2}{E}\,,
\ee
one combines the theoretical rate with the experimental measurements in the 
same momentum range of the pion to arrive at the following relation for 
$\vert V_{ub} \vert^2$, 
\be
\vert V_{ub} \vert^2=\frac{12\pi^3}{G_F^2 m_B}\, \frac{1}
{T_B(\vert p^{\rm min} \vert, \vert p^{\rm max} \vert)}
\,\int_{\vert p^{\rm min} \vert}^{\vert p^{\rm max}\vert} dp\, \frac{d\Gamma(B
\to \pi \ell \nu_\ell)}{d p}\,.
\label{fnalratio}
\ee
This avoids the need to extrapolate to higher pion momenta (or low $q^2$).
The practical problem in using (\ref{fnalratio}) is
the paucity of experimental data in low $\vert p\vert$-region, as the
differential decay rate has a kinematic suppression for low pion momenta.

Alternatively, one has to use models for $q^2$ extrapolation of the 
Lattice results to lower values of $q^2$. This is done, for example, by 
the UKQCD~\cite{Bowler:1999xn} and the APE 
collaborations~\cite{Abada:2000ty}, which make use of the 
LCSRs (discussed above) to constrain the
form factors at lower values of $q^2$.
 
Apart from this, there are also other systematic differences among the
various Lattice calculations of the $B \to \pi$ form factors, the most 
important of which is related to the fact that for the current lattice 
spacing $a$ one has $m_b a >1$. To control lattice spacing effects, one 
has to do the calculations for values of the heavy quark mass $m_Q$ much 
smaller than $m_b$ and then extrapolate from $m_Q \simeq 1 - 2$ GeV, 
where the lattice data is available, to the $b$-quark 
mass. This, for example, is done by the UKQCD~\cite{Bowler:1999xn} and
APE~\cite{Abada:2000ty} collaborations.
A different route is taken by the FNAL Lattice 
group~\cite{El-Khadra:2001rv}, in which HQET is
applied directly to the Lattice observables, using the same Wilson action    
for fermions as adopted by the other groups, but adjusting the couplings 
in the action and the normalization of the currents, so that the leading 
and the next-to-leading terms in HQET are correct. This allows to perform 
the calculations directly at $m_Q=m_b$. 
However, there are still some open issues
in this approach what concerns the non-perturbative matching of the lattice
with the continuum. Finally,  JLQCD~\cite{Aoki:2001rd} also uses the Fermilab 
NRQCD approach.  The result of the four Lattice 
groups for the $B \to \pi$ form factors, $F_+(q^2)$ and 
$F_0(q^2)$, are shown in Fig.~\ref{damirfo}, taken from the review  
by Becirevic~\cite{Becirevic:2002zp}. These  
calculations agree at the level of $\pm 20\%$, though this consistency is 
less marked for the form factor $F_0(q^2)$.

The results of the Lattice groups (APE~\cite{Abada:2000ty}, 
UKQCD~\cite{Bowler:1999xn},
FNAL~\cite{El-Khadra:2001rv}, and JLQCD~\cite{Aoki:2001rd}) have
also been fitted by a phenomenological form for $F_{+,0}(q^2)$
due to Becirevic and Kaidalov~\cite{Becirevic:1999kt}
\be
F_+(q^2)= \frac{F(0)}{(1-q^2/m_{B^*}^2)(1-\alpha_\pi q^2/m_{B^*}^2)}\,;
\qquad 
F_0(q^2)= \frac{F(0)}{1-q^2/(\beta_\pi m_{B^*}^2)},
\label{bkparam}
\ee
involving three parameters $F(0)$, $\alpha_\p$ and $\beta_\pi$. The
best-fit solution for $F_+(q^2)$ and $F_0(q^2)$ is shown by the dashed 
curve in Fig.~\ref{damirfo}. This figure also shows the prediction 
obtained by the LCSRs~\cite{Colangelo:2000dp}. The resulting theoretical 
description for the form factors from lattice QCD and LCSRs is strikingly 
consistent, albeit not better than $\pm 20\%$. Further details can be seen
elsewhere~\cite{Becirevic:2002zp}.

In future, with more CPU power at their disposal, it should be possible to
increase the pion momentum range accessible on the lattice, allowing a larger and 
statistically  improved overlap of the lattice results with the experimental data 
on $B \to \pi \ell \nu_\ell$. 
Also, by using 
the FNAL method of applying HQET directly to the Lattice observables, or
else by doing simulations at larger values of $m_Q$ than is the case right now, 
theoretical errors on the FFs can be reduced to an acceptable 
level in the quenched approximation. There are also other techniques being developed 
for treating heavy quarks on the Lattice~\cite{Sommer:2003wm}.
 The last step in this theoretical 
precision study will come with the estimates of unquenching effects using 
dynamical fermions; first unquenched results for $B \to \pi$ form 
factors are expected soon.

%
%
\begin{figure}[tbp]
\vspace{10pt}
\centerline{\includegraphics[width=10cm]{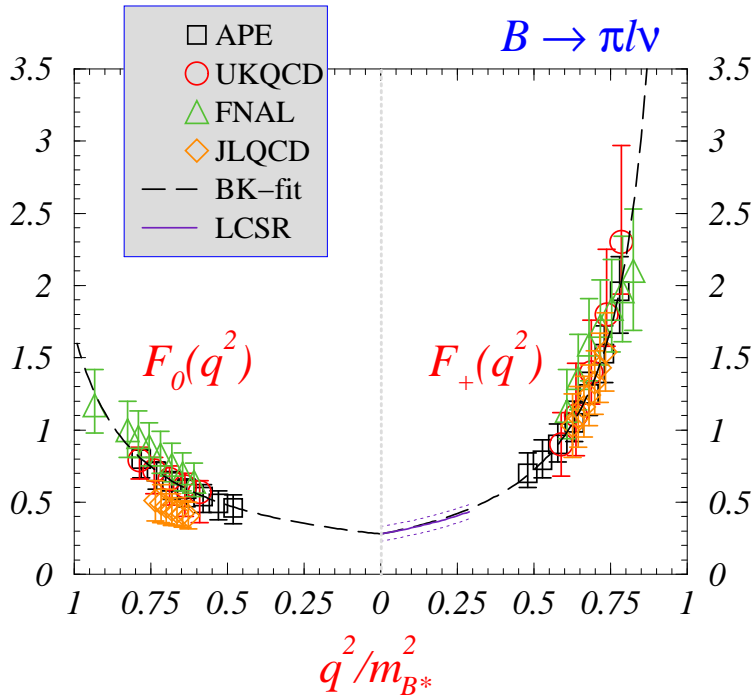}}
\vskip 0.2cm
\caption{ Summary of the current (unquenched) Lattice calculations of the $B 
\to \pi$ form factors $F_{0}(q^2)$ and $F_{+}(q^2)$ and from the Light Cone QCD 
sum  rules. The dashed curve shows a parametrization by the Becirevic-Kaidalov 
model. (Figure taken from Becirevic~\protect\cite{Becirevic:2002zp}\,.)}
%
\label{damirfo}
\vskip -0.2cm
\end{figure}
After this longish detour of the currently used theoretical framework, we
return to the extraction of $\vert V_{ub} \vert$ from exclusive 
semileptonic decays. The CLEO result~\cite{Athar:2003yg} quoted below is 
obtained by using the quenched Lattice QCD results for $q^2 > 16$ GeV$^2$ 
and the LC QCD sum rules for lower values of $q^2$,
\be
\vert V_{ub} \vert = (3.17 \pm 0.17 ^{+0.16 + 0.53}_{-0.17 -0.39} \pm
0.03) \times 10^{-3}~~[{\rm CLEO (exclusive)}]\,. 
\label{cleoexcl}
\ee
BABAR~\cite{Aubert:2003zd} uses the theoretical decay width calculated
using Lattice QCD, LC QCD sum rules, and three quark models to estimate
the form factors. The combined result is the weighted average of these 
theoretical approaches, where the weight is obtained by the theoretical
uncertainty, and an overall theoretical uncertainty is assigned by taking
it to be half of the full spread over these models. 
The result~is~\cite{Aubert:2003zd}:
\be
\vert V_{ub} \vert = (3.64 \pm 0.22 \pm 0.25^{+0.39}_{-0.56})
\times 10^{-3}~~[{\rm BABAR}]\,.
\label{babarexc}
\ee
The two measurements (\ref{cleoexcl}) and (\ref{babarexc}) are consistent with each 
other, and they have been averaged by 
Schubert~\cite{Schubert:2003} to yield a value of $\vert V_{ub} \vert$ from the 
exclusive decays,
\be
 \vert V_{ub} \vert_{\rm excl}= (3.40 ^{+0.24}_{-0.33} \pm 0.40) 
\times 10^{-3}\,.
\label{vubexclshu}
\ee
 However, this value  
lies below  $\vert V_{ub} \vert$ measured from the inclusive 
decays $B \to X_u \ell \nu_\ell$, whose current average is given in 
(\ref{vubmuheim}). The mismatch in the values of $\vert V_{ub} \vert$ from the 
inclusive and exclusive 
decays is roughly about $20\%$ and has to be resolved as more precise data 
and theory become available. The  possibility that in the quenched Lattice QCD and 
LCSR estimates, the form factor  $F_{+}(q^2)$ is estimated too high by about 
$20\%$ can not be excluded at present.

 Digressing from the discussion of $\vert V_{ub} \vert$, we remark that this trend is 
also seen in the comparison of
data on $B \to K^* \gamma$ with the LC-QCD sum rule estimates of the $B \to K^*$ 
form factors. To put this in a quantitative perspective, we recall that the current 
branching ratios for the $B \to K^* \gamma$ decays are~\cite{hfag03} (again charge 
conjugated averages are implied)
\be
{\cal B}(B^0 \to K^{*0} \gamma) =(4.17 \pm 0.23) \times 10^{-5}\,,
\qquad
{\cal B}(B^- \to K^{*-} \gamma) =(4.18 \pm 0.32) \times 10^{-5}\,.  
\label{brkstarg}
\ee
The corresponding theoretical rates have been calculated in the
NLO accuracy~\cite{bdgAP,Bosch:2001gv,Beneke:2001at} using the 
QCD-factorization framework~\cite{Beneke:1999br}.
An updated analysis based on~\cite{bdgAP} (neglecting a small isospin 
violation in the decay widths) yields
\begin{equation}
{\cal B} (B \to K^* \gamma) = (7.4 \pm 1.0) \times 10^{-5}
\left( \frac{\tau_B}{1.6~{\rm ps}} \right )\,
\left( \frac{m_{b, {\rm pole}}}{4.65~{\rm GeV}} \right )^2\,
\left( \frac{T_1^{K^*} (0, \bar m_b)}{0.38} \right )^2 \,,
\label{brkstargqcd}
\end{equation}
where the default value for the form factor $T_1^{K*}(0, \bar{m}_b)$ 
is taken from the LC-QCD sum rules~\cite{Ball:2003rd}~\footnote{The decay rates
in this approach depend on the 
 effective theory parameter, called $\xi^{K^*}(0)$, 
which is related by an $O(\alpha_s)$ relation to the $B \to K^*$ form 
factor by $T_1^{K*}(0, \bar{m}_b) \simeq C_{K^*}(\bar{m}_b)\xi^{K^*}(0)$, with 
$C_{K^*}(\bar{m}_b)=1.05$ - $1.08$~\cite{Beneke:2000wa}. To keep the discussion 
simple, we used this relation to express the rates in $T_1^{K*}(0, \bar{m}_b)$.},
and the pole mass $m_{b,~{\rm  pole}}$ is the one-loop corrected central value 
obtained from the $\overline{\rm MS}$ $b$-quark mass $\bar{m}_b(m_b)=(4.26 \pm 0.15 
\pm 0.15)$ GeV in the PDG reviews~\cite{Hagiwara:fs}.
Since the inclusive branching ratio for $B \to X_s \gamma$ in the SM agrees 
well with the current measurements of the same (discussed below), the
mismatch in the estimates of the exclusive branching ratios in 
(\ref{brkstargqcd}) and current measurements
(\ref{brkstarg}) in all likelihood has a QCD origin. Of the possible 
suspects, form factor is probably the most vulnerable link in the chain of arguments
leading to (\ref{brkstargqcd}). Interpreting the 
factorization-based QCD estimates and the data on their face value, good
agreement between the two requires $T_1^{K^*}(0,\bar{m}_b) \simeq 
0.27 \pm 0.02$. This is shown in Fig.~\ref{RatKsT1} where the ratio
\be
 R(K^*\gamma/X_s \gamma)\equiv \frac{{\cal B}(B \to K^* \gamma)}{{\cal B}(B \to X_s 
\gamma)}\,,
\label{brxskstar}
\ee
is plotted as a function of $T_1^{K^*}(0, \bar{m}_b)$. The horizontal bands show 
the current experimental value for this quantity $R(K^*\gamma/X_s \gamma)=0.125 \pm 
0.015$.
%
%
\begin{figure}[bt]
\centerline{
            \psfig{width=0.45\textwidth,file=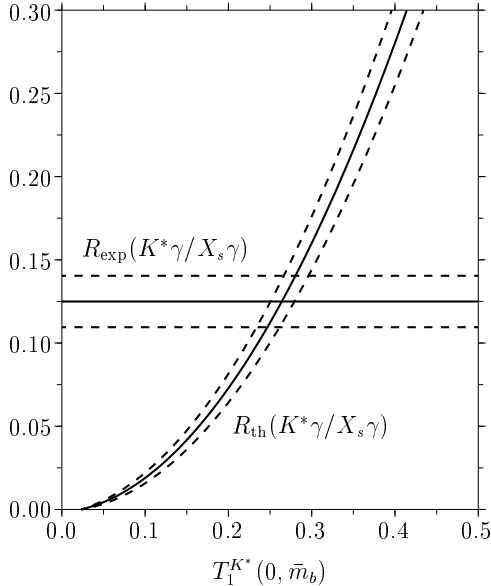}}
\caption{The ratio of the  branching ratios
 defined in Eq.~(\ref{brxskstar}),
         plotted as a function of the QCD form factor~$T_1^{K^*}
         (0, \bar m_b)$ and the current experimental measurement
         of this ratio. The solid lines are the central experimental
         and theoretically predicted values and the dotted lines
         delimit the $\pm 1 \sigma$ bands. (Figure updated from 
Ref.~\protect\cite{bdgAP}.)}
\label{RatKsT1}
\end{figure}
The allowed values of $T_1^{K^*}(0, \bar{m}_b)$ are about $25\%$ below the current 
estimates of the same from the LC-QCD approach $(=0.38 \pm 0.05$). There is a need to
do an improved calculation of this (and related) form factors. Along this direction, 
SU(3)-breaking effects in the $K$ and $K^*$ LCDA's have been recently re-estimated by 
Ball and Boglione~\cite{Ball:2003sc}. This  modifies the input value for the Gegenbauer 
coefficients in the $K^*$-LCDA and  
the contribution of the so-called hard spectator diagrams in the
decay amplitude for $B \to K^* \gamma$ is reduced, decreasing in turn the branching 
ratio by about 
$7\%$~\cite{bdgAP-2003}. 
The effect of this correction on the form factor $T_1^{K^*}(0, \bar{m}_b)$, as well 
as of some other technical improvements~\cite{Ball:2003sc}, has not yet been 
worked out. Updated  calculations of this form factor on the lattice 
are also under way~\cite{Becirevic:2003}, with preliminary results yielding values 
for $T_1^{K^*}(0, \bar{m}_b) \sim 0.27$, as suggested by the analysis 
in Fig.~\ref{RatKsT1}, and considerably smaller than the ones from the earlier 
lattice-constrained parameterizations by the UKQCD collaboration~\cite{DelDebbio:1997kr}. 
Theoretical estimates of the form factors are still in a state of flux. 
Phenomenologically, smaller values of  the form factors in $B \to \pi$ and $B \to \rho$ 
transitions  are preferred by the data, bringing $\vert V_{ub} 
\vert$ from the exclusive decays  more in line 
with the value of this matrix element measured from the inclusive decays. Smaller value 
of the $B \to K^*$ form factor would also improve the agreement between the 
QCD-factorization based estimates for ${\cal B}(B \to K^* \gamma)$ 
and~experiments.

A robust average of $\vert V_{ub} \vert$ based on current measurements is
expressly needed to determine one of  the sides ($R_b$) of the unitarity triangle 
precisely. This is, however, not yet provided by HFAG~\cite{hfag03}.
A bonafide average is difficult to undertake, as the common (and experiment-specific) 
correlated systematic errors are not at hand, as stated in some of the recent
experimental reviews on this subject~\cite{Gibbons:2003gq,Thorndike:2003}.
In any case, the dominant errors on $\vert V_{ub} 
\vert$ are theoretical. Typically, theory-related error from the inclusive 
measurements is of 
$O(15\%)$ at present (and somewhat higher from the exclusive decays),
in comparison with the experimental error (statistics and detector 
systematics) on this quantity, which is typically of $O(7\%)$.  This is 
reflected in the world averages for $\vert V_{ub} \vert$ presented by 
Stone~\cite{Stone:2003pc} and Schubert~\cite{Schubert:2003},~respectively, 
\begin{eqnarray} 
\vert V_{ub} \vert &=& (3.90 \pm  0.16({\rm exp}) \pm 0.53 ({\rm theo})) 
\times 10^{-3}\,, 
\nonumber\\[-1.5mm] 
\label{vubwavs}\\[-1.5mm] 
\vert V_{ub}
\vert &=& (3.80^{+0.24}_{-0.13}({\rm exp}) \pm 0.45 ({\rm theo})) \times
10^{-3}\,.
\nonumber 
\end{eqnarray}
Adding the errors in quadrature, the first of these leads to $\vert V_{ub} \vert = (3.90 
\pm  0.55) \times 10^{-3}$, yielding $\delta \vert V_{ub} \vert/\vert 
V_{ub} \vert= 14\%$, and a very similar range if one uses the value given in the
second. Thus, the matrix element
$\vert V_{ub} \vert$ is still considerably uncertain, and we trust that the $B$ 
factory experiments and theoretical developments  
will make a major contribution here, pushing the error down to
its theoretical limit $O(5\%)$, mentioned earlier.  

Using the current averages, $\vert V_{cb} \vert =(41.2\pm 2.0) \times 10^{-3}$
and $\vert V_{ub} \vert = (3.90 \pm  0.55) \times 10^{-3}$, we get
\be
\frac{\vert V_{ub} \vert}{\vert V_{cb} \vert}=0.095 \pm 0.014 \qquad \Longrightarrow
~R_b=\frac{1}{\lambda}\frac{\vert V_{ub} \vert}{\vert V_{cb} \vert}=0.42 \pm 0.06\,,
\label{rbcurrent}
\ee
which determines one side of the unitarity triangle.

\section{Status of the Third Row of $V_{\rm CKM}$}
Knowledge about the third row of the CKM matrix $ V_{\rm CKM}$ is crucial 
in quantifying the  FCNC  transitions $b \to s$ and $b \to d$ (as well as 
$s \to d$) and to search for physics beyond the SM. The
FCNC transitions in the SM are generally dominated  
by the  top quark contributions giving rise to the dependence 
on the matrix elements $\vert V_{tb}^*V_{ts}\vert$ 
(for $b \to s$ transitions) and 
$\vert V_{tb}^*V_{td}\vert$ (for $b \to d$ transitions). 
Of these, only the matrix element $\vert V_{tb}\vert$ has been 
measured by a tree amplitude $t \to W b$ at
the Tevatron through the ratio

\be
R_{tb}\equiv\frac{{\cal B}(t \to Wb)}{{\cal B}(t \to Wq)} =\frac{\vert 
V_{tb}\vert^2}{\vert V_{td}\vert^2 + \vert V_{ts}\vert^2 +
\vert V_{tb}\vert^2}\,.
\ee

\noindent
The current measurements yield~\cite{Hagiwara:fs}: 
$R_{tb}=0.94^{+0.31}_{-0.23}$, which in turn gives:

\be
\vert V_{tb} \vert =0.96^{+0.16}_{-0.23}~~\Longrightarrow~~\vert V_{tb} 
\vert > 0.74~~({\rm at}~95\%~{\rm CL})\,.
\ee

Thus, this matrix element is consistent with unity, expected from 
the unitarity relation $\vert V_{ub} \vert^2 + \vert V_{cb} \vert^2 + \vert V_{tb} 
\vert^2=1$, though the current  precision on the direct measurement of $\vert V_{tb} 
\vert$ is rather modest. (Unitarity gives $\vert V_{tb}\vert  \simeq 0.9992$.) 
The precision on $\vert V_{tb}\vert$ will be greatly improved, in particular, at 
a Linear Collider, such as TESLA~\cite{Aguilar-Saavedra:2001rg}, but
the corresponding measurements of $\vert V_{ts}\vert$ and $\vert V_{td}\vert$ from the 
tree processes are not on the cards. They will have to be determined by (loop)
induced processes which we discuss below.

\subsection{Status of $\vert V_{td} \vert$}
The current best measurement of $\vert V_{td} \vert$ comes from 
$\Delta M_{B_d}$, the mass difference between the two mass eigenstates
of the $B_d^0$ - $\overline{B_d^0}$ complex. This has been measured in a 
number of experiments and is known to an accuracy of $\sim 1\%$; the current
world average is~\cite{hfag03} $\Delta M_{B_d}= 0.502 \pm 0.006$ (ps)$^{-1}$. 

In the SM, $\Delta M_{B_d}$ and its counterpart $\Delta M_{B_s}$, the mass 
difference in the $B_s^0$ - $\overline{B_s^0}$ system, are calculated by box 
diagrams, dominated by the 
$Wt$ loop. Since  $(M_W, m_t) \gg m_b$, $\Delta M_{B_d}$ is governed by the
short-distance physics. The expression for 
$\Delta M_{B_d}$ taking into account the perturbative-QCD corrections reads as 
follows~\cite{Buras:2001pn}
\be
\Delta M_{B_d} = \frac{G_F^2}{6 \pi^2} \, \hat{\eta}_B \, 
\vert V_{td} V_{tb}^* \vert^2  \, 
m_{B_d}  \, (f_{B_d}^2\hat{B}_{B_d})  \, M_W^2  \, S_0(x_t)\,.
\label{buras1990fn}
\ee
The quantity $\hat{\eta}_B$ is the NLL perturbative QCD renormalization of the
matrix element of the $(\vert \Delta 
B\vert =2, \Delta Q=0)$ four-quark operator, 
whose value is $\hat{\eta}_B=0.55 \pm 
0.01$ ~\cite{Buras:1990fn}; $x_t=m_t^2/M_W^2$ 
and $S_0(x_t)=x_tf_2(x_t)$ is an 
Inami-Lim function~\cite{Inami:1980fz}, with
\be
f_2(x)=\frac{1}{4} +\frac{9}{4}\frac{1}{(1-x)} -\frac{3}{2}\frac{1}{(1-x)^2}
-\frac{3}{2}\frac{x^2 \ln x}{(1-x)^3}\,.
\label{inamilimf2}
\ee
The quantity $f_{B_d}^2\hat{B}_{B_d}$ enters through the hadronic 
matrix element 
of the four-quark box operator,~defined~as:
\be
\langle \bar{B}_q^0 | (\bar{b} 
\gamma_\mu(1-\gamma_5)q)^2 |B_q^0 \rangle \equiv \frac{8}{3}f_{B_q}^2 B_{B_q} M_{B_q}^2\,,
\label{bqhatfb}
\ee
with $B_q =B_d$ or $B_s$. With $\Delta M_{B_d}$ and $\hat{\eta}_B$ known to a very 
high accuracy, and the current value of the top quark mass, defined in the
$\overline{\rm MS}$ scheme, $\bar{m}_t(m_t) =(167 \pm 5)$ GeV,  known to an 
accuracy of $\sim 3\%$, leading to $\delta S_0(x_t)/S_0(x_t)\simeq 
4.5\%$, the combined uncertainty on $\vert V_{td} \vert$ from all these factors 
is about 3\%. This is 
completely negligible in comparison with the current theoretical uncertainty on the 
matrix element $f_{B_d}\sqrt{\hat{B}_{B_d}}$. For example, 
$O(\alpha_s)$-improved
calculations in the QCD sum rule approach yield~\cite{Jamin:2001fw}
$f_{B_d}=(210\pm 19)$ MeV and~\cite{Penin:2001ux} $f_{B_d}=(206\pm 20)$ MeV,
whereas $\overline{B}_{B_d}$ in the $\overline{\rm MS}$ scheme in this approach 
is estimated as~\cite{Korner:2003zk}
$\overline{B}_{B_d} =1$ to within 10\%, yielding for the renormalization group 
invariant quantity $\hat{B}_{B_d} \simeq 1.46$, and an accuracy of 
about $\pm 15\%$ on $f_{B_d}\sqrt{\hat{B}_{B_d}}$.

Lattice calculations of $f_{B_d}\sqrt{\hat{B}_{B_d}}$ are uncertain due to
the chiral extrapolation. This is shown in Fig.~\ref{JLQCD-chi_fB} from the JLQCD 
collaboration~\cite{Aoki:2003xb}, in which the
quantity $\Phi_{f_{B_{q}}}\equiv f_{B_q}\sqrt{M_{B_q}}$ is plotted 
for $q=d,s$ as a  function of the pion mass squared, with both axes normalized 
with the  Sommer scale $r_0$  determined from the heavy quark potential at each 
sea quark mass. The  lattice calculations in this figure are done with two flavours of 
dynamical quarks $u$ and $d$,
for the $u$ and $d$ quark masses in the range $(0.7$ - $2.9)m_s$, with $m_s$  
being the strange quark mass. The solid line represents a linear plus quadratic 
fit in $(r_0m_\pi)^2$, which describes the lattice data well. This fit, however, 
does not contain the chiral logarithmic term, predicted by the Chiral perturbation 
theory~\cite{Grinstein:1992qt}
\be
\frac{f_{B_d}\sqrt{M_{B_d}}}{(f_{B_d}\sqrt{M_{B_d}})^{(0)}}
= 1 -\frac{3(1+ 3 g^2)}{4} \, 
\frac{m_\pi^2}{(4\pi f_\pi)^2} \,  \ln \frac{m_\pi^2}{\mu^2}
+...\,,
\label{chiralform}
\ee
where terms regular in $m_\pi^2$ are omitted, $f_\pi =130$ MeV, and $g$ 
is the
$B^*B\pi$ coupling in chiral perturbation theory. A recent lattice
calculation~\cite{Abada:2003un} 
gives $g_{B^* B \pi}=0.58 \pm 0.06 \pm 0.10$. A related quantity
$g_{D^* D\pi}$ has been 
determined from $D^* \to D \pi$ decay~\cite{Anastassov:2001cw}, $g_{D^* D\pi}=0.59 \pm 
0.01 \pm 0.07$. The value of $g$ is fixed at 
$g=0.6$ in drawing the
three curves with the chiral behaviour (\ref{chiralform}) with the three values of 
the hard chiral cutoff: $\mu= 300$ MeV (dotted curve), $\mu=500$ MeV (thin dashed 
curve) and $\mu =\infty$ (thick dashed curve). Lattice data with the currently 
used high values of the dynamical quarks is not able to verify the chiral 
logarithm. The data are not inconsistent with such a behaviour either.

 The chiral behaviour of the bag constant is given by the  expression 
\be
\frac{B_{B_d}}{B_{B_d}^{(0)}}
= 1 -\frac{(1- 3 g^2)}{2} \, 
\frac{m_\pi^2}{(4\pi f_\pi)^2} \, \ln \frac{m_\pi^2}{\mu^2}
+...\,,
\label{chiralform2}
\ee
the coefficient $(1-3g^2)/2$ is numerically small ($ =-0.04$), 
as opposed to the
coefficient in $\Phi_{f_{B_{d}}}$ for which $3(1+3g^2)/4=-1.56$, 
with $g=0.6$. Hence, extrapolation of the lattice data to the small 
quark masses poses no problems for the bag parameters.  

Taking this into account, the 
unquenched lattice QCD calculation from the 
JLQCD Collaboration~yields~\cite{Aoki:2003xb}
\be
f_{B_d}\sqrt{\hat{B}_{B_d}}=215(11)(^{+0}_{-23})(15)~{\rm MeV}\,,
\label{aoki2003xb}
\ee
where the first error is statistical, the second (asymmetric) 
is the uncertainty from the chiral extrapolation and the last is the 
systematic error from finite lattice spacing. The largest error is from 
the chiral extrapolation in $f_{B_d}$,
which in the conservative estimate of the JLQCD collaboration 
could be as large as -10\%, with $f_{B_d}=191(10) (^{+0}_{-19})(12)$~MeV. 
For further discussion,
see the recent reviews by Becirevic~\cite{Becirevic:2003hf},
Kronfeld~\cite{Kronfeld:2003} and Wittig~\cite{Wittig:2003cf}.
%
%
%
\begin{figure}[tbp]
  \centering
  \includegraphics*[width=12cm,clip=true]{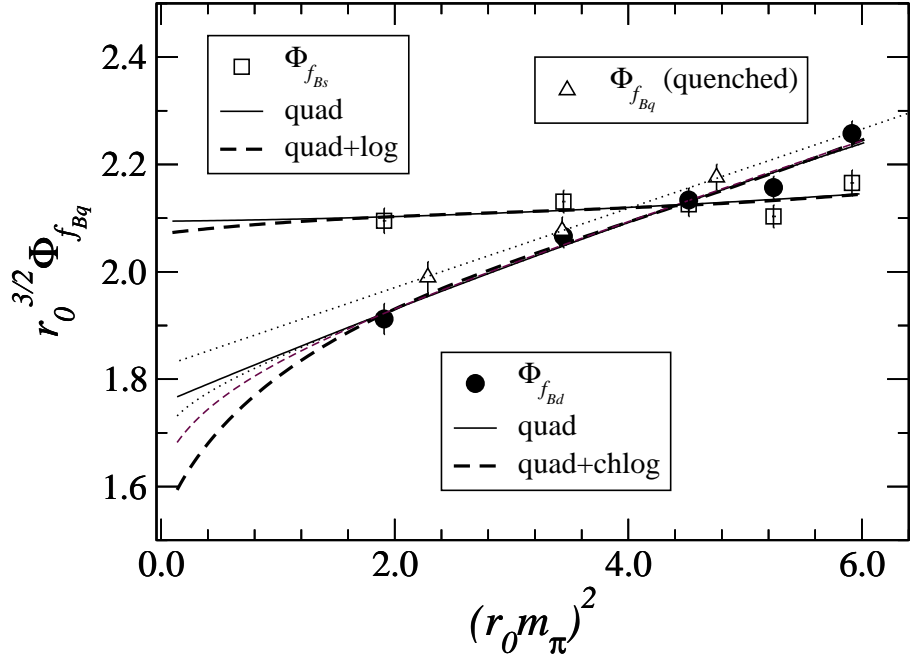}
  \caption{
    Chiral extrapolation of $\Phi_{f_{B_d}}$ (filled circles)
    and $\Phi_{f_{B_s}}$ (open squares).
    The quadratic extrapolation is shown by solid lines,
    while the fits with the hard cutoff chiral logarithm are
    shown for $\mu$ = 300 MeV (dotted curve), 500 MeV (thin dashed
    curve) and $\infty$ (thick dashed curve).
    Quenched results are also shown (triangles). (Figure taken from
    the JLQCD collaboration~\protect\cite{Aoki:2003xb}\,.)
  }
  \label{JLQCD-chi_fB}
\end{figure}

We shall use the unquenched lattice result in (\ref{aoki2003xb}) by 
adding the errors in quadrature and symmetrizing the errors, getting
\be 
f_{B_d}\sqrt{\hat{B}_{B_d}}=210 \pm 24~{\rm MeV}\,.
\label{fbdsqrtbd}
\ee
The dependence of $\vert V_{td} \vert$ on the various input parameters 
can be expressed through the following numerical~formula
\be
\vert V_{td} \vert = 8.5 \times 10^{-3} \bigg[\frac{210~{\rm 
MeV}}{f_{B_d}\sqrt{\hat{B}_{B_d}}}\bigg]
 \bigg[\frac{\Delta M_{B_d}}{0.50/{\rm ps}}\bigg]^{0.5}
\sqrt{\frac{0.55}{\hat{\eta}_B}} \sqrt{\frac{2.40}{S_0(x_t)}}\,,
\label{buras2003jf}
\ee
where the default value of $S_0(x_t)$ corresponds to $\overline{m_t}(m_t) =167$ 
GeV, and the dependence of this function on $m_t$~\cite{Buras:2003jf} is $S_0(x_t) 
\sim x_t^{1.52}$.
To get the $\pm 1 \sigma$ range for $\vert V_{td} \vert$, we vary the 
input parameters within their respective $\pm 1 \sigma$ range and add the errors
in quadrature. This exercise yields
\be
\vert V_{td} \vert = (8.5 \pm 1.0) \times 10^{-3}\,.
\label{vtdmass}
\ee
 In future, unquenched
lattice results for $f_{B_d}$ and other quantities will be available at smaller values of 
the dynamical quark masses
(than is the case in the current JLQCD calculation), allowing to check the chiral
logarithmic behaviour of $f_{B_d}$, or at least reduce the error associated with
this extrapolation.

Knowing $\vert V_{td} \vert$ from (\ref{vtdmass}), and
$\vert V_{cb} \vert$, and $\lambda$ from previous sections, one can determine 
the other side of the UT, which has the following central value:
\be
R_t \equiv \frac{1}{\lambda}\frac{\vert V_{td} \vert}{\vert V_{cb} \vert} = 0.93 
\bigg[\frac{\vert V_{td} \vert}{8.5 \times 10^{-3}}\bigg] \bigg[\frac{0.041}
{\vert V_{cb} \vert}\bigg]\,.
\ee
Taking into account the errors (and taking symmetric errors on $\vert V_{td} 
\vert$), we get $R_t=0.93 \pm 0.12$.
\subsection{ $\vert V_{td} \vert$ from $B \to \rho \gamma$ decays}
Independent information on $\vert V_{td} \vert$ (more 
precisely on $\bar{\rho}$ and 
$\bar{\eta}$) will soon be available from the radiative decays $B \to (\rho,\omega) 
\gamma$. There is quite a lot of theoretical interest lately in this process,
starting from the earlier papers a decade ago~\cite{Soares:1993dx,Ali:vd}, 
where the potential impact of these decays on the CKM phenomenology was first worked out 
using the leading  order estimates for the penguin amplitudes. 
Since then, annihilation 
contributions have been estimated in a number of   
papers\cite{Ali:1995uy,Khodjamirian:1995uc,Grinstein:2000pc}, 
and the next-to-leading 
order corrections to the decay amplitudes have also been 
calculated~\cite{bdgAP,Bosch:2001gv}. 
Deviations from the SM estimates in the branching 
ratios, isospin-violating asymmetry $\Delta^{\pm 0}$ and 
CP-violating asymmetries
${\cal A}_{\rm CP}(\rho^\pm \gamma)$ and 
${\cal A}_{\rm CP}(\rho^0 \gamma)$ have also
been worked out in a number of theoretical  
scenarios~\cite{Ali:2000zu,Ali:2002kw,Ko:2002ee}.
These CKM-suppressed radiative penguin decays were searched for by the CLEO 
collaboration~\cite{Coan:1999kh}, and the searches have been set 
forth at the $B$ 
factory experiments BELLE~\cite{Nakao:2003qt} and BABAR~\cite{Aubert:2003me}. 
The current upper limits at $\cl{90}$ (averaged over the charge 
conjugated modes) are given in Table \ref{ulbdgamma}.
%
\begin{table}[htbp]
\caption{$90\%$ confidence level upper limits on the branching ratios
(in units of $10^{-6}$) for the 
decays $B \to \rho \gamma$ and $B \to \omega \gamma$ from the 
CLEO~\protect\cite{Coan:1999kh},
BELLE~\protect\cite{Nakao:2003qt} and BABAR~\protect\cite{Aubert:2003me} 
collaborations.}
\label{ulbdgamma}

\vspace*{3mm}
\begin{center}
\renewcommand{\arraystretch}{1.3} 
\begin{tabular}{l c c c}\hline \hline\\[-4mm]
 & $\hspace*{5mm}$ ${\cal B}(B^+ \to \rho^+ \gamma)$ $\hspace*{5mm}$ 
& $\hspace*{5mm}$ ${\cal B}(B^0 \to \rho^0 \gamma)$ $\hspace*{5mm}$ 
& $\hspace*{5mm}$ ${\cal B}(B^0 \to \omega \gamma)$
\\[1.5mm] \hline \\[-4mm]
CLEO (9.1~fb$^{-1}$) & $13.0$ & $17.0$ & $9.2$\\
BELLE (78~fb$^{-1}$) & $2.7$ & $2.6$ & $4.4$\\
BABAR (78~fb$^{-1}$) & $2.1$ & $1.2$ & $1.0$ \\[2mm] 
\hline \hline
\end{tabular}
\end{center}
\end{table}

The BABAR upper limits on ${\cal B}(B^+ \to \rho^+ \gamma)$ 
and ${\cal B}(B^0 \to 
\rho^0 \gamma)$ have been combined using isospin symmetry
to yield an improved upper limit~\cite{Aubert:2003me}
\beq
{\cal B}(B \to \rho \gamma) < 1.9 \times 10^{-6}\; .
\label{brhobabar}
\eeq
Together with the current measurements of the branching
ratios for $B \to K^* \gamma$ decays, studied earlier, this yields
a $\cl{90}$ upper limit on the isospin-weighted and charge-conjugate 
averaged ratio~\cite{Aubert:2003me}
\be
{\bar R}(\rho \gamma/K^*\gamma) \equiv  \frac{{\cal B}(B \to \rho
\gamma)}{{\cal B}(B\to K^* \gamma)} < 0.047 \; .
\label{rhogamaexp}
\ee

The branching ratios for $B \to \rho \gamma$ have been 
calculated in the SM at next-to-leading 
order~\cite{bdgAP,Bosch:2001gv} in the QCD factorization 
framework~\cite{Beneke:1999br}.  
As the absolute values of the form factors in this decay and in $B \to K^* \gamma$ decays
discussed earlier are quite uncertain, it is advisable to calculate
instead the ratios of the branching ratios
\begin{equation}
R^\pm(\rho \gamma/K^*\gamma) \equiv \frac{{\cal B} (B^\pm \to \rho^\pm
\gamma)}{{\cal B} (B^\pm \to K^{*\pm} \gamma)},
\label{rpm}
\end{equation}
\begin{equation}
 R^0(\rho \gamma/K^*\gamma) \equiv \frac{{\cal B} (B^0 \to
\rho^0 \gamma)}{{\cal B} (B^0 \to K^{*0} \gamma)}.
\label{rzero}
\end{equation}
The results in the NLO accuracy can be
expressed as \cite{bdgAP}:
\bea
R^\pm (\rho \gamma/K^* \gamma) &=&  \left| V_{td} \over V_{ts} \right|^2
 \frac{(M_B^2 - M_\rho^2)^3}{(M_B^2 - M_{K^*}^2)^3}~\zeta^2
(1 + \Delta R^\pm(\epsilon_{\rm A}^\pm, \bar{\rho}, \bar{\eta})) \; , 
\nonumber\\[-1.5mm] 
\label{rapp}\\[-1.5mm] 
R^0 (\rho \gamma/K^* \gamma) &=& {1\over 2} \left| V_{td}
 \over V_{ts} \right|^2
 \frac{(M_B^2 - M_\rho^2)^3}{(M_B^2 - M_{K^*}^2)^3}~\zeta^2
(1 + \Delta R^0 (\epsilon_{\rm A}^0, \bar{\rho}, \bar{\eta})) \; ,
\nonumber
\eea
where
$\zeta=T_1^{\rho}(0)/T_1^{K^*}(0)$, with $T_1^{\rho}(0)$ and $T_1^{K^*}(0)$
being the form factors evaluated at $q^2=0$ in the decays $B \to \rho \gamma$ 
and $B \to K^* \gamma$, respectively. The functions
 $\Delta R^\pm (\epsilon_{\rm A}^\pm, \bar{\rho}, \bar{\eta})$
 and $\Delta R^0 (\epsilon_{\rm A}^0, \bar{\rho}, \bar{\eta})$, 
appearing on the r.h.s. of the above equations encode
both the $O(\alpha_s)$ and annihilation contributions, 
and they have a non-trivial dependence on the CKM 
parameters $\bar{\rho}$ and $\bar{\eta}$~\cite{bdgAP,Bosch:2001gv}.
Updating them, incorporating also a shift in the quantity
called $\lambda_B^{-1}$, related to an integral over the $B$-meson LCDA,
$\lambda_B^{-1}=\int_0^\infty dk/k ~\phi_{+}(k,\mu)$, which has been evaluated
in the QCD sum rule approach recently by Braun and Korchemsky~\cite{Braun:2003wx},
$\lambda_{B}^{-1} =(2.15 \pm 0.50)$ GeV$^{-1}$, the result for the functions in 
(\ref{rapp})
is~\cite{bdgAP-2003}
\be
\Delta R^\pm =0.056 \pm 0.10\,, \qquad \Delta R^0=-0.010 \pm 0.064\,,
\label{apupdate1}
\ee
where the uncertainties reflect also the variations in the CKM parameters $\bar{\rho}$
and $\bar{\eta}$, for which the ranges $\bar{\rho}=0.21 \pm 0.09$ and $\bar{\eta}=0.34 
\pm 0.05$ have been used.

Theoretical uncertainty in the evaluation of the ratios $R^\pm (\rho \gamma /K^* 
\gamma)$ and $R^0 (\rho \gamma /K^* \gamma)$ is dominated by the imprecise knowledge 
of the quantity $\zeta$. In the SU(3) limit $\zeta=1$; SU(3)-breaking corrections 
have been calculated in several approaches, including the QCD sum rules and Lattice QCD.
In the earlier calculations of the ratios, the following ranges were used
$\zeta=0.76 \pm 0.06$ (by Ali and Parkhomenko~\cite{bdgAP}), $\zeta=0.76 \pm 0.10$
(Ali and Lunghi~\cite{Ali:2002kw}) and $1/\zeta=1.33\pm 0.13$, leading to $\zeta=0.75 \pm 0.07$ 
(Bosch and Buchalla~\cite{Bosch:2003ckm}). These ranges reflect
the  earlier estimates of this quantity in the QCD sum rule approach
\cite{Ali:1995uy,Ball:1998kk,Narison:1994kr,Melikhov:2000yu}, and indicates
substantial SU(3) breaking in the $B \to V$ form factors.  Now there 
exists
an improved Lattice estimate of this quantity, with the result~\cite{Becirevic:2003}
$\zeta=0.9 \pm 0.1$, which is within $1\sigma$ compatible with no SU(3)-breaking!
We conclude that $\zeta$ is at present poorly determined. It is essential to
calculate it precisely if the measurement of $\bar{R} (\rho\gamma/K^*\gamma)$ is to make 
an impact on the CKM phenomenology.  

To illustrate the impact of the current bound on $\bar{R} (\rho\gamma/K^*\gamma)$,
we use the following estimate
\be
\zeta= 0.85 \pm 0.10\,.
\label{zetarange}
\ee
 Including the uncertainties from other
input parameters, the updated results are~\cite{bdgAP-2003}
\bea
R^\pm (\rho \gamma /K^* \gamma) &=& 0.033 \pm 0.012 \,,
\nonumber\\[-1.5mm] 
\\[-1.5mm]
R^0 (\rho \gamma /K^* \gamma) &=& 0.016\pm 0.006 \,.
\nonumber
\eea
Combining these with the measured values of the branching ratios ${\cal B}(B^\pm \to 
K^{*\pm} \gamma)$ and ${\cal B}(B^0 \to K^{*0} \gamma)$, the
predictions for ${\cal B}(B^\pm \to \rho^{\pm} \gamma)$ and ${\cal B}(B^0 \to 
\rho^{0} \gamma)$ are as follows:
\bea
{\cal B}(B^\pm \to \rho^{\pm} \gamma)&=&(1.36 \pm 0.49[{\rm th}] 
\pm 0.10 [{\rm exp}])\times 10^{-6}\,,
\nonumber\\[-1.5mm]
\\[-1.5mm]
{\cal B}(B^0 \to \rho^{0} \gamma) &=& (0.64 \pm 0.23 [{\rm th}] \pm 0.04
 [{\rm exp}]) \times 10^{-6}\,,
\nonumber
\eea
and ${\cal B}(B^0 \to \rho^{0} \gamma) = {\cal B}(B^0 \to \omega \gamma)$
for the stated range of theoretical uncertainty. Comparing these predictions with the 
present experimental bounds given in Table \ref{ulbdgamma}, we expect that all 
these 
branching ratios lie within a factor 2 of the current experimental bounds, and hence will 
be measured soon.

The isospin-weighted and charge-conjugation averaged 
ratio $\bar{R}(\rho\gamma/K^*\gamma)$
is given by the following expression in the SM~\cite{bdgAP}
\begin{eqnarray}
\bar R (\rho\gamma/K^*\gamma) & = & \frac{\lambda^2 \zeta^2}{2} \,
\frac{(M_{B_d}^2 - m_\rho^2)^3}{(M_{B_d}^2 - m_{K^*}^2)^3} \,
\bigg \{
\left [ 1 - \bar \rho + \varepsilon_A^{\pm} \bar \rho \right ]^2 
+\left ( 1 - \varepsilon_A^{\pm} \right )^2 \bar \eta^2 
\nonumber \\
& + & {\rm Re} \left [ G_1 (\bar \rho, \varepsilon_A^{\pm}) 
+ \bar \eta^2 \,  G_2 (\varepsilon_A^{\pm}) \right ] + (\varepsilon_A^{\pm}
\to \varepsilon_A^{0})
\bigg \} ,
\label{rhokstarvtd}
\end{eqnarray}
where the NLO contribution are introduced through the functions:
\begin{eqnarray}
G_1 (\bar \rho, \varepsilon_A) & = & \frac{2}{C_7^{(0) {\rm eff}}}
\left \{ (1 - \bar \rho)^2
\left [ A_{\rm sp}^{(1)\rho} - A_{\rm sp}^{(1)K^*} \right ]\right.
+
\left.\bar \rho (1 - \bar \rho)
\left [ A^u + \varepsilon_A A^{(1)t} \right ]
+ \bar \rho^2 \varepsilon_A  A^u
\right \} ,
\nonumber \\[-1.5mm] 
\label{auxiliar}\\[-1.5mm] 
G_2 (\varepsilon_A) & = & \frac{2}{C_7^{(0) {\rm eff}}}
\left \{ A_{\rm sp}^{(1)\rho} - A_{\rm sp}^{(1)K^*} - A^u
+ \varepsilon_A \left [ A^u - A^{(1)t} \right ]
\right \} .
\nonumber 
\end{eqnarray}
Here, $\epsilon_A$ represents the annihilation contribution, estimated as
$\epsilon_A^\pm \simeq 0.3 \pm 0.07$~\cite{Ali:1995uy} with $\epsilon_A^0 \ll 
\epsilon_A^\pm$ due to being colour- and electric charge suppressed,
and the other quantities in (\ref{rhokstarvtd}) and (\ref{auxiliar}) 
can be seen in the literature~\cite{bdgAP}.
 
The dependence of  the ratio $\bar{R}(\rho\gamma/K^*\gamma)$ on $\vert 
V_{td}\vert/\vert V_{ts}\vert$  is shown in~Fig.~\ref{rhogamackm}. Note, 
that this deviates from a quadratic dependence, which holds only if one 
neglects the annihilation and $O(\alpha_s)$ corrections. Including these 
corrections, we have given the dependence of 
$\bar{R}(\rho\gamma/K^*\gamma)$ on $\bar{\rho}$ and $\bar{\eta}$
explicitly. The solid curve corresponds to the central values of
the input parameters, and the dashed curves are obtained by taking
into account the $\pm 1 \sigma$ errors on the individual input parameters in 
$\bar{R}(\rho\gamma/K^*\gamma)$ and adding the errors in quadrature.
The current experimental upper limit on $\bar{R}(\rho\gamma/K^*\gamma)$, 
given in (\ref{rhogamaexp}), is also shown. Taking the least restrictive 
of the three theoretical curves, the current experimental upper limit yields
$\vert V_{td}\vert/\vert V_{ts}\vert < 0.28$, to be compared with the SM range
$\vert V_{td}\vert/\vert V_{ts}\vert =0.21 \pm 0.03$, shown as the region 
between the two dotted horizontal lines. Conversely, constraining the
ratio $\vert V_{td}\vert/\vert V_{ts}\vert$ in the SM range, we get 
$\bar{R}(\rho\gamma/K^*\gamma)=0.032 \pm 0.008$. 
%
%
\begin{figure}[htbp] 
\centerline{
            \psfig{width=0.45\textwidth,file=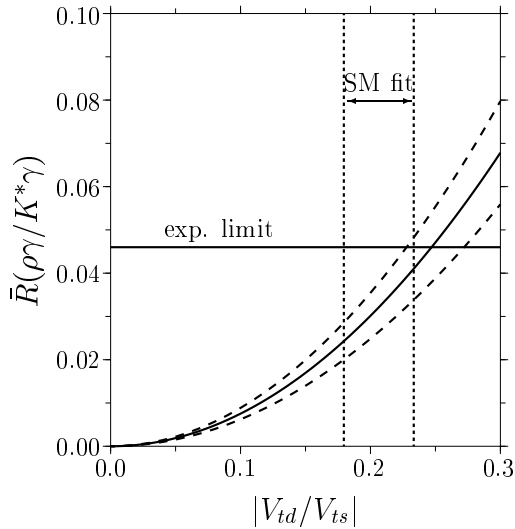}}
\caption{Ratio of the branching ratios for the decays $B \to \rho \gamma$
         and $B \to K^* \gamma$, $\bar{R}(\rho\gamma/K^*\gamma)$,  
         plotted as a function of  $|V_{td}/V_{ts}|$.
         The current experimental upper limit $\bar{R}(\rho\gamma/K^*\gamma) < 0.47$ 
         is shown as the horizontal line. The
         vertical dotted lines demarcate the $\pm 1 \sigma$ range of
         $|V_{td}/V_{ts}|$ from the SM-based unitarity fits.
         The solid curve
         corresponds to the central values of the input parameters and
         the dotted curves delimit the $\pm 1 \sigma$ range in the NLO corrected
         calculations for $\bar{R}(\rho\gamma/K^*\gamma)$ in the SM.
         (Figure updated from Ref.~\protect\cite{bdgAP}.)}
\label{rhogamackm}
\end{figure}

\subsection{Present status of $\vert V_{ts} \vert$}
There are two measurements at present which yield indirect information on
$\vert V_{ts} \vert$, namely the lower bound on the $B_s^0$ - $\overline{B_s^0}$
mass difference $\Delta M_{B_s}$ and the
measurement of the branching ratio ${\cal B} (B \to X_s \gamma)$. We discuss them 
in turn below.

The expression for $\Delta M_{B_s}$ in the SM can be obtained from the one for
$\Delta M_{B_d}$ in (\ref{buras1990fn}) by the replacements: $B_d \to B_s$, $V_{td} \to 
V_{ts}$. 
However, as opposed to $\Delta M_{B_d}$, $\Delta M_{B_s}$ is not yet measured, and the 
current lower bound (at 95\% C.L.) 
is \cite{hfag03} $\Delta M_{B_s} > 14.4~{\rm (ps)}^{-1}$.
This can be converted into a lower bound on $\vert V_{ts} \vert$, knowing
$f_{B_s}\sqrt{\hat{B}_{B_s}}$. 
The unquenched lattice QCD calculation from the JLQCD 
Collaboration for $f_{B_s}$ and 
$f_{B_s}\sqrt{\hat{B}_{B_s}}$~yields~\cite{Aoki:2003xb}
\bea
f_{B_s}&=&215(9)(^{+0}_{-2})(13)(^{+6}_{-0})~{\rm MeV}\,, 
\nonumber\\[-1.5mm]
\label{aoki2003fbs}\\[-1.5mm]
f_{B_s}\sqrt{\hat{B}_{B_s}}&=&255(10)(^{+3}_{-2})(17)(^{+7}_{-0})~{\rm MeV}\,,
\nonumber 
\eea
where the errors have the same origin as in (\ref{aoki2003xb}), and the 
additional error here is due to the ambiguity in the determination of the $s$-quark mass.
An average of the JLQCD, MILC and CP-PACS data gives~\cite{Becirevic:2003hf} 
$f_{B_s}\sqrt{\hat{B}_{B_s}} =254 \pm 13 \pm 14 \pm 13$ MeV.
However, a recent
calculation of the coupling constants $f_{B_s}$ and $f_{D_s}$ in the 
unquenched lattice QCD including the effects of one strange sea quark 
and two light sea quarks by Wingate {\it et al.}~\cite{Wingate:2003gm} yields,
\bea
f_{B_s} &=& 260 \pm 7 \pm 26 \pm 8 \pm 5~{\rm MeV}\,,
\nonumber\\[-1.5mm]
\label{wingateetal}\\[-1.5mm]
f_{D_s} &=& 290 \pm 20 \pm 29 \pm 29 \pm 6~{\rm MeV}\,,
\nonumber 
\eea
where the errors are respectively due to statistics and fitting, 
perturbation theory, relativistic corrections, and discretization effects. 
The result for $f_{B_s}$ in~(\ref{wingateetal}) is typically 20\% higher 
compared to the one from the JLQCD collaboration given in
(\ref{aoki2003fbs}). As both of these calculations are based on the NRQCD 
framework for heavy quarks, the difference between the two lies in the 
details of the lattice simulations, such as the dynamical quark masses 
used and $n_f$, which are different in the two approaches. Based on these 
comparisons, we conclude that the current lattice precision on 
$f_{B_s}\sqrt{\hat{B}_{B_s}}$ is of order $20\%$.  

It has become customary to use the ratio of the mass 
differences $\Delta M_{B_d}/\Delta M_{B_s}$ to constrain $\vert V_{td} 
\vert/\vert V_{ts} 
\vert$ from the SM relation~\cite{Ali:1978kk}:
\be
\frac{\Delta M_{B_s}}{\Delta M_{B_d}}=\xi \, \frac{M_{B_s}}{M_{B_d}} \, 
\frac {\vert V_{tb}^*V_{ts}\vert^2}{\vert V_{tb}^*V_{td}\vert^2}\,,
\label{deltamsd}
\ee
where 
\be
\xi \equiv\frac{f_{B_s}\sqrt{\hat{B}_{B_s}}}{f_{B_d}\sqrt{\hat{B}_{B_d}}}\,.    
\label{xidef}
\ee
Theoretical  uncertainty in $\xi$ 
is arguably smaller compared to the one in $f_{B_s}\sqrt{\hat{B}_{B_s}}$, as in the 
SU(3) limit $\xi=1$, and the uncertainty is
really in the estimate of SU(3)-breaking corrections. Thus, $\delta \xi \simeq 10\%$
is not an unrealistic error on $\xi$. 
Current estimates can be exemplified by 
the unquenched lattice calculations of $\xi$ from the JLQCD collaboration
~\cite{Aoki:2003xb}
\be
\xi = 1.14(3)(^{+13}_{-0})(2)(^{+3}_{-0})\,.
\label{aokixi}
\ee
 The single largest uncertainty in $\xi$ is due to the chiral extrapolation - 
the same source as in the estimates  of $f_{B_d}\sqrt{\hat{B}_{B_d}}$. Symmetrizing 
the errors, JLQCD result yields
\be
\xi =1.19 \pm 0.09\,.
\label{xijlqcd}
\ee
It should be remarked that the quantity $\Xi\equiv\xi f_\pi/f_K$ is useful to
estimate $\xi$, as the chiral logs largely cancel in $\Xi$~\cite{Becirevic:2002mh}.
Using the JLQCD data ~\cite{Aoki:2003xb}, Kronfeld~\cite{Kronfeld:2003} quotes 
$\xi=1.23 \pm 0.06$, with the error increasing if one includes the preliminary
HPQCD results~\cite{Wingate:2003ni}, yielding $\xi=1.25 \pm 0.10$. The errors in
this and the one in (\ref{xijlqcd}) are almost the same and about $8\%$.
Thus, $\xi$ has a value in the range $1.1 \leq \xi \leq 1.3$. 

The constraint that a measurement of (or equivalently a bound on) $\Delta M_{B_s}$ 
provides on $R_t$ can be expressed as follows 
\be
R_t=0.90 \, \bigg[ \frac{\xi}{1.20}\bigg] \, 
\sqrt{\frac{17.3/{\rm ps}}{\Delta M_{B_s}}} \, 
\sqrt{\frac{\Delta M_{B_d}}{0.50/{\rm ps}}}\,,
\label{deltamscon}
\ee
where the default value of $\Delta M_{B_s}$ is the best-fit value from the CKM 
unitarity fits, discussed in detail later~\footnote{I acknowledge the help
provided by Enrico Lunghi and Alexander Parkhomenko in updating the CKM unitarity 
fits.}.
The present bound  $\Delta M_{B_s} > 14.4~({\rm ps})^{-1}$ yields
\be
\frac{\vert V_{td}\vert}{\vert V_{ts} \vert} > 0.22 ~~\Longrightarrow~~
\vert V_{tb}^* V_{ts} \vert > 0.034\,,
\ee
where the last inequality follows from using $\vert V_{tb}^* V_{td} \vert
> 7.5 \times 10^{-3}$, determined earlier.
This is to be compared with the CKM-unitarity constraint $\vert V_{tb} V_{ts}^* 
\vert= \vert 
V_{cb}V_{cs}^*\vert + O(\lambda^2)$, which for the central value of $\vert 
V_{cb}\vert =0.041$, predicts $\vert V_{ts} \vert=0.04$.

\subsection{Determination of $\vert V_{ts} \vert$ from $\Gamma (B \to X_s 
\gamma)$}
We now discuss the determination of $\vert V_{ts} \vert$ from 
${\cal B}(B \to X_s 
\gamma)$. The effective Hamiltonian
which governs the transition $B \to X_s \gamma$ in the SM is 
\be \label{Heff}
{\cal H_{\rm eff}} =  -\frac{4 G_F}{\sqrt{2}} \, V_{ts}^* V_{tb} \, 
 \sum_{i=1}^8 C_i(\mu) O_i(\mu),
\ee
which is obtained by integrating out all the particles that are
much heavier than the $b$-quark. The operators $O_i$ can be seen, for
example, in Ref.~\cite{Ali:2002jg}, and unitarity of the CKM matrix is used
to factorize the CKM-dependence of ${\cal H_{\rm eff}}$ in the multiplicative product 
$V_{ts}^* V_{tb}$.

The decay rate for $B \to X_s \gamma$ is calculated by taking into account the 
QCD perturbative and power corrections. What concerns the perturbative corrections,
there are two effects: (i) renormalization of the Wilson coefficients $C_i(M_W) \to
C_i(\mu_b)$, where $\mu_b \sim O(m_b)$, and (ii) perturbative corrections to the matrix 
elements of the operators $\langle O_i \rangle$. From step (i), one has a 
perturbative expansion for $C_i(\mu_b)$~\cite{Chetyrkin:1996vx}
\be \label{cmb}
C_i(\mu_b) = C_i^{(0)}(\mu_b)+ \frac{\alpha_s(\mu_b)}{4\pi}
C_i^{(1)}(\mu_b) + \ldots.
\ee
In the leading order, i.e., without the QCD corrections, $C_i^{(0)}(\mu_b) = 
C_i(M_W)$
and $C_i^{(1)}(\mu_b)=0$, and of the Wilson coefficients $C_i (M_W)$ only 
$C_7(M_W)$, corresponding to the electromagnetic penguin
operator $O_7= e m_b/g_s^2 (\bar{s}_L \sigma^{\mu \nu} b_R)F_{\mu \nu}$, is relevant
for the decay $b \to s \gamma$. As $C_7(M_W)$  is dominated by the (virtual) top 
quark contribution, the amplitude ${\cal M}(b \to s \gamma)$ is proportional to 
$\lambda_t=V_{tb}V_{ts}^*$. In this order, the contributions
from the intermediate up and charm quarks are negligible, being power suppressed
due to the GIM mechanism~\cite{Glashow:gm}. So, in the leading order, the decay width
$\Gamma (B \to X_s \gamma)$ depends 
quadratically on~$\vert V_{ts}^* V_{tb}\vert$.
 
However, including  QCD corrections, the power-like GIM-suppression of the 
intermediate up and charm quarks is no longer operative. The reason for this is that QCD 
forces a very significant operator-mixing between the
operator $O_2= (\bar{s}_L \gamma_\mu c_L)(\bar{c}_L \gamma^\mu b_L)$,
whose Wilson coefficient $C_2$ is of order 1, and the electromagnetic penguin
operator $O_7$,
whose Wilson coefficient is much smaller, $C_7(\mu) \ll 1$. Hence the
contribution from the intermediate charm state
becomes numerically very important bringing with it the 
dependence of the decay amplitude on $\lambda_c=V_{cb}V_{cs}^*$.
The contribution from the intermediate $u$-quark can always be expressed in
terms of $\lambda_t$ and $\lambda_c$, using the unitarity relation
$\Sigma_{u,c,t}\lambda_i=0$. However, on noting that  $\lambda_u/\lambda_t 
=O(\lambda^2)$, $\lambda_u$ can be dropped, to an excellent approximation, 
yielding $\lambda_c=-\lambda_t$, and the electromagnetic penguin amplitude $b \to s 
\gamma$ factorizes in $\lambda_t$ ($=-\lambda_c$).

Following this line of argument,  one  
fixes the value of  $\lambda_t=-\lambda_c =(41.0 \pm 2.0) \times 
10^{-3}$ in  calculating the SM decay rate for $B \to X_s
\gamma$. Thus, for example,
in the $\overline{\rm MS}$ scheme, taking into account the next-to-leading 
order (in $\alpha_s)$ and leading order (in $1/m_c^2$ and
$1/m_b^2$) power corrections, the SM branching 
ratio~is~\cite{Gambino:2001ew,Buras:2002tp}:
\be
{\cal B}(B \to X_s \gamma) = (3.70 \pm 0.30) \times 10^{-4}\,,
\label{gambino}
\ee
to be compared with the current experimental world average (based on the CLEO, 
ALEPH, BELLE and BABAR measurements) of this quantity~\cite{Jessop02}
\be
{\cal B}(B \to X_s \gamma) = (3.34 \pm 0.38) \times 10^{-4}\,.
\label{bsgamwav}
\ee
The consistency of the two implies that  the CKM unitarity, implemented through 
the unitarity relation
\be
V_{ub}V_{us}^* + V_{cb}V_{cs}^* + V_{tb}V_{ts}^* =0\,,
\label{bstriangle}
\ee
holds within experimental and theoretical precision. Note, that this is a different 
unitarity relation than the one given in (\ref{vudtriangle}) and shown in
Fig.~\ref{triangle}. While it does not provide any information on the CKM parameters
$\bar{\rho}$ and $\bar{\eta}$, or for that matter on $\alpha$, $\beta$ and $\gamma$,
it involves the CKM matrix element $V_{ts}$, apart from the other known ones, and
its best use is to determine $\vert V_{ts} \vert$ and its argument $\delta \gamma_s$.
 
We address the question how to convert the information on ${\cal B}(B \to X_s \gamma)$ 
to determine $\vert V_{ts} \vert$. To do that, we have to keep the 
individual contributions from the intermediate $u$, $c$ and $t$ 
quarks with their respective CKM dependencies, $\lambda_u$, $\lambda_c$ and 
$\lambda_t$, and {\it not} invoke unitarity, in calculating the decay width $\Gamma(B 
\to X_s \gamma)$. This is just a different book keeping of the partial contributions 
to the decay amplitude in the SM.
Dropping small numerical contributions $(< 2.5\%)$, current measurements of ${\cal 
B}(B \to X_s \gamma)$  yield the following relation~\cite{alimisiak:2003}
\be
\vert 1.69 \lambda_u + 1.60 \lambda_c + 0.60 \lambda_t \vert =(0.94 \pm 0.07) 
\vert V_{cb} \vert\,.
\label{alimisiak}
\ee
Note the much larger coefficient of $\lambda_c$ compared to the coefficient of
$\lambda_t$, reflecting the large $O_2$ - $O_7$ mixing under QCD renormalization.
Note also that the relative signs of $\lambda_c$ and $\lambda_t$ are opposite,
which means that a destructive interference between the charm and top quark 
contributions is absolutely essential to explain the observed branching ratio for
$B \to X_s \gamma$. Solving this equation with the known values of the CKM factors,
$\lambda_c= (41.2 \pm 2.0) \times 10^{-3}$ and  $\lambda_u$ (which is complex) from 
the discussions earlier (or from PDG) yields~\cite{alimisiak:2003}
\be
\lambda_t = V_{tb} V_{ts}^*=-(47 \pm 8) \times 10^{-3}\,.
\label{vtsbsg}
\ee
The reason for the large error on $\lambda_t$ is, apart from the experimental
precision on ${\cal B}(B \to X_s \gamma)$, the relatively small coefficient of
this term in (\ref{alimisiak}).   
Within measurement errors, this determination of $\lambda_t$ 
is consistent with the CKM unitarity expectations $\vert V_{ts} \vert = \vert 
V_{cb} \vert +O(\lambda^2)$, though it is less precise at present than $\vert V_{cb} 
\vert$. However, due to the interference of the terms proportional to
$\lambda_c$ and $\lambda_t$, ${\cal B}(B \to X_s \gamma)$  determines the
relative sign of $\lambda_t$ and $\lambda_c$. The sign in (\ref{vtsbsg}) is in 
accord with the Wolfenstein parametrization given in (\ref{CKM-W}), which has
$\lambda_t = -A\lambda^2$. Hence, $A$ is positive definite.

\section{Summary of the Current Status of $V_{\rm CKM}$ and the Weak 
Phases}
The current knowledge of the magnitudes of the CKM matrix 
elements that we have discussed in the previous sections is summarized 
in Table~\ref{vijmeas}. Note, that these are direct measurements 
in the sense that
unitarity has not been used in arriving at these entries. The 
corresponding values  obtained on using the unitarity constraints are lot 
tighter, as also discussed above for some specific matrix elements.
%
%
\begin{table}[htbp]
\caption{Summary of current measurements of $\vert V_{ij} \vert$.
\label{vijmeas}}

\vspace*{2mm}
\begin{center}
\renewcommand{\arraystretch}{1.3} 
\begin{tabular}{l c c }\hline \hline\\[-4mm]
$\vert V_{ij} \vert$ 
& Value 
& $\delta \vert V_{ij} \vert/\vert V_{ij} \vert$\\[1.5mm] 
\hline \\[-4mm]
$\vert V_{ud} \vert $&$ 0.9739 \pm 0.0005 $& $5 \times 10^{-4}$\\
$\vert V_{us} \vert $&$ 0.2224 \pm 0.0020 $& $9 \times 10^{-3}$\\
$\vert V_{ub} \vert $&
$\hspace*{15mm}$ $ (3.90 \pm 0.55) \times 10^{-3}$ $\hspace*{15mm}$ 
& $ 14\%$\\
$\vert V_{cd} \vert $&$ 0.224 \pm 0.016 $& $ 7\%$\\
$\vert V_{cs} \vert $&$ 0.97 \pm 0.11 $& $ 11\%$\\
$\vert V_{cb} \vert $&$ 0.041 \pm 0.002 $& $ 5\%$\\
$\vert V_{td} \vert $&$ (8.5 \pm 1.0) \times 10^{-3}$ & $ 12\%$\\
$\vert V_{ts} \vert $&$ 0.047 \pm 0.008 $& $ 17\%$\\
$\vert V_{tb} \vert $&$ 0.96^{+0.16}_{-0.23} $& $ 20\%$\\[2mm] 
\hline \hline
\end{tabular}
\end{center}
\end{table}

This information can also be expressed in terms of the 
Wolfenstein parameters and the sides of the unitarity triangle in the SM:
\bea
& & A = 0.83 \pm 0.04\,, 
\nonumber\\[0.8mm]
& & \lambda = 0.2224 \pm 0.0020\,, 
\nonumber\\[-1.9mm] 
\label{wolfparams}\\[-1.9mm] 
& & R_t = 0.93 \pm 0.12\,, 
\nonumber\\[0.8mm]
& & R_b = 0.42 \pm 0.06\,,
\nonumber 
\eea
where the range of $R_t$ given above is coming from $\Delta M_{B_d}=0.503 \pm 0.006$
(ps)$^{-1}$. The current lower bound on $\Delta M_{B_s}$ also gives a constraint on 
$R_t$, which already is quite effective in restricting 
the allowed $\bar{\rho}$ - $\bar{\eta}$ space in the SM. Finally, 
the quantity called $\bar{R}(\rho \gamma/K^* \gamma)$ also constrains
$\bar{\rho}$ and $\bar{\eta}$, but  the current upper limit is less
effective than either $\Delta M_{B_d}$ or $\Delta M_{B_s}$.

So far, we have seen that the unitarity of the CKM matrix, as determined through the
magnitudes of the matrix elements in each row or each column, holds with deviations
which are statistically not significant. In the rest of this section, we use the 
information on  the CP-violating asymmetries in the Kaon and $B$-meson systems to see 
first the consistency of the data in terms of the sides and the angles of the unitarity 
triangle, and then, as the final step, we undertake a fit of
all the relevant data to determine the apex of the unitarity triangle shown in 
Fig.~\ref{triangle}. This will allow us to update the predictions for some interesting 
quantities which have either not been measured yet, such as $\Delta M_{B_s}$, or not 
precisely enough, such as the angles $\alpha$ and $\gamma$.

\subsection{Precise tests of the CKM theory including CP-violating phases}
In addition to the constraints that we have already given 
in~(\ref{wolfparams}), (\ref{deltamscon}) (for $\Delta M_{B_s}$), 
the theoretical expression for $\bar{R}(\rho \gamma/K^*\gamma)$ 
given in (\ref{rhokstarvtd}) and
(\ref{auxiliar}), and  the current experimental bound on this ratio in 
(\ref{rhogamaexp}), there are three precise measurements involving 
CP violation in the $K$- and $B$-meson sectors providing constraints on 
the CKM parameters. We discuss them briefly here.

The observed CP-Violation in the $K_L \to \pi \pi$ and $K_S \to 
\pi \pi$ decays  yield the following information on the quantities
$\vert \epsilon_K \vert$ and 
${\rm Re}(\epsilon^\prime/\epsilon)$~\cite{Hagiwara:fs} 
\bea
& & \vert \epsilon_K \vert = (2.280 \pm 0.013) \times 10^{-3}\,, 
\nonumber\\[-1.5mm] 
\label{cpinput}\\[-1.5mm] 
& & {\rm Re}(\epsilon^\prime/\epsilon) = (16.6 \pm 1.6) \times 10^{-4}\,.
\nonumber 
\eea
The value quoted for ${\rm Re}(\epsilon^\prime/\epsilon)$ is the world average  
from the  NA48~\cite{Lai:2001ki} and KTEV~\cite{Alavi-Harati:1999xp} 
collaborations, including also the  earlier results from their forerunners, NA31 and E731, 
respectively. While  ${\rm Re}(\epsilon^\prime/\epsilon)$ is a benchmark measurement in 
flavour physics,  as so far this
is the only well established example of CP violation in decay amplitudes, unfortunately its
impact on the CKM phenomenology is muted due to the imprecise knowledge of the hadronic
quantities  needed to extract the information on the CKM parameters quantitatively. Given the 
hadronic uncertainties, 
which admittedly are not small, the measured value of ${\rm Re}(\epsilon^\prime/\epsilon)$ is in 
agreement with the SM estimates.
 For further details and discussions, the interested reader is referred to
a recent review on this subject by Buras and Jamin~\cite{Buras:2003zz}, where 
references to the original theoretical papers in the analysis of ${\rm 
Re}(\epsilon^\prime/\epsilon)$ can 
also be found.
 
Recently, time-dependent CP asymmetries in $B \to J/\psi K_S$ and related decays have been 
measured. Denoting these asymmetries generically by $a_{\psi K_S}(t)$, one can measure
$\sin 2 \beta$ (or $\sin 2 \phi_1$) from the time dependence of the asymmetry 
\be
a_{\psi K_S}(t) \equiv a_{\psi K_S}\,\sin (\Delta M_{B_d} t)
=\sin 2\beta ~\sin (\Delta M_{B_d} t)\,.
\label{apsiksdef}
\ee
 As opposed to the theoretical predictions for $\vert \epsilon_K \vert$ and 
${\rm Re}(\epsilon^\prime/\epsilon)$, the CP asymmetry $a_{\psi K_S}(t)$ is free of hadronic 
uncertainties~\cite{Carter:tk}, which allows to write down the above expression. 
Current measurements of $a_{\psi K_S}$ are
dominated by the  BABAR~\cite{Aubert:2002ic} ($a_{\psi K_S} 
=0.741 \pm 0.067 \pm 0.033$) and BELLE~\cite{Abe:2003yu} ($a_{\psi 
K_S}=0.733
\pm 0.057 \pm 0.028$) measurements, and the world average~\cite{Browder:2003,hfag03},
\be
 a_{\psi K_S}= 0.736 \pm 0.049\,,
\label{apsiksinput}
\ee
is in good agreement with the predictions in the SM~\cite{Hagiwara:fs}. 
We shall also quantify this agreement below.

 We now discuss the constraints on the CKM parameters in the SM that 
follow from the measurements of $\vert \epsilon_K \vert$ and $a_{\psi K_S}$.
The expression for $\vert \epsilon_K \vert$ in the SM, including NLO 
corrections~is~\cite{Buras:1984pq}  
\bea
\vert \epsilon_K \vert = C_K \, \hat{B}_K  \, (A^2 \lambda^6 
\bar{\eta}) \, \bigg(x_c \, \left ( {\eta}_3 
f_3(x_c,x_t) -{\eta}_1\right )
+ {\eta}_2 \,x_t  \,f_2(x_t) \, A^2  \,\lambda^4  \,(1-\bar{\rho})\bigg)\,,
\label{epsilonksm}
\eea
where $C_K=G_F^2f_K^2m_K M_W^2/6\sqrt{2} \pi^2 \Delta M_K$,
with~\cite{Hagiwara:fs} $\Delta M_K=(3.490 \pm 0.006) \times 10^{-12}$ MeV  
the $K^0$ - $\overline{K^0}$
mass difference and $f_K=(159.8 \pm 1.5)$ MeV the $K$-meson coupling constant; 
 $x_i=m_i^2/M_W^2$, and  $f_2(x)$ and $f_3(x,y)$ are the Inami-Lim 
functions~\cite{Inami:1980fz}, of which we have already given $f_2(x)$ in
(\ref{inamilimf2}), and $f_3(x,y)$ is given by the following expression (for $x\ll y$) 
\be
f_3(x,y) = \ln\frac{x}{y} - \frac{3y}{4(1-y)}\bigg(1 +\frac{y}{1-y} \ln y \bigg).
\label{inamilimf3}
\ee
The ${\eta}_i$ are perturbative renormalization constants calculated in NLO
accuracy~\cite{Buras:1990fn,Herrlich:1993yv}. They depend on the definition of the
quark masses, and the values $\eta_2=0.57 \pm 0.01$ and $\eta_3=0.46 \pm 0.05$  that will 
be used for numerical analysis below correspond to the
definitions $m_t=m_t(m_t)$ and $m_c=m_c(m_c)$, which represent the quark masses in the 
$\overline{\rm MS}$ scheme at their indicated scales. With this definition, the residual 
$m_t$-dependence of $\eta_2$, and that of $\eta_3$ on $m_t$ and $m_c$, are negligible, but
the dependence of $\eta_1$ on $m_c$ is significant. Following
Gambino and Misiak in the CERN-CKM proceedings~\cite{Battaglia:2003in}, we shall use
\be
\eta_1 = (1.32 \pm 0.32)\left (\frac{1.30}{m_c(m_c)}\right )^{1.1}\,,
\label{eta1}
\ee
and take $m_c(m_c)=1.25 \pm 0.1$ GeV. The quantity
$\hat B_K$ is the bag parameter, for which lattice QCD estimates yield
$\hat B_K= 0.86(6)(14)$~\cite{Lellouch:2002nj}. In working out the constraints from
$\epsilon_K$ numerically, we shall take~$\hat B_K= 0.86 \pm 0.15$.

A useful numerical expression showing the constraints that the current 
value of $\vert \epsilon_K \vert$ provides  
on the CKM parameters is as follows~\cite{Buras:2003jf}:
\be
\bar \eta \left [ (1 - \bar \rho) \, A^2  \, {\eta}_2  \, S_0 (x_t)
+ P_c (\varepsilon) \right ]  A^2  \, \hat B_K = 0.187,
\label{epsilonksmbur}
\ee
where $S_0 (x_t)=x_t f_2(x_t)\simeq 2.40 \, ( \bar m_t/167~{\rm GeV})^{1.52}$,
and $P_c (\varepsilon)$ summarizes the contribution from the 
first row in (\ref{epsilonksm}), which depends on both $m_c$ and $m_t$. Taking into 
account the dependence of ${\eta}_1$ on $m_c$, this quantity is estimated
as~\cite{Herrlich:1993yv} $P_c (\varepsilon)=0.29\pm 0.07$.

 Measurements of $a_{\psi K_S}=\sin 2 \beta$~\cite{Carter:tk}  translate into the following 
constraints on the Wolfenstein parameters 
\be
\sin 2 \beta = \frac{2 (1 - \bar \rho) \bar \eta}
                      {(1 - \bar \rho)^2 + \bar \eta^2},
\qquad {\rm or} \qquad
\bar \eta = \frac{1 \pm \sqrt{1 - \sin^2 2 \beta}}
                 {\sin 2 \beta} \, (1 - \bar \rho)\,.
\label{sintwobeta}
\ee
The constraints resulting from the five quantities $(a_{\psi K_S}, \vert \epsilon_K \vert, 
\Delta M_{B_d}, \Delta M_{B_s}$, and $R_b)$ on the CKM parameters $\bar{\rho}$ and $\bar{\eta}$
are shown in Fig.~\ref{figEtaRho}. Of these, the allowed bands correspond to  $\pm 1\sigma$
errors on $a_{\psi K_S}$, $\vert \epsilon_K \vert$, $\Delta M_{B_d}$ and $R_b$,  
and the constraint shown for $\Delta M_{B_s}$ is for $\xi=1.28$, the maximum value in the
$\pm 1 \sigma$ range given in (\ref{xijlqcd}), with $\Delta M_{B_s} > 14.4$ (ps)$^{-1}$.
Of the two solutions shown for $a_{\psi K_S}$, only one is compatible
with the measurement of $R_b$, i.e. with the SM, and for this solution we have a
consistent  description of all the data indicated in this figure; the resulting
allowed region is shown as a shaded area. Note, that this is not a fit, but a
simple consistency check of the CKM theory with a large number of measurements.
The three dotted curves labelled as $\bar{R}(\rho \gamma/K^*\gamma)$ 
refer to the 90\% C.L. bound on this ratio, and show the current theoretical 
uncertainty in the interpretation of this bound, which is dominated
 by the imprecise knowledge of $\zeta$: $\zeta =0.75$ (the leftmost curve), 
$\zeta=0.85$ (the central curve), and $\zeta=0.95$~(the rightmost curve).

 Fig.~\ref{figEtaRho} is quite instructive. 
%
%
\begin{figure}[htbp] 
\centerline{\psfig{width=0.95\textwidth,file=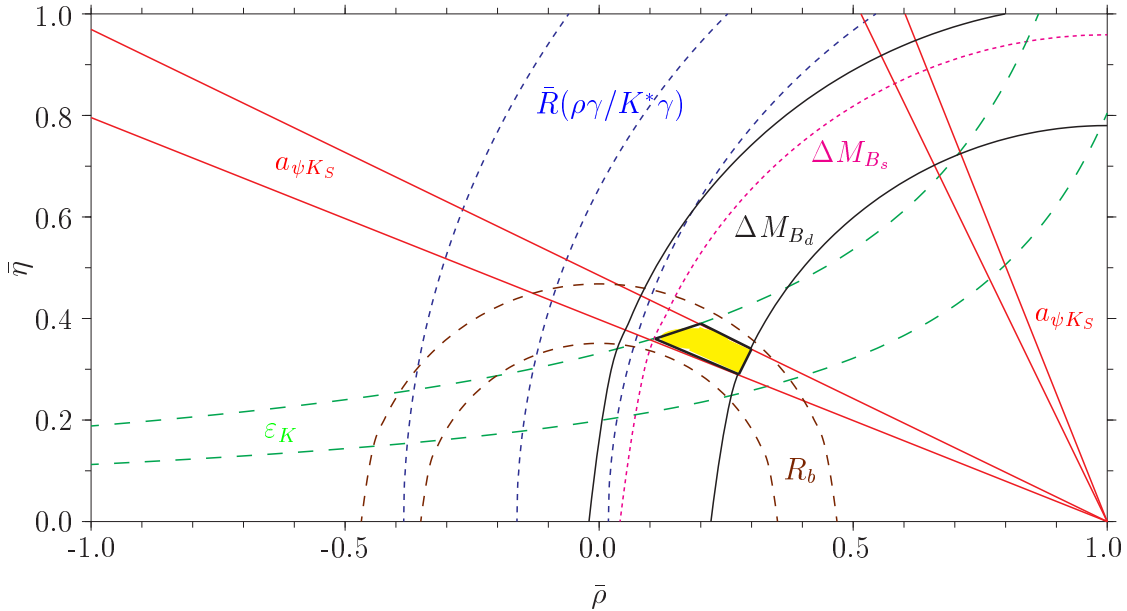}}
\caption{The constraints resulting from the measurements of $a_{\psi K_S}$,
         the ratio $R_b$, $\Delta M_{B_d}$, 
         $\varepsilon_K$, and the upper bounds on 
         $\Delta M_{B_s}$ and the isospin-weighted and
         charged-conjugate averaged ratio $\bar R (\rho\gamma /
         K^*\gamma)$, in the $\bar{\rho}$ - $\bar{\eta}$ plane. The overlap
         region is indicated by shaded area.}
\label{figEtaRho}
\end{figure}
It shows that the knowledge of $R_b$ is required to distinguish between 
the two allowed solutions for $a_{\psi K_S}$. However, for the solution 
compatible with the SM, the allowed region in the 
($\bar{\rho}$, $\bar{\eta}$) plane is now largely determined by 
the measurement of $a_{\psi K_S}$ and $\Delta M_{B_d}$, and the bound 
on $\Delta M_{B_s}$. The constraint from $\vert \epsilon_K \vert$ is still 
required and the compatibility of $\vert \epsilon_K \vert$ and 
$a_{\psi K_S}$ is an important consistency check of the CKM theory.  
However, as the bound on $\Delta M_{B_s}$ becomes stronger, more so if 
$\Delta M_{B_s}$ is measured which we anticipate soon, the allowed 
unitarity triangle could be determined entirely from the $B$-meson 
data. This is potentially good news, as the uncertainty on $\hat{B}_K$ is 
still quite substantial. This figure also reveals that the  theoretical 
uncertainty on the ratio $\bar{R}(\rho \gamma/K^* \gamma)$ 
has to  decrease by at least a factor 2 for it to be competitive with the 
constraints from $\Delta M_{B_d}$ and $\Delta M_{B_s}$. This requires a 
dedicated effort from the Lattice community, which is already under 
way~\cite{Becirevic:2002zp}.  
We hope that a robust calculation of SU(3) symmetry breaking in $\zeta$ will 
soon be undertaken, as  experiments are fast 
approaching the SM-sensitivity in $\bar{R}(\rho \gamma/K^* \gamma)$.

\begin{table}[htbp]
\caption{Input parameters used in the CKM-unitarity fits. 
Values of the other parameters 
are taken from the recent PDG~review.~\protect\cite{Hagiwara:fs}
\label{inputparms}}

\vspace*{3mm}
\begin{center}
\renewcommand{\arraystretch}{1.3} 
\begin{tabular}{c c c}\hline \hline\\[-4mm]
$\hspace*{3mm}$ Parameter $\hspace*{3mm}$ & $\hspace*{12mm}$
& Input Value\\[1.5mm] \hline \\[-4mm]
$\lambda $ 
& & $\hspace*{5mm}$ $0.2224 \pm 0.002~({\rm fixed})$ $\hspace*{5mm}$ \\ 
$\vert V_{cb} \vert $
& & $ (41.2 \pm 2.1) \times 10^{-3} $\\
$\vert V_{ub} \vert$
& & $(3.90 \pm 0.55) \times 10^{-3} $ \\
$a_{\psi K_S}$
& & $0.736 \pm 0.049 $\\
$\vert \epsilon_K \vert$
& & $(2.280 \pm 0.13) \times 10^{-3} $ \\
$\Delta M_{B_d} $& & $ 0.503 \pm 0.006~{\rm (ps)}^{-1}$\\
${\eta}_1(m_c(m_c)=1.30~{\rm GeV})$ 
&  & $ 1.32 \pm 0.32$\\
${\eta}_2  $ &  & $ 0.57 \pm 0.01 $\\
${\eta}_3$ &  & $ 0.46 \pm 0.05$\\
$m_c(m_c)$ &  & $ 1.25 \pm 0.10~{\rm GeV}$\\
$m_t(m_t)$ &  & $ 167 \pm 5~{\rm GeV}$\\
$\hat{B}_K$ &  & $ 0.86 \pm 0.15$\\
$f_{B_d} \sqrt{B_{B_d}}$ &  & $ (210 \pm 24)~{\rm MeV}$\\
$\eta_B$ &  & $ 0.55 \pm 0.01$\\
$\xi$ &  &  $ 1.19 \pm 0.09$\\
$\Delta M_{B_s} $ & & $ < 14.4~{\rm (ps)}^{-1}$ at 95\% C.L.
\\[2mm]
\hline \hline
\end{tabular}
\end{center}
\end{table}

\subsection{A Global fit of the CKM parameters and predictions for $\alpha$, 
$\gamma$ and~$\Delta M_{B_s}$}
To conclude this section, we give here the allowed ranges 
of the CKM-Wolfenstein parameters 
and the angles of the unitarity triangle obtained from a 
global fit of the data. 
Several input quantities that enter in the fits have evolved with time 
and their current values differ (see, Table~\ref{inputparms}) from the 
ones given in the CERN  CKM-Workshop proceedings~\cite{Battaglia:2003in}, and also from those 
used in the popular fits of the CKMfitter group~\cite{Hocker:2001xe}. Hence, a consistent
update is not out of place. 

First, a few words about the fits.
We follow the Bayesian analysis method in fitting the data. However, it should be pointed 
out that the debate on the Bayesian vs.~Non-Bayesian interpretation of data is a
lively subject and it has implications for the CKM fits. In the present context the
issues involved and the quantitative differences in the resulting profiles of the unitarity 
triangle are discussed in depth in the  
literature~\cite{Hocker:2001xe,Battaglia:2003in,Dubois-Felsmann:2003jd}. These differences 
are not crucial for our discussion, but the input values of the parameters are indeed
important. To incorporate the constraint from $\Delta M_{B_s}$, we have used the
modified-$\chi^2$ method (as described in the CERN CKM Workshop 
proceedings~\cite{Deltams:ckm}), which makes use of the ''Amplitude Technique''
introduced by Moser and Roussarie~\cite{Moser:1996xf}, in which the 
time-dependent oscillation probabilities are modified to have the dependence
$P(B_s^0(0) \to B_s^0 (t)) \propto ( 1 +{\cal A} \cos \Delta M_{B_s} t)$ and
$P(B_s^0(0) \to \overline{B_s^0} (t)) \propto ( 1 -{\cal A} \cos \Delta M_{B_s} t)$.
 The contribution to $\chi^2$ of the
fit from the analysis of $\Delta M_{B_s}$ is obtained using the following 
expression
\be
\chi^2= 2 \, \left [ {\rm Erfc}^{-1}\bigg(\frac{1}{2} {\rm Erfc} 
(\frac{1-A}{\sqrt{2}\sigma_A})\bigg) \right ]^2\,,
\label{modchisq}
\ee
where $A$ and $\sigma_A$ are the world average amplitude and the error,
respectively. Relegating the details to a forthcoming publication~\cite{bdgAP-2003}, the main 
results are summarized below.

The constraints in the ($\bar{\rho}$,$\bar{\eta}$) plane resulting from the 
five individual input quantities
($R_b$, $\epsilon_K$, $\Delta M_{B_d}$, $\Delta M_{B_s}$, and 
$a_{\psi K_S}$) are shown in Fig.~\ref{fitcontours} and  correspond to 95\% 
C.L., in contrast to the ones shown in Fig.~\ref{figEtaRho}.
The resulting 95\% C.L. fit contour  is shown in Fig.~\ref{fitcontours}. The apex of 
the triangle for the best-fit
solution ($\chi^2=0.57$ for two variables) is shown by the black dot and corresponds to 
the values $(\bar{\rho}, \bar{\eta})=(0.17, 0.36)$, with the 68\% C.L. 
range being
\be
0.11 \leq \bar {\rho} \leq 0.23\,, \qquad 0.32 \leq \bar{\eta} \leq 0.40 \,.
\ee
 We also show the constraint
from the 90\% C.L. upper bound on $\bar{R}(\rho \gamma/K^* \gamma)$ for
the value $\zeta=0.75$, though we have not used this input in the fits.
The 68\% C.L. ranges of the Wolfenstein parameters $A$, $\bar{\rho}$ and $\bar{\eta}$,
together with the corresponding ranges for the CP-violating phases
$\alpha, \beta, \gamma$, and the mass difference $\Delta M_{B_s}$  
are given in Table \ref{fitvalues}. Note that the fit-range for $\sin 2 \beta$ coincides 
practically with the experimental measurement $\sin 2 \beta = 0.736 \pm 0.049$. If we take 
away the input value of $\sin 2 \beta$ from the fits, and instead determine
$\sin 2 \beta$ from the unitarity fit, we get $\sin 2 \beta =0.730 \pm 0.085$, which is 
in remarkable agreement with the experimental value, but less precise.
In fact, similar estimates of $\sin 2 \beta$ from the CKM unitarity constraints were
obtained in the SM by several groups long before its measurements from the CP asymmetries 
~\cite{Ali:1999we,Mele:1998bf,Atwood:2001jr,Ciuchini:2000de,Plaszczynski:1999xs,Grossman:1997gw}.
Now, that the measurement of $a_{\psi K_S}$ is quite precise and its translation into
$\sin 2 \beta$ is free of hadronic uncertainties, this input 
has reduced the allowed parameter space in ($\bar{\rho}, \bar{\eta})$-plane - a feature already 
noted in the 
literature~\cite{Hocker:2001xe,Ali:2002kw,Battaglia:2003in,Dubois-Felsmann:2003jd}.

Prediction of $\Delta M_{B_s}$ from the fits deserves a discussion. First of all,
if we take away the bound on $\Delta M_{B_s}$ given in Table~\ref{inputparms} from the 
fits, the allowed range for
$\bar{\rho}$ becomes lot larger though $\bar{\eta}$ remains essentially unchanged, yielding
$0.08 \leq \bar{\rho} \leq 0.27$ and $0.31 \leq \bar{\eta} \leq 0.41$ at 68\% C.L.
The corresponding allowed range for $\Delta M_{B_s}$ in this case is $13.0 \leq \Delta 
M_{B_s}\leq 21.6$ (ps)$^{-1}$, with the 95\% C.L. interval being $[8.6, 26.2]$ (ps)$^{-1}$. This 
range is
to be compared with the corresponding one $[10.2, 31.4]$ (ps)$^{-1}$ from the CKMfitter
group~\cite{Hocker:2001xe}. The reason for this apparent 
mismatch lies in the values of the input
parameters $\xi$ and $f_{B_d} \sqrt{B_{B_d}}$, 
for which we take the currently updated
Lattice values $1.19 \pm 0.09$ and $(210 \pm 24)$ MeV, as opposed to
the values $1.21 \pm 0.04 \pm 0.05$ and $(228 \pm 30 \pm 10)$ MeV, 
respectively, used 
by the CKMfitter group.  With their input values,  
the resulting 95\% C.L. range for $\Delta M_{B_s}$ from our fits becomes
$[10.7, 30.3]$ (ps)$^{-1}$, which is almost identical to theirs. 
Hence, it is important
to know $\xi$ and $f_{B_d} \sqrt{B_{B_d}}$ more precisely.
 Including the $\Delta M_{B_s}$ bound in the fits, we get at 68\% C.L. 
$14.4 \leq \Delta M_{B_s} \leq 20.5$ (ps)$^{-1}$ 
with the 95\% C.L. range being $[14.4, 23.7]$ (ps)$^{-1}$.

\begin{table}[htbp]
\caption{68\% C.L. ranges for the Wolfenstein parameters,
CP-violating phases and $\Delta M_{B_s}$ from the 
CKM-unitarity fits.}
\label{fitvalues}

\vspace*{2mm}
\begin{center}
\renewcommand{\arraystretch}{1.3} 
\begin{tabular}{c c c}\hline \hline\\[-4mm]
$\hspace*{2mm}$ Parameter $\hspace*{2mm}$ 
& $\hspace*{15mm}$ & 68\% C.L. Range
\\[1.5mm] \hline \\[-4mm]
$\bar{\rho} $& & $ 0.11$ -- $ 0.23$\\
$\bar {\eta} $ & & $ 0.32$ -- $0.40 $ \\
$A$& & $ 0.79 $ -- $ 0.86 $\\
$\sin 2 \beta$ & &  $ 0.68$ -- $0.78 $\\
$\beta$ & &  $ 21.6^\circ$ -- $ 25.8^\circ$\\
$\sin 2 \alpha $ & &  $-0.44$ -- $0.30 $ \\
$ \alpha$ & &  $ 81^\circ$ -- $103^\circ$\\
$\sin 2 \gamma $ & &  $ 0.49$ -- $0.95$ \\
$\gamma $ & &  $53^\circ $ -- $75^\circ$\\
$\Delta M_{B_s}$ & & $ 14.4$ -- $20.5~{\rm (ps)}^{-1}$ \\[2mm] 
\hline \hline
\end{tabular}
\end{center}
\end{table}

$\hspace*{3mm}$

%
%
\begin{figure}[htb]
\centerline{\psfig{width=0.95\textwidth,file=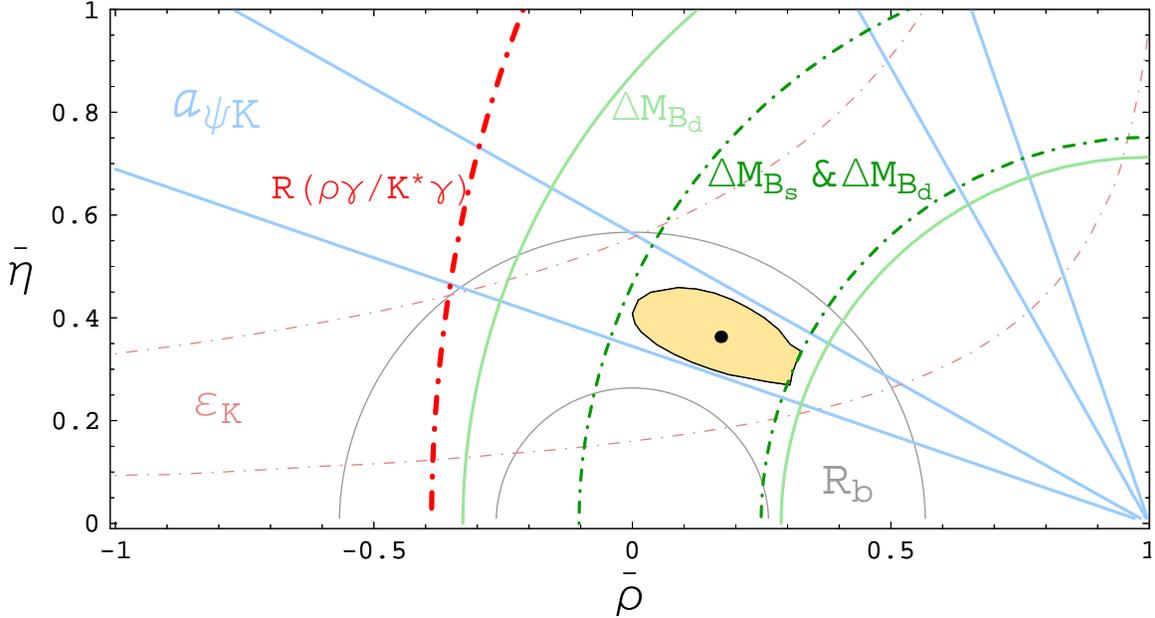}}
\caption{Constraints on the $\bar{\rho}$ - $\bar{\eta}$ plane from the
         five measurements as indicated. Note that the curves labelled as 
         $R(\rho \gamma/K^*\gamma)$
         and $\Delta M_{B_s}$ are obtained from their 95\% C.L. upper limits
         0.047 and 14.4 (ps)$^{-1}$, respectively. The fit contour corresponds 
         to 95\% C.L., with the fitting procedure explained in the text. The 
         dot shows the best-fit value. (Figure updated from 
         Ref.~\protect\cite{Ali:2002kw}.)}
\label{fitcontours}
\end{figure}

Concluding this section, we remark that
the overall picture that has emerged from the current knowledge of the CKM parameters
and hadronic quantities is that the CKM theory describes all data on 
flavour physics remarkably consistently, including $\vert \epsilon_K\vert$
and $a_{\psi K_S}$. Hence, it is very likely that CP violation in hadronic physics is 
dominated by the Kobayashi-Maskawa phase.  Despite this
impressive synthesis of flavour physics, a clean bill of
health for the CKM theory still awaits a number of benchmark measurements. These include
quantitative determination of the other two angles $\alpha$ (or $\phi_2)$ and $\gamma$ 
(or $\phi_3)$, and $\Delta M_{B_s}$. In all these cases, experiments have well-defined 
targets to shoot at, as the steady progress in the knowledge of the CKM parameters, and quite 
importantly the precise measurement of $\sin 2 \beta$, have reduced the allowed parameter 
space substantially. The 95\% C.L. ranges for these quantities resulting from 
the fits described earlier are: 
\be
 70^\circ \leq \alpha \leq 114^\circ\,, \qquad  
43^\circ \leq \gamma \leq 86^\circ\,, \qquad  
14.4~({\rm ps})^{-1}  \leq \Delta M_{B_s} \leq 23.7~({\rm ps})^{-1}\,.
\label{alilunghi-03}
\ee 
The long run to these goal posts has already started. We anticipate significant 
measurements of all three within the next couple of years in experiments at 
the $B$ factories and Tevatron, but definitely in experiments planned
at the hadron colliders (LHC-B, ATLAS, CMS and BTEV), which will measure
all three angles $(\alpha, \beta, \gamma)$ accurately, as well as $\Delta M_{B_s}$
and a number of rare $B$-decays, such as $B_s \to \mu^+ \mu^-$. Present 
situation together with some theoretical suggestions is reviewed in the next section.

\section{CP Violation in $B$-Meson Decays}
In the preceding section, we briefly discussed
the CP asymmetries $\vert \epsilon_K\vert$, ${\rm Re}~(\epsilon^\prime/\epsilon)$
and $a_{\psi K_S}$, representing three different ways in which CP violation 
has been measured so far in the weak decays of the hadrons. However, CP 
violation is a rich and diverse phenomenon~\cite{Branco:1999}. This is illustrated 
in Fig.~\ref{figbtofcp}, showing its various manifestations  
in the decays of a neutral meson ($P^0$) into a final CP eigenstate $(f_{\rm 
CP}$). Here, $P^0$ stands for any of the mesons $K^0, D^0, B_d^0, B_s^0$.
%
%
\begin{figure}[tbp]
\centerline{\includegraphics*[width=0.6\columnwidth]{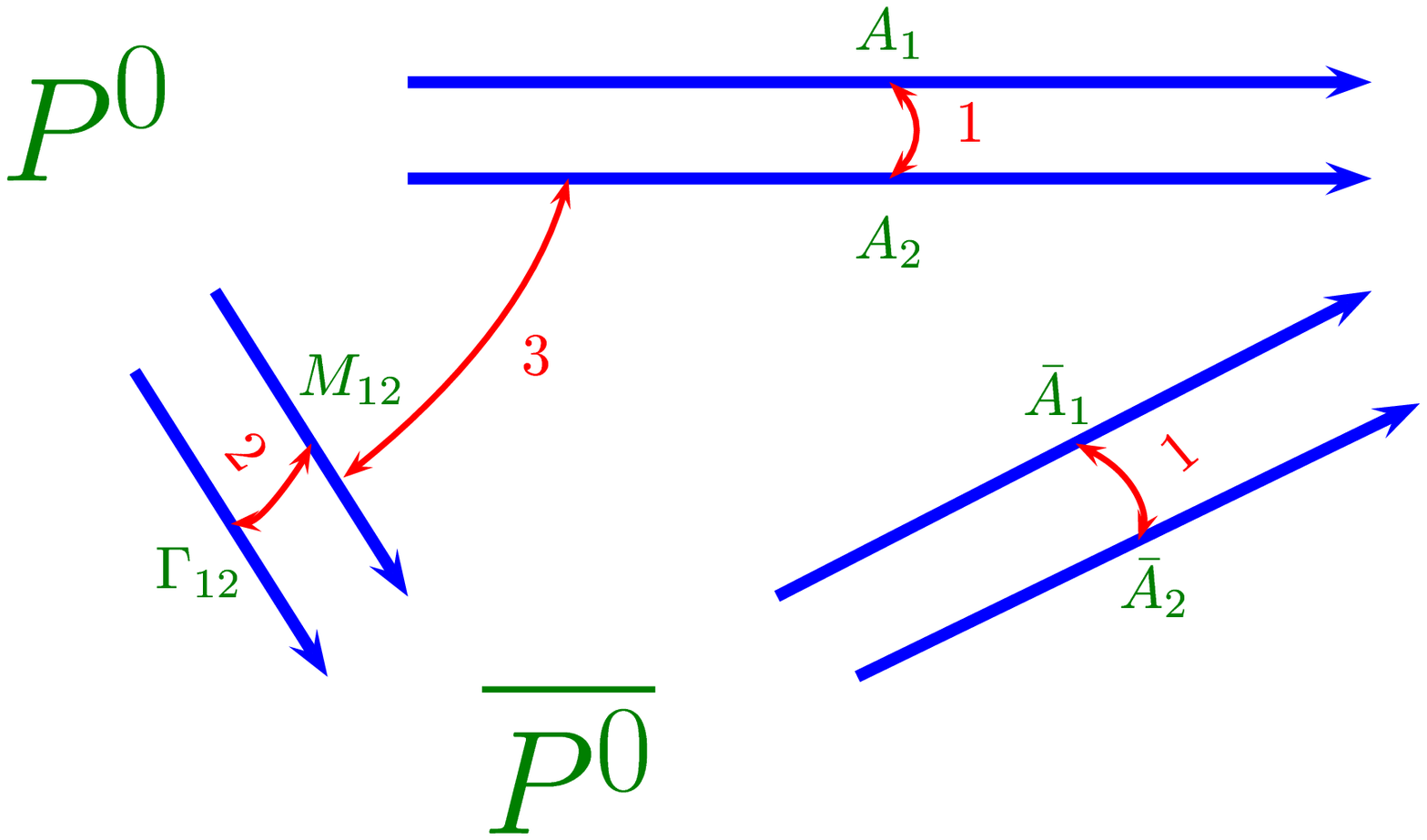}}
\caption{CP violation in neutral mesons decaying into a final CP eigenstate.
Each directed line represents an amplitude with its own weak phase,
and the connecting double arrows between these lines indicate possible interference 
patterns involving these amplitudes. (Figure taken from
Nir~\protect\cite{Nir:2002gu})}
\label{figbtofcp}
\end{figure}
As CP asymmetries arise due to the interference of two {\it different} amplitudes
with their own weak phases, 
this figure shows that there are three generic manifestations of CP violation in $P^0$ 
decays.
\begin{enumerate}
\item Interference between the two decay amplitudes, called $A_1$ and $A_2$.
\item Interference between the mass $(M_{12})$ 
and width $(\Gamma_{12})$ parts of the
 $P^0$ -- $\overline{P^0}$ mixing amplitude, giving rise to a non-vanishing relative weak
phase.
\item Interference of the decay amplitudes with and without mixing, 
 $P^0 \to \overline{P^0} \to f_{\rm CP}$ and $P^0 \to f_{\rm CP}$.
\end{enumerate}   
In the decays of charged mesons $P^\pm$ (such as $B^\pm$, $D^\pm$, $D_s^\pm$) and baryons,  
CP violation can take place only through the interference in the decay amplitudes. 

 Assuming CPT invariance, time reversal violation, or T-violation, implies CP 
violation. In the KM theory, CP- and T-violation have a common origin, namely the KM-phase 
of the CKM matrix. T-violation has been established in the $K^0$-$\overline{K^0}$ 
system. The measured T-violating asymmetry $A_T=(6.6 \pm 1.0) \times 10^{-3}$ is
consistent with the measurements of the CP-violating parameter  
$\epsilon_K$~\cite{Hagiwara:fs}, hence with the KM phase.
T-violations have also been searched for in flavour conserving transitions. The 
current best upper limits on the T-violating
electric dipole moments (EDMs) have been obtained for the neutron~\cite{Harris:jx},
Thallium-205~\cite{Regan:ta} and Mercury-199~\cite{Romalis:2000mg}:
\bea
& & \vert d_n \vert < 6.3 \times 10^{-26}~{\rm e~cm}~~(90\%~{\rm C.L.})\,, 
\nonumber \\
& & \vert d_e({\rm Tl-205}) \vert < 1.6 \times 
10^{-27}~{\rm e~cm}~~(95\%~{\rm C.L.})\,, 
\label{edms} \\
& & \vert d_e({\rm Hg-199})  \vert < 2.1 \times 
10^{-28}~{\rm e~cm}~~(95\%~{\rm C.L.})\,,
\nonumber
\eea 
where the last two are to be interpreted as limits on the EDM of the electron.  
Judging from the current upper limit on $d_n$ 
and the prediction of the KM theory $d_n =O(10^{-33})$ e cm~\cite{Shabalin:rs},
T-violations in flavour conserving transitions do not provide any information on the 
KM phase. As a measurement of $d_e$ is not foreseen in the SM either, a positive  
result in any of the three EDMs  will be a proof that additional weak phases are 
operative in the weak interactions of the hadrons and leptons. In particular, 
Supersymmetry has a myriad of weak phases, including the phases of the $A$ and $\mu$ 
terms,  which in some parts of the supersymmetric parameter space are 
large enough to yield values of the EDMs of the neutron, Hg-199 and Tl-205 
atoms just below the present experimental 
upper bound~\cite{Demir:2003js}.  In some cases, supersymmetric weak phases will 
also manifest themselves in CP violation in flavour changing transitions in the $B$- and 
$K$-meson sectors. For example, such theories may lead to CP-violating effects in 
the radiative decay
$B \to X_s \gamma$~\cite{Kagan:1998bh}. Recent analyses incorporating 
the constraints from $d_n$~\cite{Goto:2003iu} predict CP asymmetry in this decay  
close to the current experimental bounds~\cite{Nakao:2003gc}
\bea
{\cal A}_{\rm CP}(X_s\gamma) &=& 0.004 \pm 0.051 \pm 0.038\, \, \Longrightarrow 
-0.107 \leq {\cal A}_{\rm CP}(X_s\gamma) \leq 0.099\, \, (90\%\, {\rm C.L.})\, ,\\
\label{acpbsgamaexp}
{\cal A}_{\rm CP}(K^*\gamma) &=& (-0.5 \pm 3.7)\% \, \, \Longrightarrow
-0.066 \leq {\cal A}_{\rm CP}(K^*\gamma) \leq 0.056\, \, (90\% \, {\rm C.L.})\, .
\eea
The bounds on ${\cal A}_{\rm CP}(K^*\gamma)$ are stronger but they still allow this 
CP asymmetry to be of $O(5\%)$, much larger than the SM-based 
expectations~\cite{Ali:1998rr,Hurth:2003dk} $\vert {\cal A}_{\rm CP}(X_s \gamma)\vert \leq 
0.5\%$. 
It is entirely conceivable that  precision studies of CP violation 
in $B$- and $K$-decays may require the intervention of some of the weak phases in 
supersymmetric theories. The search for non-KM weak phases is an important part of the
research programme at the $B$-factories and later at the hadron colliders and 
has to be pursued vigorously. However, we will not discuss these scenarios here, as 
the principal focus of this review is on the CKM phenomenology and current data
show no significant deviations from the CKM theory.

In this section, we first discuss each of the three classes of CP asymmetries
depicted in Fig.~\ref{figbtofcp} and define the underlying physical quantities which 
are being sensitively probed in each one of them. This will be followed by a summary of 
the existing results on CP asymmetries from BABAR and BELLE in some of the main decay modes 
of  interest for our discussion, such as $J/\psi K_S, \phi K_S, \eta^\prime K_S, \pi \pi, K 
\pi$. In the SM, these asymmetries provide information on the weak phases $\alpha, 
\beta, \gamma$ (or $\phi_1, \phi_2, \phi_3$), and we review their current knowledge 
from the $B$-factory experiments.

\subsection{CP violation in decay amplitudes}
As opposed to $K$-mesons, direct CP violation in the $B$-meson sector
is potentially a very prolific phenomenon as the number of decay channels 
available in the latter is enormous~\cite{Ali:1997nh}. In practice, however, 
measuring  
some of these asymmetries, which are estimated at several percent or somewhat higher 
level~\cite{Ali:1998gb} 
in decay modes with typical branching ratios of order $10^{-5}$, requires 
$O(10^9)$ $B\bar{B}$ pairs (or more). 
With the present integrated luminosity of $O(10^8)$ $B\bar{B}$ mesons recorded by the 
BABAR and BELLE detectors, there 
are now indications in the data that these experiments are observing  
the first direct CP asymmetry in $B$ decays. The case in point is ${\cal  
A}_{\rm 
CP}(K^+\pi^-)=(-9.5 \pm 2.9)\%$~\cite{hfag03} having a 3$\sigma$ significance.

To study direct CP asymmetry, we note that unitarity can be used to write the decay 
amplitudes $A(B \to f)$ and its CP-conjugate
$A(\bar{B} \to \bar{f})$ as:

\bea
A(B \to f) &=& \vert A_1 \vert {\rm e}^{+i\theta_1}{\rm e}^{+i\delta_1}
            + \vert A_2 \vert {\rm e}^{+i\theta_2}{\rm e}^{+i\delta_2}\,, 
\nonumber\\[-1.5mm] 
\label{directcp}\\[-1.5mm] 
A(\bar{B} \to \bar{f}) &=& \vert A_1 \vert {\rm e}^{-i\theta_1}
{\rm e}^{+i\delta_1} 
            + \vert A_2 \vert {\rm e}^{-i\theta_2}{\rm e}^{+i\delta_2}\,,
\nonumber
\eea
where $\theta_1$ and $\theta_2$ are the weak phases, and the strong 
interaction amplitudes $\vert A_i \vert {\rm e}^{+i\delta_i}$ $(i=1,2)$ 
contain the CP-conserving strong phases $\delta_i$. With the help of these 
amplitudes, the CP-rate asymmetry can be written as  
\bea
 {\cal A}_{\rm CP}(f) \equiv \frac{\Gamma(\bar{B} \to \bar{f}) - \Gamma(B 
\to f)}
{\Gamma(\bar{B} \to \bar{f}) + \Gamma(B \to f)}
=  \frac{2 \vert A_1 \vert \vert A_2 \vert \sin 
(\delta_1 - \delta_2) \sin (\theta_1-\theta_2)}{\vert A_1 \vert^2 + 2 \vert 
A_1 \vert \vert A_2 \vert \cos (\delta_1 - \delta_2) \cos (\theta_1-\theta_2)
+ \vert A_2 \vert^2}\,.
\eea
The goal is to extract the weak phase difference
$\theta\equiv\theta_1 - \theta_2$ from the measured partial rate asymmetries, for which 
knowledge of the  strong interaction amplitudes $\vert  A_{1}\vert$, $\vert  A_{2}\vert$ 
and the strong-phase difference $\delta\equiv\delta_1 - \delta_2$ is essential.
The required strong phase difference involving a tree and penguin amplitudes $\delta= 
\delta_P -\delta_T$, or involving two penguin amplitudes with different strong phases 
$\delta= \delta_P
-\delta_{P^\prime}$, are generated by the so-called Bander-Silverman-Soni
mechanism~\cite{Bander:jx}. In addition, final state interactions also generate
strong phases, which have to be estimated or measured.

 From the CP-averaged decay rate $(\Gamma(B \to f) + \Gamma(\bar{B} \to \bar{f}))/2$
and the CP asymmetry ${\cal A}_{\rm CP}(f)$, one has two equations 
with four unknowns which we take as $r\equiv\vert A_2 \vert/\vert A_1 \vert$, $\delta$, 
$\theta$, and $\vert A_1 \vert$, with $r<1$. Hence, one has to
use additional experimental input and assumptions. For the decays $B \to h_1 
h_2$, with $h_i$ belonging to a $U(3)$ nonet of (usually vector or 
pseudoscalar) mesons, an
approximate SU(3) symmetry is invoked which allows to express 
$\vert A_1\vert$ in terms of a known branching ratio dominated by 
this amplitude. This then leaves only three unknowns with two 
constraints, which is still not sufficient to determine all the parameters,
but allows to derive bounds on the weak phase $\theta$ which could be useful
if data is benevolent.

While the phenomenology of direct CP violation is rich~\cite{Ali:1998gb}, a theoretically 
robust description is difficult and not yet at hand. Determining the weak phases 
from 
the observed asymmetries and decay rates requires a good control over the 
ratios of the amplitudes $\vert A_P/A_T\vert$, $\vert A_{P^\prime}/A_T\vert$, and
$\vert A_P/A_{P^\prime}\vert$ etc., and the corresponding strong-phase differences 
$\delta_P - \delta_T$, $\delta_{P^\prime} - \delta_T$ and $\delta_P - 
\delta_{P^\prime}$, where we are assuming that there are two different penguin 
amplitudes, one generated by the exchange of gluons and the other by electroweak
bosons ($\gamma$ and $Z$). Of course, direct CP asymmetries also arise when two
different tree amplitudes interfere, such as in $A_{\rm CP}(DK)$. Anyway, 
calculating them from 
first principles with a non-perturbative technique, such as Lattice QCD, is simply not on 
the cards, as the amplitudes $\vert A_i\vert {\rm e}^{i \delta_i}$ depend on  $\langle h_1 h_2 
|{\cal O} | B\rangle$, where ${\cal O}$ is a four-quark operator.
We know all too well that the vengeance of strong interactions is merciless in the
analysis of ${\rm Re}(\epsilon^\prime/\epsilon)$ in $K$-meson decays
involving the matrix elements $\langle \pi \pi|{\cal O} | K_{L,S}\rangle$, which has hindered
the extraction of any information on the CKM parameters from this~measurement.

 One would like to argue that strong interaction effects  
are tractable in $B$-decays due to the large mass of the $b$-quark, and hence calculable 
using 
techniques based on $1/m_b$-expansion and perturbative QCD~\cite{Beneke:1999br,Keum:2000ph}.
However, the dynamical scale in $B \to h_1 h_2$ decays is not set 
by the inverse $b$-quark mass, $1/m_b$, but by a lower value, of order $1$ 
-$2$ GeV - the typical virtuality of a gluon nested somewhere in the Feynman diagrams.
This raises the question if perturbative methods are adequate at such low scales.
Jury is still out on this issue, but when the jury returns one should not be surprised
at an unfavourable verdict for the perturbative-QCD inspired factorization models.

 Some help in extracting the weak phases from data can certainly be 
sought by using arguments based on the isospin and flavour SU(3) symmetries.
The most celebrated use of the isospin symmetry is in the extraction of the phase $\alpha$
(or $\phi_2)$ from the $B \to \pi \pi$ decays~\cite{Gronau:1990ka}.
While isospin is a good symmetry, there are still missing experimental links to complete 
the chain of
arguments to determine the weak phase $\alpha$ in a model-independent way. However, use of 
the SU(3) 
symmetry is more problematic. We have seen that the issue of SU(3)-breaking in simpler 
quantities, such as the ratios $\xi$ and $\zeta$, is far from being 
settled. In the much more complicated situation in non-leptonic decays, 
SU(3) symmetry breaking effects are at best modeled, often based on the {\it assumed} 
properties of the factorized amplitudes. This lack of a robust theoretical basis
and/or data introduces hadronic uncertainties in the extraction of the weak phases from 
data. We shall discuss applications of these methods in the context of $B \to \pi \pi$ and 
$B \to K \pi$ decays to quantify some of the issues involved.

\subsection{Indirect CP violation involving $B^0$ - $\overline{B^0}$ mixing amplitudes}
Indirect CP asymmetries involve an interference between the absorptive and dispersive 
parts of
the amplitudes in the $B_d^0$ - $\overline{B_d^0}$ and $B_s^0$ -
$\overline{B_s^0}$ mixings. Their
experimental measures are the charge asymmetries ${\cal A}_{\rm SL}(B_d^0)$ and ${\cal 
A}_{\rm SL}(B_s^0)$:
\bea
{\cal A}_{\rm SL}(B^0) &\equiv& \frac{\Gamma(\overline{B^0} \to \ell^+ X) -
\Gamma(B^0 \to \ell^- X)}{\Gamma(\overline{B^0} \to \ell^+ X) +
\Gamma(B^0 \to \ell^- X)}= \frac{1-\vert q/p \vert^4}{1+\vert q/p \vert^4}\,,
\label{chasym}
\eea
where the ratio $q/p$ is defined as follows
\be
\left (\frac{q}{p} \right )^2 \equiv \frac{2M_{12}^* - 
i \Gamma_{12}^*}{2M_{12} - i \Gamma_{12}}\,,
\ee
and the off-diagonal elements
$M_{12}$  and $\Gamma_{12}$ govern the mass difference ($\Delta M_B$) and the
width-difference ($\Delta \Gamma_B$) between the two mass eigenstates, respectively. Thus,

\be
{\cal A}_{\rm SL}(B_d^0) ={\rm Im}\bigg(\frac{\Gamma_{12}(B_d)}
{M_{12}(B_d)}\bigg)\,, \qquad 
{\cal A}_{\rm SL}(B_s^0)= {\rm Im}\bigg(\frac{\Gamma_{12}(B_s)}
{M_{12}(B_s)}\bigg)\,.
\label{aslb}
\ee
Parameterizing the ratio $\Gamma_{12}(B_q)/M_{12}(B_q)=r_q e^{i \zeta_q}$, SM 
estimates are
\be
{\cal A}_{\rm SL}(B_d^0) = r_d \sin \zeta_d \simeq {\cal O}(10^{-3})\,,
\qquad
{\cal A}_{\rm SL}(B_s^0) = r_s \sin \zeta_s \simeq {\cal O}(10^{-4})\,.
\ee
Present measurements yield~\cite{hfag03}  ${\cal A}_{\rm SL}(B_d^0)=(0.1
\pm 1.4) \times 10^{-2}$.
%
The bound on ${\cal A}_{\rm SL}(B_d^0)$ does not probe the SM, but constrains some
beyond-the-SM (BSM) scenarios~\cite{Laplace:2002ik}. Combining the data on the direct CP 
asymmetry ${\cal A}_{\rm CP}(B^+ \to J/\psi K^+)$, one can get an improved bound on 
$\vert q /p \vert$~\cite{Nir:2002gu}, and we shall discuss it later. However,
for  all practical purposes discussed here, one can set $\vert q/p 
\vert=1$ for  the $B_d^0$
-$\overline{B_d^0}$ system. Currently, there is no experimental bound on ${\cal A}_{\rm
SL}(B_s^0)$.

It is worth pointing out that the  analogue of the charge asymmetry (\ref{chasym})
in the $K^0$ - $\overline{K^0}$ system, namely the CP-violating asymmetry in the
semileptonic decays $K_L \to \pi^\pm \ell^\mp \nu_\ell$, defined as,
\be
\delta_\ell \equiv \frac{\Gamma(K_L \to \pi^- \ell^+ \nu_\ell) - \Gamma(K_L \to \pi^+ \ell^-
\nu_\ell)}
{\Gamma(K_L \to \pi^- \ell^+ \nu_\ell) + \Gamma(K_L \to \pi^+ \ell^- \nu_\ell)}\,,
\label{deltal}
\ee
is a well measured quantity~\cite{Hagiwara:fs} $\delta _\ell =(3.27 \pm 0.12) \times
10^{-3}$. This value is consistent with the relation $\delta_\ell=2{\rm Re} 
(\epsilon_K)/(1+\vert \epsilon_K\vert^2) \simeq 2{\rm Re} (\epsilon_K)$,
and the experimental value~\cite{Hagiwara:fs} of ${\rm Re} (\epsilon_K)$. Hence, the
 asymmetry $\delta_\ell$ arises entirely
from the imaginary part of $M_{12}(K)$, which in the SM is given by 
${\rm Im}(V_{ts}^* V_{td})$.

Current and planned experiments at the $e^+e^-$ and hadronic $B$ factories are not 
anticipated  to reach the SM-sensitivity in ${\cal A}_{\rm
SL}(B_d^0)$ (even less so in ${\cal A}_{\rm SL}(B_s^0)$), and, hence, a measurement 
of any of these asymmetries will be a sure signal of BSM physics.
\subsection{Interplay of mixing and  decays of $B^0$ and $\overline{B^0}$ to CP 
eigenstates}
As already mentioned, this class of CP asymmetries involves an interference 
between the decays $B^0 \to f_{\rm CP}$ and 
$B^0 \to \overline{B^0} \to f_{\rm CP}$. Due to mixing 
these CP asymmetries are time-dependent, and the two time-dependent functions
$\cos (\Delta M_{B} t)$ and $\sin (\Delta M_{B} t)$ can be measured, 
allowing to extract their coefficients $C_f$ and $S_f$, respectively:
\bea 
 {\cal A}_{\rm f CP}(t) \equiv \frac{\Gamma(\overline{B^0} (t) \to f) - 
\Gamma(B^0(t) 
\to  f)}{\Gamma(\overline{B^0} (t) \to f) + \Gamma(B^0(t) \to  f)} 
= -C_f \cos (\Delta M_B t) + S_f \sin (\Delta M_B t)\,,
\label{scdeltam}
\eea
where 
\be
 C_f=-{\cal A}_f=\frac{1-\vert \lambda_f \vert^2}{1+\vert \lambda_f \vert^2}\,,
 \qquad S_f= \frac{ 2 {\rm Im}\lambda_f}{1+\vert \lambda_f \vert^2}\,,
\label{cfsf}
\ee
and the dynamical quantity $\lambda_f$ is given by:
\be
\lambda_f = (q/p) \bar{\rho}(f)\,;  \qquad 
\bar{\rho}(f)= \frac{\bar{A}(f)}{A(f)}\,.
\ee
Concentrating now on $B_d^0$ and $\overline{B_d^0}$ decays, the amplitudes 
$A(f)$, $\bar{A}(f)$ and $q/p$ are defined as follows~\cite{Hagiwara:fs}:
\be
A(f)= \langle f|H |B_d^0\rangle\,; \qquad 
\bar{A}(f)= \langle f|H |\overline{B_d^0}\rangle\,;  \qquad 
q/p= \frac{V_{tb}^* V_{td}}{V_{tb}V_{td}^*} 
={\rm e}^{-2i\phi_{\rm mixing}}={\rm e}^{-2i\beta}\,.
\ee
If only a single decay amplitude is dominant, then one can write:
\be
 \bar{\rho}(f)= \eta_f {\rm e}^{-2i \phi_{\rm decay}} \qquad \Rightarrow~~\vert 
\bar{\rho}(f)\vert=1\,,
\ee
where $\eta_f=\pm 1$ is the intrinsic CP-Parity of the state $f$.
In this case, $\vert \lambda_f \vert=1$, and one has
\be
S_f= {\rm Im} (\lambda_f)= -\eta_f \sin 2 (\phi_{\rm mixing} + \phi_{\rm decay})\,,
~~C_f=0\,,
\ee
and the CP asymmetry ${\cal A}_{\rm f CP}$
has a very simple interpretation:
\be
{\cal A}_{\rm f CP} (t) = S_f \sin (\Delta M_B t)\,.
\ee
If, in addition, $\phi_{\rm decay}=0$, which is the case for the transitions
$b \to c\bar{c}s$, $b \to s \bar{s} s$ and $b \to d \bar{d} s$, then a measurement of $S_f$ 
is a measurement of $\sin 2 \theta_{\rm mixing}$ (modulo the sign $-\eta_f$).
The current thrust of the BABAR and BELLE experiments in CP 
asymmetry  measurements is indeed in extracting $S_f$ and $C_f$ for a large number of 
final states.

\subsection{Current status of CP asymmetries in $b \to c\bar{c}s$ and $b \to s\bar{s} s$ 
decays}

We now discuss the coefficients $S_f$ and $C_f$ for 
$f=J/\psi K_S, \phi K_S, \eta^\prime K_S, K^+ K^- K_S$.
Current data on these and some other final states are summarized in Table \ref{cfsfasym}. The 
values quoted for
$f=J/\psi K_S$ are averages over several decay modes of the quark decay $b \to c \bar{c} 
s$ and include also the final states 
$J/\psi K_L$ and $J/\psi K^*$, with the latter 
angular momentum analyzed into states with well-defined 
CP-parities and taking into account the intrinsic CP-parity, and some other related 
states~\cite{Browder:2003}. Also, the state $K^+ K^- K_S$ is not a CP eigenstate and an 
angular analysis like the one carried out for the $B \to J/\psi K^*$ case has not yet been 
undertaken due to limited  statistics. However, using arguments based on branching ratios of 
related modes, BELLE collaboration~\cite{Abe:2003yt} concludes that the $K^+ K^- K_S$
state  is predominantly a CP-even eigenstate, with a fraction $1.03 \pm 0.15 \pm 
0.05$.
 
In the SM, $S_{J/\psi K_S}$ and $C_{J/\psi K_S}$ are determined by the tree amplitude
$b \to c\bar{c}s$ with the penguin amplitude suppressed by $\lambda^2$. In any case, both the 
$T$ and $P$ amplitudes have the same phase,  and hence in the SM $S_{J/\psi K_S}=\sin 2 
\beta$ (or $\sin 2 \phi_1)$ and $C_{J/\psi K_S}=0$ (or $\vert \lambda_{J/\psi K_S}\vert=1$). 
Comparison of  $S_{J/\psi K_S}$ with the indirect estimates of the same discussed earlier shows that
the agreement with the SM expectations in this decay mode holds quantitatively. This is a
great triumph for the KM mechanism of CP violation.

The present measurement $C_{J/\psi K_S}=0.052^{+0.048}_{-0.046}$ is in agreement with no 
direct CP 
violation in the decays $B^\pm \to J/\psi K^\pm$. The current bound~\cite{hfag03}
\be
{\cal A}_{\rm CP}(B^+ \to J/\psi K^+) =\frac{\vert \bar{A}/A\vert^2 -1}
{\vert \bar{A}/A\vert^2 +1}=-0.007 \pm 0.019\,,
\ee
yields $\vert \bar{A}/A\vert=0.993 \pm 0.018$. Combined with $\vert q/p \vert 
=0.9996^{+0.0068}_{-0.0067}$ from $A_{\rm SL}(B_d^0)$, this yields 
$\vert \lambda_{J/\psi K_S}\vert=0.992 \pm 0.019$, in  precise agreement with $\vert 
\lambda_{J/\psi 
K_S}\vert=1$. 

 Going down the entries in Table \ref{cfsfasym}, we note that the final 
states $\phi K_S$ and $K^+ 
K^- K_S$ are determined by the penguin transition 
$ b \to s \bar{s} s$, which has the weak phase $\pi$. 
The final state $ \eta^\prime K_S$
receives contributions both from the penguin $b \to s \bar{s} s$ and $b \to s 
\bar{d} d$ amplitudes and the tree amplitude $b \to u \bar{u} s$, 
due to the $u \bar{u}$ content of the 
$\eta^\prime$-meson wave function. The tree amplitude, however, is CKM 
suppressed. So, to a good approximation, also this final state is dominated 
by the penguin transitions $b \to s \bar{s} s$ and $b \to s \bar{d} d$, 
and has the weak phase $\pi$ in the decay amplitude. Thus, in the SM, 
we expect that for
these decays ($f=J/\psi K_S, \phi K_S, \eta^\prime K_S, K^+ K^- K_S$)
\be
-\eta_{f}S_{f} =  \sin 2\beta \,,  \qquad  C_{f} =  0\,.
\label{sfdiff} 
\ee
The current measurements of $S_f$ and $C_f$ are summarized 
in~Table \ref{cfsfasym}.
\begin{table}[htbp] 
\caption{The coefficients $S_{f}$ and $C_{f}=-{\cal A}_{f}$ 
from time-dependent CP asymmetry ${\cal A}_{\rm f CP}(t)$, 
taken from the HFAG listings~\protect\cite{hfag03} 
[BE(BA) stands for a BELLE (BABAR) entry].}
\label{cfsfasym}

\vspace*{2mm}
\begin{center}
\renewcommand{\arraystretch}{1.3} 
\begin{tabular}{c c c}\hline \hline\\[-4mm]
$S_{f}$ & $\hspace*{12mm}$ & $C_{f}$\\[1.5mm] \hline \\[-3mm]
 $S_{J/\psi K_S} = 0.736 \pm 0.049$ & &
$C_{J/\psi K_S} = 0.052^{+0.048}_{-0.046}$\\ [2.3mm]
\hline \\[-3mm]
$ S_{\phi K_S}=-0.14 \pm 0.33$ & &
$C_{\phi K_S} = -0.04 \pm 0.24 $ \\
$ S_{\eta^\prime K_S}=+0.27 \pm 0.21$ & &
$C_{\eta^\prime K_S} = 0.04 \pm 0.13$ \\
$ S_{K^+ K^- K_S}[{\rm BE}]=-0.51 \pm 0.26 \pm 0.05$ & &
$C_{K^+ K^- K_s}[{\rm BE}] = 0.17 \pm 0.16 \pm 0.05$ \\[2.3mm]
\hline\\[-3mm]
$ S_{J/\psi \pi^0}=-0.40 \pm 0.33$ & &
$C_{J/\psi \pi^0} =  0.13 \pm 0.24$  \\
$ S_{D^*\overline{D^*}}[{\rm BA}]= 0.06 \pm 0.37
\pm 0.13$ & & $C_{D^*\overline{D^*}}[{\rm BA}]
 = 0.28 \pm 0.23 \pm 0.02 $ \\
$ S_{+-}(D^{*+}D^{-})[{\rm BA}]= -0.82 \pm 0.75 \pm 0.14
$ & & $C_{+-}(D^{*+}D^{-})[{\rm BA}]
 = -0.47 \pm 0.40 \pm 0.12 $ \\
$ S_{-+}(D^{*-}D^{+})[{\rm BA}]= -0.24 \pm 0.69 \pm 0.12
$ & & $C_{-+}(D^{*-}D^{+})[{\rm BA}]
 = -0.22 \pm 0.37 \pm 0.10 $  \\[2.3mm]
\hline \\[-3mm]
$ S_{\pi \pi}=-0.58 \pm 0.20$ & & $C_{\pi \pi}
= -0.38 \pm 0.16$\\[2mm] 
\hline \hline
\end{tabular}
\end{center}
\end{table}
The data are from BELLE~\cite{Abe:2003yt} and 
BABAR~\cite{Aubert:2003bq}, updated by
Browder~\cite{Browder:2003} at the Lepton-Photon Conference and the Summer 
2003 updates by HFAG~\cite{hfag03}. Note that 
$\eta_{J\psi K_S}=\eta_{\phi K_S}=\eta_{\eta^\prime 
K_S}=-1$, but $\eta_{K^+ K^- K_S}=+1$ due to the dominance of the 
$CP=+1$ eigenstate in this mode. One sees that within the experimental 
errors, SM predictions $C_f=0$ for the final 
states specified in (\ref{sfdiff}) are in agreement with the data, 
though the errors in the individual modes are still quite large. 
On the other hand,
$S_{\phi K_S}$ and to a lesser extent also
$S_{\eta^\prime K_S}$, appear to be out 
of  line with the SM expectations (\ref{sfdiff}). However, it 
should be noted that the two experiments BABAR and BELLE are
not consistent with each other~\cite{Browder:2003,Abe:2003yt}, with
 $S_{\phi K_S}({\rm BELLE})=-0.96 \pm 0.50^{+0.09}_{-0.11}$ and
$S_{\phi K_S}({\rm BABAR})=+0.45 \pm 0.43 \pm 0.07$, differing 
by 2.1$\sigma$. Following the PDG rules~\cite{Hagiwara:fs}, one has to 
scale the error in $S_{\phi K_S}$ given in Table \ref{cfsfasym} by this 
factor, yielding
$S_{\phi K_S}= -0.14 \pm 0.69$, which is only $1.3\sigma$ away from
$S_{J/\psi K_S}=0.736 \pm 0.049$, and hence the difference between the 
two is not all that compelling.

As the current statistics is low, it is helpful to combine the three 
($b \to s \bar{s} s)$ penguin-dominated final states. Defining 
$-\eta_f S_f \equiv \sin2 \beta_{\rm eff}$, and averaging over the 
three final states gives~\cite{hfag03} 
\bea
\langle \sin2 \beta_{\rm eff}\rangle&=& 0.24 \pm 0.15~~({\rm C.L.} =0.11)\,, 
\nonumber\\[-1.5mm] 
\label{smequs}\\[-1.5mm] 
\langle C_{\rm eff}\rangle &=& 0.07 \pm 0.09~~({\rm C.L.} =0.76)\,.
\nonumber
\eea
The two experiments (BABAR and BELLE) are still hardly compatible with 
each other in 
$ \langle \sin2 \beta_{\rm eff}\rangle$; 
in contrast the measurements of $\langle C_{\rm 
eff}\rangle$ are quite consistent.
 The  current average of $\langle C_{\rm 
eff}\rangle$ within errors also agrees with
the corresponding quantity measured in the $b \to c \bar{c} s$ 
transitions, with
$C_{J/\psi K_S}- \langle C_{\rm eff}\rangle= 0.018 \pm 0.10$. However,
 the current measurements yield~\cite{Browder:2003,hfag03}
\be
\sin 2 \beta  - \langle \sin2 \beta_{\rm eff} \rangle = 0.50 \pm 0.16\,,
\ee
which is a 3.1$\sigma$ effect on the face value. 
Taking into account the scale 
factor in $ \langle \sin2 \beta_{\rm eff}\rangle$ increases the error, 
yielding 
$\sin 2 \beta  - \langle \sin2 \beta_{\rm eff} \rangle = 0.50 \pm 
0.25$, which differs from 0 by $2\sigma$. 
This difference is more significant than
$S_{J/\psi K_S} - S_{\phi K_S}$, but still does not have 
the statistical weight
to usher us into a new era of CP violation. We have to wait for more data.

While there are already quite a few suggestions in the recent 
literature explaining the
difference $S_{\phi K_S}-S_{J/\psi K_S}$ in terms of physics beyond the SM, 
we will not discuss them as the experimental significance of the effect is 
marginal. However, a more pertinent question to ask is: How well are the 
SM equalities given in (\ref{sfdiff})
satisfied? This point has been investigated recently by Grossman {\it et 
al.}~\cite{Grossman:2003qp}. Following their notation,
the SM amplitudes in these decays can be parametrized as:
\be
  A_f\equiv A(B^0 \to f)=V_{cb}^* V_{cs} a_f^c + V_{ub}^* V_{us} a_f^u\,,
\ee
where $a_f^c$ is dominated by
the $b \to s \bar{s} s$ gluonic penguin diagrams and  $a_f^u$ gets
contributions from both penguin and $b \to u \bar{u} s$ tree diagrams. 
The second term is  
CKM-suppressed compared to the first:
\be
{\rm Im} \left( \frac{V_{ub}^* V_{us}}{V_{cb}^* V_{cs}}\right)
= \left \vert \frac{V_{ub}^* V_{us}}{V_{cb}^* V_{cs}} \right \vert \sin \gamma
=O(\lambda^2) \sin \gamma\,.
\ee
It is conceivable though not very likely that the CKM-suppression is offset by 
a dynamical enhancement of the ratio $a_f^u/a_f^c$.
Note that  $\vert a_f^u/a_f^c \vert \sim 1$ (from penguins), but
$\vert a_f^u/a_f^c \vert $ (from tree) could be $\gg 1$. To quantify this,
Grossman {\it et al.}~\cite{Grossman:2003qp} define
\be
\xi_f \equiv \frac{V_{ub}^* V_{us}a_f^u}{V_{cb}^* V_{cs}a_f^c}
~\Rightarrow ~A_f=V_{cb}^* V_{cs}a_f^c(1+ \xi_f)\,.
\ee
SU(3) allows to put bounds on $\vert \xi{_f}\vert$:
\bea
-\eta_f S_f - \sin 2 \beta &=& 2 \cos 2 \beta \sin \gamma \cos
\delta_f \vert \xi_f \vert\,, 
\nonumber\\[-1.5mm] 
\\[-1.5mm] 
C_f &=& -2 \sin \gamma \sin \delta_f \vert \xi_f \vert\,,
\nonumber
\eea
where $\delta_f={\rm arg}(a_f^u/a_f^c)$ and $\xi_f$ also characterizes the
size of $C_f$. Note that $\delta_f$ can be determined from
$\tan \delta_f=(\eta_f S_f + \sin 2 \beta)/(C_f \cos 2 \beta)$.
However, present bounds are not very restrictive due to lack of information
on some decays and additional assumptions are required to be more 
quantitative. Typical estimates are~\cite{Grossman:2003qp}:
\bea 
\vert \xi_{\eta^\prime K_S}\vert &<& 0.36~~[{\rm SU(3)}]; 
~~0.09~~[{\rm SU(3)} + {\rm Leading}~N_c]\,, 
\nonumber\\
\vert \xi_{\phi K_S}\vert &<& 0.25~~[{\rm SU(3)} + {\rm Non-cancellation}]\,, 
\label{niretal} \\
\vert \xi_{K^+ K^- K_S}\vert &<& 0.13~~[U-{\rm Spin}]\,,
\nonumber
\eea
where the various assumptions in arriving at the numerical inequalities have been
specified. More data and further theoretical analysis are required to 
further restrict $\vert \xi_f \vert$ in a model-independent way.
Hence, at this stage, one should use values of $\xi_f$  indicated in (\ref{niretal}).

\subsection{Current status of the CP asymmetries in $b \to c\bar{c}d$ decays}
CP asymmetries in some of the final states which are induced by the transition $b \to 
c\bar{c}d$ have also been studied by the BABAR and BELLE collaborations. 
 The coefficients $S_f$ and $C_f$ for $f=J/\psi \pi^0,~D^* \overline{D^*},~D^{*+} 
D^-,~D^{*-} D^+$ measured through the time-dependent CP-asymmetry in these channels
are given in the lower part of Table \ref{cfsfasym}.
Note that the entry for $f=J/\psi \pi^0$ is the
average of the BABAR~\cite{Aubert:2003ut} and BELLE~\cite{Abe:2003ps} experiments, but
the entries for $f=D^{*+} D^-,~f=D^{*-} D^+$ and $f=D^*\overline{D^*}$ are from the 
BABAR collaboration~\cite{Aubert:2003ca,Aubert:2003uv} alone with the BELLE 
results not yet available. Just like
the $b \to c\bar{c}s$ case, the tree amplitude in the $b \to c \bar{c} d$ transition does not 
have a weak 
phase. 
However, as opposed to the $b \to c\bar{c}s$ case, where the penguin amplitude has the weak 
phase 
$\pi$, now there could be a non-negligible contribution from the
$b \to d$ penguin amplitude, which carries the weak phase $\beta$. In terms of the CKM 
factors, both the T and P amplitudes are of order $\lambda^3$. The tree-penguin 
interference
with different weak phases may lead to direct CP violation, giving rise to $C_f \neq 0$,
or $\vert \lambda_f \vert \neq 1$. Also, in this case, the equality
$-S_{J/\psi \pi^0} =S_{J/\psi K_S}$ or $-S_{D^*\overline{D^*}}(+)=S_{J/\psi K_S}$ etc. can
be violated. Here the sign$(+)$ stands for the CP$=+1$ component of the Vector-Vector 
final state, which for $D^*\overline{D^*}$ dominates the decay, with the CP-odd component
given by~\cite{Aubert:2003uv} $R_\perp=D^*\overline{D^*}({\rm P~wave})/ 
D^*\overline{D^*}=0.063 \pm 0.055 \pm 0.009$. Present data are consistent with the SM 
and no direct CP violation in these decays is observed. Also, the measured 
coefficients $S_f$ for these decays do not show significant deviations from
the SM expectations (ignoring penguins) $-\eta_f S_f=S_{J/\psi K_S}$.
This brings us to the entry in the last row in  
Table \ref{cfsfasym} for $f=\pi \pi$, which we discuss below together with the two
methods most discussed in the literature to determine the angle $\alpha$ (or 
$\phi_2$). 

\subsection{Measurements of the weak phase $\alpha$ (or $\phi_2)$ in $B$-meson 
decays}
In the previous section we have given the 95\% C.L. range for the weak phase $\alpha$ 
obtained from the CKM unitarity fits: $70^\circ \leq \alpha \leq 115^\circ$. This
phase will be measured through CP violation in the $B \to \pi \pi$ and $B \to \rho 
\pi$ decays. To eliminate the hadronic uncertainties in the determination of
$\alpha$, an isospin analysis of these final states (as well as an angular analysis in 
the $\rho \pi$ case) will be necessary. However, at a less rigorous level,
data from $B \to K \pi$ decays may be combined with the available data on the $B \to 
\pi \pi$ decays to extract $\alpha$ from the current data. We discuss both of these 
methods below. 

The decay $B^0 \to \pi^+ \pi^-$ involves tree and penguin 
contributions with different strong and weak phases. Denoting the
strong phase difference by $\delta\equiv \delta_P-\delta_T$, the amplitudes can be written 
as
\bea 
- A(B^0 \to \pi^+ \pi^-) &=& \vert T\vert {\rm e}^{i\gamma}
+\vert  P \vert {\rm e}^{i \delta}\,, 
\nonumber\\[-1.5mm] 
\label{pipiampl}\\[-1.5mm] 
-A(\overline{B^0} \to \pi^+ \pi^-) &=& \vert T\vert {\rm e}^{-i\gamma}
+\vert  P \vert {\rm e}^{i \delta}\,,
\nonumber
\eea
where the T and P components have the CKM dependence (using the Gronau-Rosner 
convention~\cite{Gronau:2002gj}) given by $V_{ub}^* V_{ud}$ and 
$V_{cb}^* V_{cd}$, respectively. The time-dependent CP asymmetry is 
given by the expression
\be
{\cal A}_{\pi \pi}(t) =-C_{\pi \pi} \,\cos (\Delta M_{B_d}t) + 
S_{\pi \pi} \,\sin (\Delta M_{B_d}t)\,,
\label{pipicpt}
\ee
and the coefficients $C_{\pi \pi}$ and $S_{\pi \pi}$ are defined 
as in~(\ref{cfsf}), with
\be
\lambda_{\pi \pi}= \eta_{\pi \pi}{\rm e}^{-2i(\beta + \gamma)} \, 
\frac{ 1+ \vert P /T \vert
{\rm e}^{i(\delta + \gamma)}}{1+ \vert P /T \vert e^{i(\delta - \gamma)}}\,.
\ee
Thus, apart from $\beta$ (or $\phi_1$), which is now well measured, 
we have three more variables, the ratio $\vert P /T \vert$, $\delta$ and 
$\gamma$ (or $\phi_3$). As this expression stands, it gives information 
on $\gamma$ and not on $\alpha$!
However, if the penguin contribution were absent (or small), then
using the relation $\alpha + \beta + \gamma =\pi$ and $\eta_{\pi \pi}=+1$, 
one has $\lambda_{\pi \pi}=
{\rm e}^{-2i(\beta + \gamma)}={\rm e}^{2i\alpha}$, yielding 
$C_{\pi \pi}=0$ and $S_{\pi \pi}=\sin 2\alpha$.
 Hence, the folklore: $S_{\pi \pi}$ measures $\sin 2 \alpha$.
Now, with strong hints from data that $\vert P/T \vert$ is significant
(for example, T-dominance would have
 $2\Gamma(\pi^+\pi^0)/\Gamma(\pi^+\pi^-)=1$,
but the latest BELLE data~\cite{Chao:2003ue} gives 
$2.10 \pm 0.58 \pm 0.25$ for this ratio) 
one interprets
$S_{\pi \pi}$ as $S_{\pi \pi}=\sqrt{1-C_{\pi \pi}^2} \sin 2 
\alpha_{\rm eff}$, where
both $C_{\pi \pi}$ and $\alpha_{\rm eff}$ involve non-trivial hadronic physics.
Hence, a transcription of $\alpha_{\rm eff}$ into $\alpha$ (or $\gamma$) is 
not easy to accomplish  in a model-independent~way.

To get a model-independent determination of $\alpha$ (or $\phi_2)$, 
one has to carry out 
the isospin analysis of the $B \to \pi \pi$ decays suggested by Gronau and 
London~\cite{Gronau:1990ka}. Defining the various amplitudes as follows:
\bea
A^{+-} &\equiv& A(B^0 \to \pi^+ \pi^-); \qquad 
\bar{A}^{+-} \equiv  A(\overline{B^0} \to \pi^+ 
\pi^-)\,, \nonumber\\
A^{00} &\equiv& A(B^0 \to \pi^0 \pi^0); \qquad 
\bar{A}^{00} \equiv   A(\overline{B^0} \to \pi^0
\label{pipiampldef} \\
A^{+0} &\equiv& A(B^+ \to \pi^+ \pi^0); \qquad 
\bar{A}^{-0} \equiv  A(B^- \to \pi^-
\pi^0)\,,
 \nonumber 
\eea
isospin-symmetry leads to the following triangular relations~\cite{Gronau:1990ka}
\bea
\frac{1}{\sqrt{2}} A^{+-} + A^{00} &=&A^{+0}\,, \qquad~\frac{1}{\sqrt{2}} \bar{A}^{+-} + 
\bar{A}^{00} =\bar{A}^{-0}\,, \nonumber\\
\vert A^{+0} \vert &=& \vert \bar{A}^{-0} \vert\,.
\label{glisospin}
\eea
Here, the last equality results from the observation that the amplitudes $A^{+0}$
and $\bar{A}^{-0}$ describe decays into pure isospin-2 states and do not receive 
contributions from the QCD penguins, and electroweak penguins  
may be ignored, as their contribution in the $\pi \pi$ system is 
not expected to exceed a few percent~\cite{Deshpande:1994pw}. One can also include 
the contribution of the electroweak penguins in this analysis by using
isospin symmetry~\cite{Gronau:1998fn}, which relates their contribution to the tree 
amplitudes and  eliminate any residual hadronic uncertainty in the determination of 
the weak phase $\alpha$. 

 The two triangles written in the first line in (\ref{glisospin}) have the 
same base (due to the second line in (\ref{glisospin})) and 
the mismatch in the apex of the two triangles then determines the difference $2\theta\equiv 
2(\alpha_{\rm eff} -\alpha)$. The determination of $2\alpha$ goes along the following 
lines: From the relative phase of the amplitudes $A^{+-}$ and 
$\bar{A}^{+-}$ one gets $2 \alpha_{\rm eff}$. From the relative orientation of
the amplitudes $A^{+0}$ and $A^{+-}$ one gets an angle $\Phi$, and finally from
the relative orientation of the amplitudes $\bar{A}^{-0}$ and $\bar{A}^{+-}$ one gets an
angle $\bar{\Phi}$. The angle $2 \alpha$ is then obtained from the difference
$2 \alpha=2 \alpha_{\rm eff} -\Phi -\bar{\Phi}$.

 From the magnitudes $\vert 
A^{ij}\vert$ and $\vert \bar{A}^{ij}\vert$ of the six amplitudes given in the isospin 
relations 
(\ref{glisospin}), the only missing pieces in the experiments are $\vert A^{00}\vert$ 
and $\vert \bar{A}^{00}\vert$. However, through 
the measurement of the charge-conjugate averaged branching ratio $B^0/\overline{B^0} 
\to \pi^0 \pi^0$, the 
combination $\vert A^{00}\vert^2 + \vert \bar{A}^{00}\vert^2$ is now known. This 
branching ratio together with some of the other $B \to \pi \pi$ and $B \to K \pi$ 
branching  ratios is given in Table \ref{brasym}, where the entries are from the
Lepton-Photon 2003 conference review by Fry~\cite{Fry:2003}. The 
Gronau-London  
isospin analysis can not be carried out for the time being. However,
theoretical bounds on $\theta$ (or $\alpha_{\rm eff}$) have been proposed. 
For example, the Grossman-Quinn bound~\cite{Grossman:1997jr} 
$\vert \sin (\alpha-\alpha_{\rm eff}) \vert 
\leq \sqrt{\bar{\Gamma}(\pi^0 \pi^0)/\bar{\Gamma}(\pi^+\pi^-)}$, 
with the branching ratios given in Table \ref{brasym} yields 
$\vert \alpha-\alpha_{\rm eff}\vert < 
48^\circ$ at 90\% C.L.~\cite{Jawahery:2003}, 
and hence currently not very helpful. For other suggestions, see recent 
papers by London, Sinha and Sinha~\cite{London:2000dn} and 
by Gronau~{\it et~al.}~\cite{Gronau:2001ff}.

%
\begin{table}[htbp]
\caption{Summary of branching fractions $(\times 10^{-6})$ 
and ${\cal A}_{\rm CP}(h_1 h_2)$
(Source: Lepton-Photon 2003 review~\protect\cite{Fry:2003}).}
\label{brasym}

\vspace*{1mm}
\begin{center}
\renewcommand{\arraystretch}{1.3} 
\begin{tabular}{c c c}\hline \hline\\[-4mm]
Decay Mode  
& $\hspace*{14mm}$ ${\cal B}(h_1 h_2)$ $\hspace*{14mm}$ 
& ${\cal A}_{\rm CP}(h_1 h_2)$ $(\%)$
\\[1.5mm] \hline \\[-4mm]
$K^+\pi^-$ & $18.2 \pm 0.8$ & $-9.5\pm 2.9$\\
$K^0\pi^+$ & $21.8 \pm 1.4$ & $-1.6\pm 5.7$\\
$K^+\pi^0$ & $12.8 \pm 1.1$ & $0\pm 7$\\
$K^0\pi^0$ & $11.9 \pm 1.4$ & $3\pm 37$\\
$\pi^+\pi^-$ & $4.6 \pm 0.4$ & -- \\
$\pi^+\pi^0$ & $5.3 \pm 0.8$ & $-7 \pm 14$\\
$\pi^0\pi^0$ & $1.90 \pm 0.47$ & --\\[2mm] 
\hline \hline
\end{tabular}
\end{center}
\end{table}

We now discuss the information that can be obtained on the phase $\alpha$
by invoking SU(3) relations between the $B \to \pi \pi$ tree and penguin 
amplitudes ($T$ and $P$) and the corresponding amplitudes in the 
$B \to K \pi$ decays ($T^\prime$ and $P^\prime$)~\cite{Zeppenfeld:1980ex}. 
In the present context, one may relate the amplitudes in the decays 
$B^+ \to K^0 \pi^+$ and $B_d^0 \to \pi^+ \pi^-$. Writing

\be
-A(B^+ \to K^0 \pi^+) =\vert P^\prime \vert \, {\rm e}^{i \delta} = 
\vert P \vert \, {\rm e}^{i\delta} \frac{f_K}{f_\pi \tan \theta_c}\,,
\label{pipikpirel}
\ee

\noindent
where a small term with the weak phase $\gamma$ has been neglected and 
factorization of the decay amplitudes is assumed. With the known values 
of $f_K$, $f_\pi$, $\tan \theta_c$ and 
${\cal B}(B_d^0 \to \pi^+ \pi^-)/{\cal B}(B^+ \to K^0 \pi^+)=0.23 \pm 0.03$
(from Table \ref{brasym}), one gets $\vert P/T \vert \simeq 0.3$.
 This allows to constrain $\alpha$ from the present measurements of 
$S_{\pi \pi}$ and $C_{\pi \pi}$~\cite{Charles:1998qx,Gronau:2002gj}. 
Another variation on the same theme~\cite{Fleischer:1999jv} is to use the 
data from the $B_d^0 \to K^\pm \pi^\mp$ decays instead of $B^+ \to K^0 \pi^+$. 
Again, using flavour SU(3) symmetry and
dynamical assumptions, one can  extract $\alpha$ from the current 
data on $S_{\pi \pi}$, $C_{\pi \pi}$ and 
${\cal B}(B_d^0 \to \pi^+ \pi^-)/{\cal 
B}(B_d^0 \to K^\pm \pi^\mp)$. 
These methods actually give information on $\gamma$ (or $\phi_3$),
as discussed above, and the consequences of the current measurements 
have been recently worked out by Fleischer~\cite{Fleischer:2003yk}, 
getting a value of $\gamma$ in the SM-ball 
park with $\beta \simeq 24^\circ$.

With the average $S_{\pi \pi}=-0.58 \pm 0.20$ and 
$C_{\pi \pi}=-0.38 \pm 0.16$, as given in Table \ref{cfsfasym}, we see that
within measurement
errors, $-S_{\pi \pi}=S_{J/\psi K_S}$ is not violated, which implies no direct CP 
violation, but $C_{\pi \pi}$ deviates from 0 by about 2.4 $\sigma$, which is a 
signature of direct CP violation. So, at present, the inferences from $C_{\pi \pi}$
and $S_{\pi \pi}$ are not quite equivocal. The relation $-S_{\pi \pi}=S_{J/\psi K_S}$ is 
expected to be  violated in the SM, as the phases $\alpha$ and $\beta$ are numerically 
quite different. A large value of $\vert C_{\pi \pi}\vert$ would also imply a large strong 
phase $\delta$, which would put to question the validity of the QCD factorization 
framework in $B \to \pi \pi$ decays~\cite{Beneke:1999br}.
However,  it should be emphasized that,
just like the coefficient $S_{\phi K_S}$, also $S_{\pi \pi}$ comes out very 
different in the BABAR~\cite{Jawahery:2003} and BELLE~\cite{Abe:2003ja} 
measurements:
 $ S_{\pi \pi}({\rm BELLE})=-1.23 \pm 0.41 
^{+0.08}_{-0.07}$ and $S_{\pi \pi}({\rm BABAR})= -0.40 \pm 0.22 \pm 0.03$,
and the average has a C.L. of 0.047~\cite{hfag03}.
Scaling the error by the PDG scale factor, we get $S_{\pi \pi}=-0.58 \pm 
0.34$ and $C_{\pi \pi}=-0.38 \pm 0.27$. With the scaled errors, $C_{\pi \pi}$ differs from
0 by only 1.4$\sigma$. We hope that with almost a factor two more data on tapes,
the experimental isuues in $B \to \pi \pi$ decays  will soon be settled.

Having stated the caveats (theory) and pitfalls (experiments),
we use (\ref{pipikpirel}) to illustrate what values of $\alpha$ are implied by the 
current averages $S_{\pi \pi}=-0.58 \pm 0.20$ and $C_{\pi \pi}=-0.38 \pm 0.16$.
To take into account the uncertainty in the
SU(3)-breaking and non-factorizing effects, we vary the ratio $\vert P /T \vert$ in the 
range~\cite{Xiao:2003bc} $0.2
\leq \vert P /T \vert \leq 0.4$, which amounts to admitting a $\pm 30\%$ uncertainty on the
central value of the magnitude obtained in the factorization approach. However, the strong 
phase $\delta$ is varied in the entire allowed range $-\pi \leq \delta \leq \pi$.
 The results of this analysis are shown in 
Fig.~\ref{fig:B-pipi} for four values of $\alpha$ which lie in the 95\% C.L. allowed range 
from 
the unitarity fits: $\alpha=80^\circ$ (upper left frame), $\alpha=93^\circ$ (upper right 
frame), corresponding to the nominal central value of the CKM fit, $\alpha=105^\circ$ (lower left 
frame), and $\alpha=115^\circ$ (lower right frame), corresponding to the 95\% C.L. upper
limit on $\alpha$ from our fits. We do not show the plot for $\alpha=70^\circ$, which
is the 95\% C.L. lower value of $\alpha$ from the unitarity fits, as already the case 
$\alpha=80^\circ$ is disfavoured by the current $B \to \pi \pi$ data. In each figure, the
outer circle corresponds to the constraint $S_{\pi \pi}^2 + C_{\pi \pi}^2=1$; the
current average of the BABAR and BELLE data satisfies this constraint as shown by the 
data point with (unscaled) error.  The two inner contours
correspond to the values $\vert P /T \vert=0.2$ and $\vert P /T \vert=0.4$. The points 
indicated on 
these contours represent the values of the strong phase $\delta$ which is varied in the 
interval $-\pi \leq \delta \leq \pi$. We see that the $B \to \pi \pi$ data is in 
comfortable  agreement with the value of $\alpha$ in the range $90^\circ \leq \alpha \leq 
115^\circ$, with $\alpha=80^\circ$ outside the $\pm 1 \sigma$ range.
Also, as $C_{\pi \pi}$ is negative, current data favours
a rather large strong phase, typically $ -90^\circ \leq \delta  \leq -30^\circ $.
This is in qualitative agreement with
the pQCD framework~\cite{Keum:2000ph}, which yields~\cite{Keum:2003qi} 
$P/T=0.23^{+0.07}_{-0.05}$
and $-41^\circ \leq \delta \leq -32^\circ$. (See, also Xiao {\it et 
al.}~\cite{Xiao:2003bc}).

 The case for significant final state rescattering, and 
hence large strong phase differences, in the decays $B \to PP$ has also been 
advocated by Chua, Hou and Yang~\cite{Chua:2002wk}, motivated in part by the 
discovery of the colour-suppressed decays~\cite{Abe:2001zi,Coan:2001ei} $\bar{B^0} 
\to D^0 h^0$, where $h^0 =\pi^0,\eta$ or $\omega$, found significantly over the 
factorization-based estimates. An analysis of the current measurements in $B \to P P$,
decays with $P=\pi,K,\eta$, done with the help of an SU(3) formalism to take into
account $8\otimes 8 \to 8 \otimes 8$ rescattering~\cite{Hou:2003rr}, shows 
marked improvement in the quality of the fit.

 Thus, the study of the CP asymmetry in the $B \to \pi
\pi$ decays is potentially the most exciting game in town, as precise measurements in this
decay mode will not only determine the weak phase $\alpha$ but will also decide several 
important theoretical issues, such as the adequacy or not of the perturbatively generated
strong phases. To this list of interesting decays, one should add the decays
$B \to K \pi$ and $B \to D K$, which we discuss below.

%
\begin{figure}[htb]
\centerline{
\psfig{width=0.47\textwidth,file=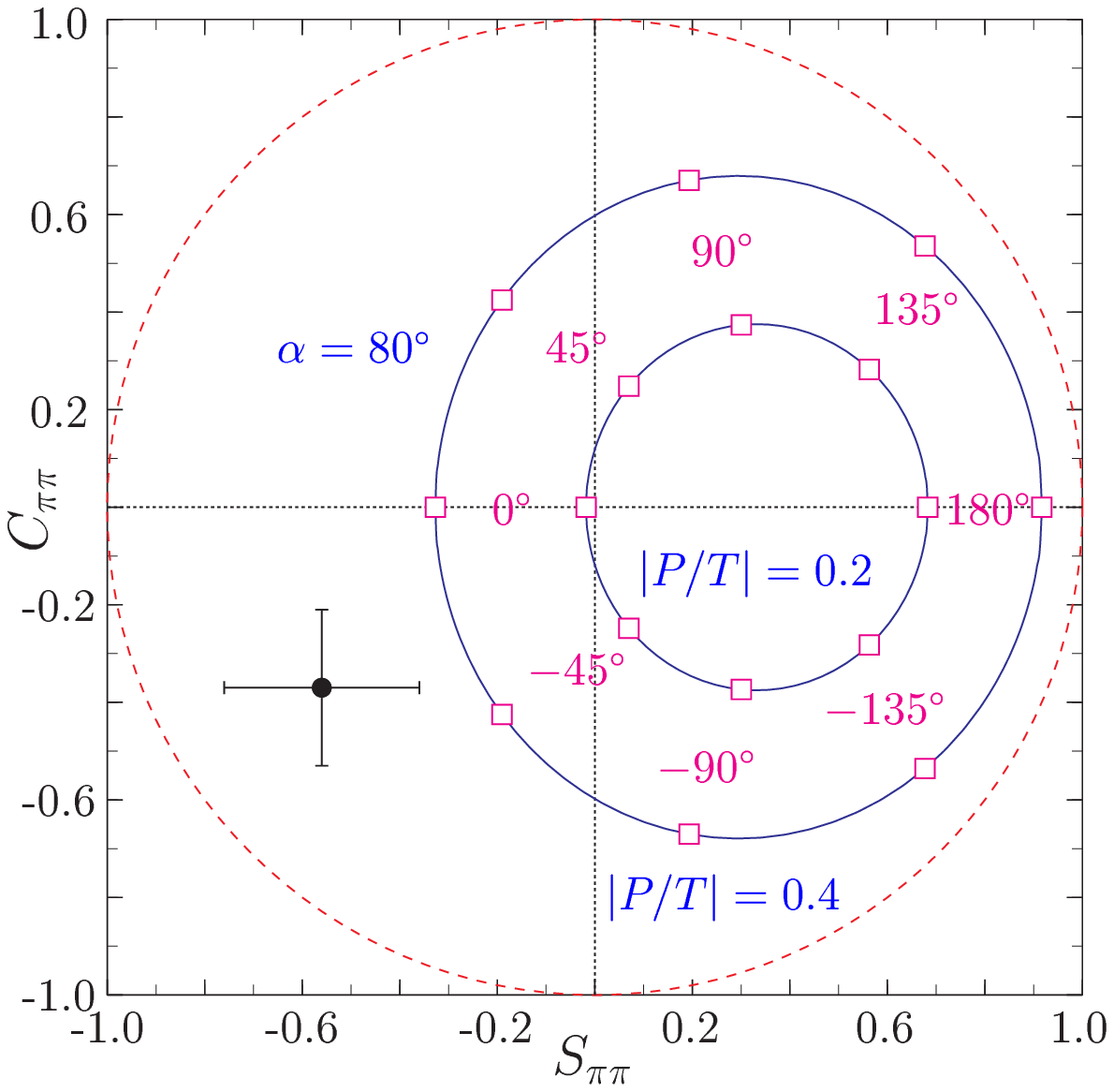}
\quad
\psfig{width=0.47\textwidth,file=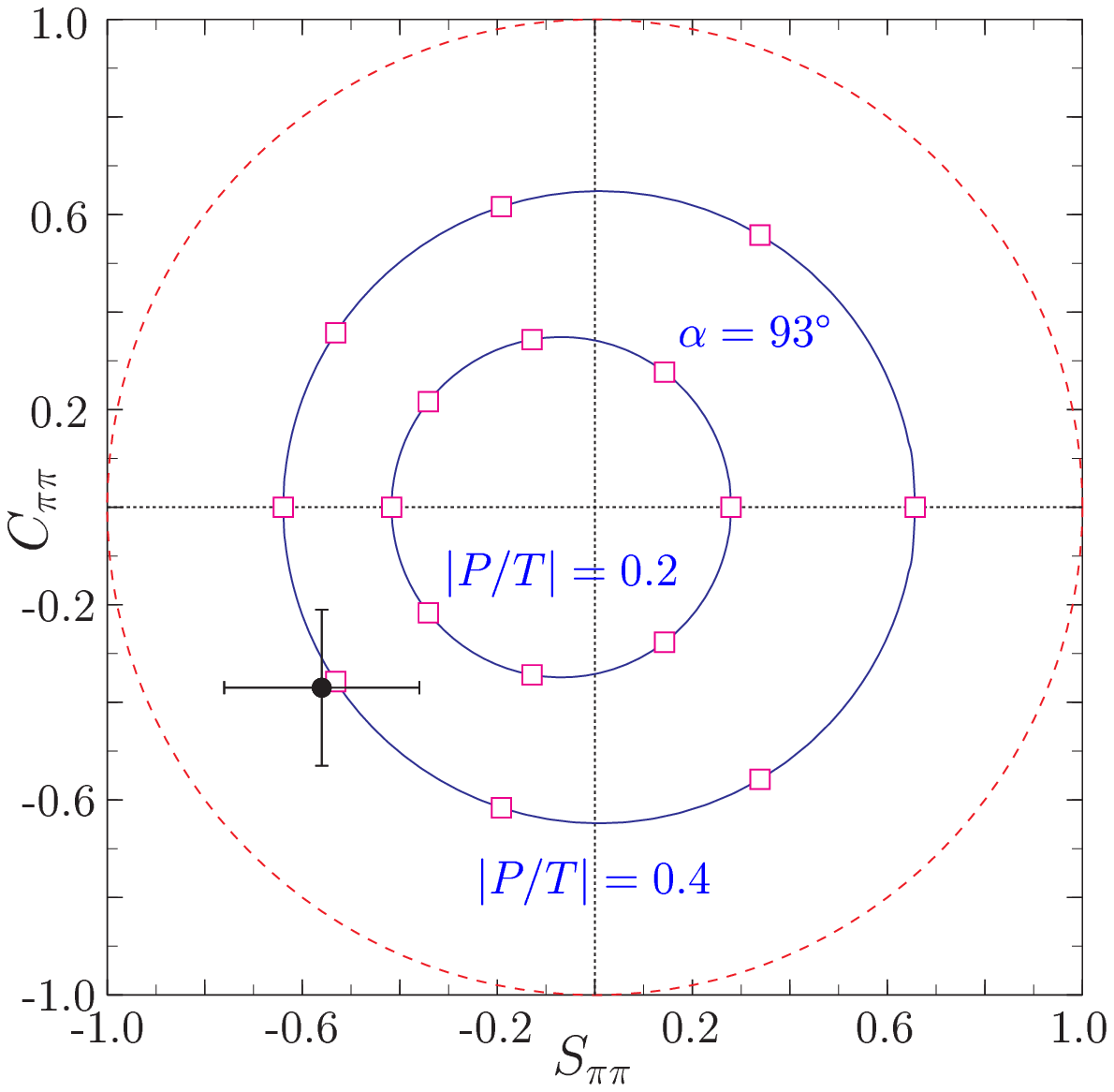}
}
\bigskip
\centerline{
\psfig{width=0.47\textwidth,file=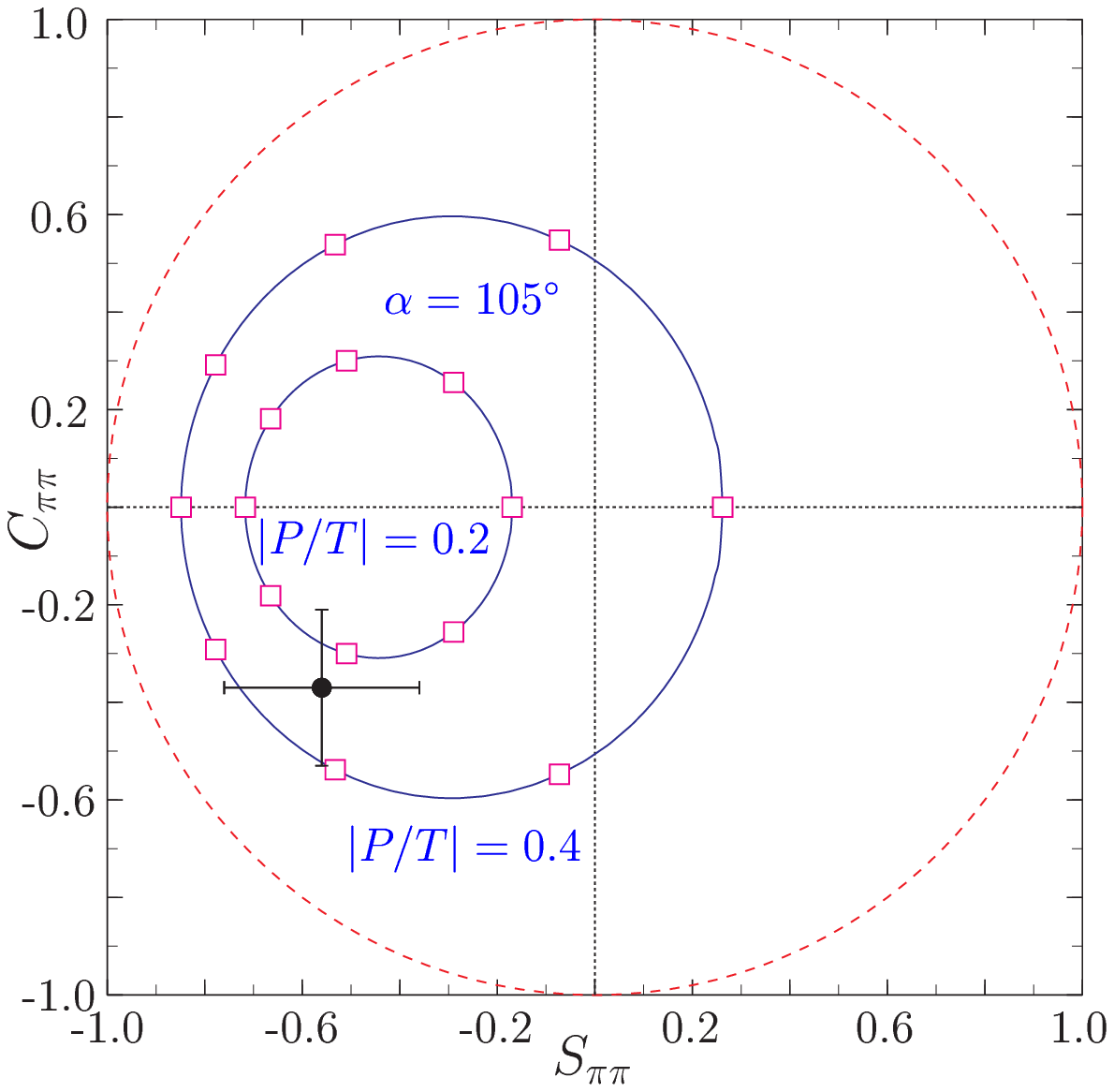} 
\quad
\psfig{width=0.47\textwidth,file=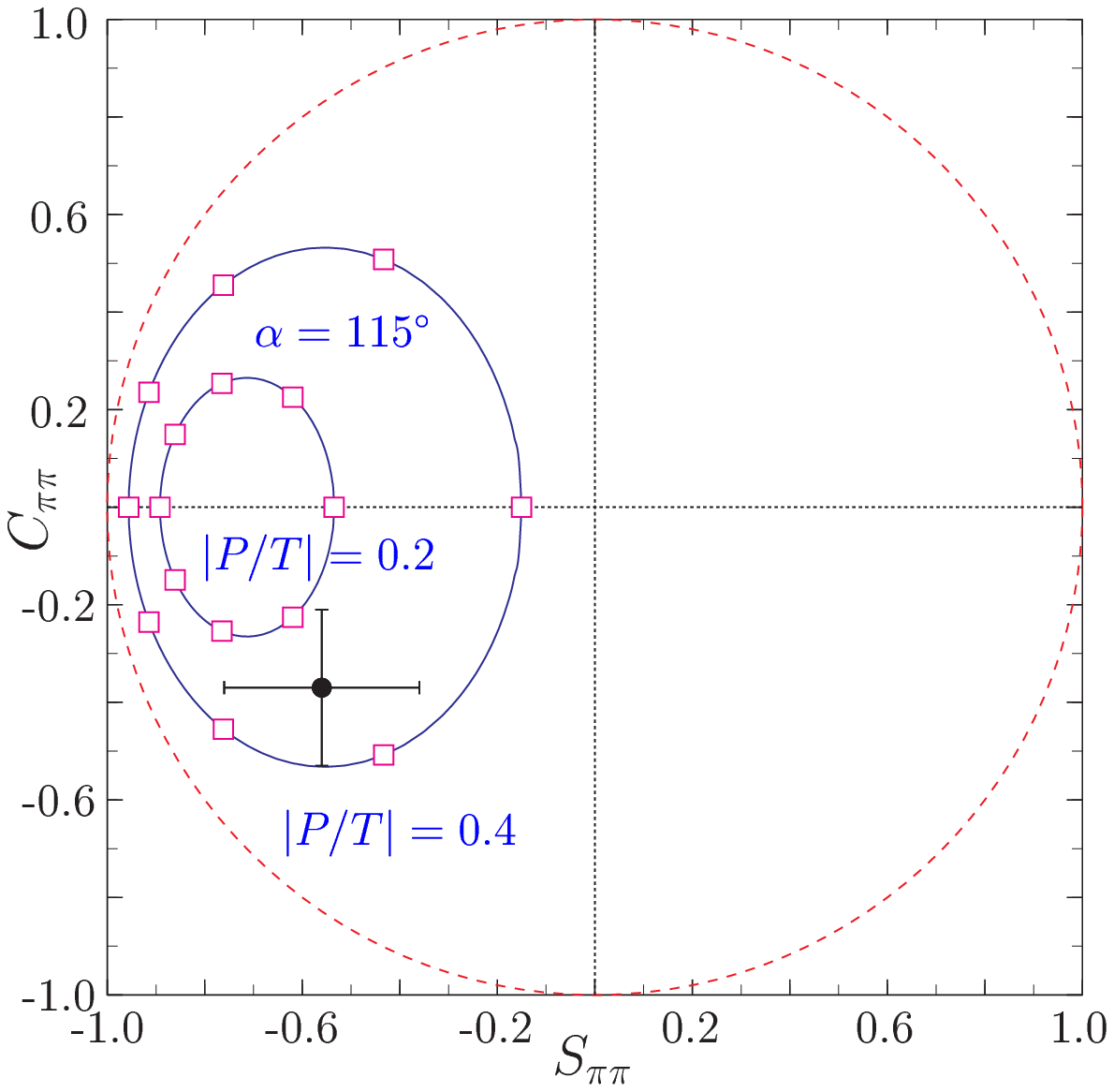}
}
\caption{Constraints on $\alpha$ from the current measurements $S_{\pi \pi}=-0.58 \pm 
0.20$ and $C_{\pi \pi}=-0.38 \pm 0.16$ (shown as the crossed bars).
The outer circle in each frame corresponds to the unitarity condition
$\vert C_{\pi \pi}\vert^2 + \vert S_{\pi \pi}\vert^2=1$.
Theoretically 
estimated range for $\vert P/T \vert$ and the strong phase $\delta$ varied in the full
range $-\pi \leq \delta \leq \pi$ are indicated. In the SM, measurements of
$C_{\pi \pi}$ and $S_{\pi \pi}$ have to lie in the region between the two inner contours. 
This figure shows that values $\alpha \leq 80^\circ$ are disfavoured by the
current $\pi \pi$ data (upper left frame). The other frames show three allowed 
values of $\alpha$  within the 95\% C.L. range from the unitarity fits,
which are in agreement with the $B \to \pi \pi$ data.
}
\label{fig:B-pipi}
\end{figure}
\subsection{Present bounds on the phase $\gamma$ from $B$ decays}
The classic method for determining the phase $\gamma$ (or 
$\phi_3$)~\cite{Gronau:1991dp,Gronau:1990ra,Atwood:1996ci,Atwood:2000ck}
involves the interference of the tree amplitudes 
$b \to u W^- \to u \bar{c}s \to D^0 K^-$ and $b 
\to c W^- \to c \bar{u} s \to \overline{D^0} K^-$. 
These decay amplitudes can interfere
if $D^0$ and $\overline{D^0}$ decay into a common hadronic final state. 
Noting that the CP$=\pm 1$
eigenstates $D^0_{\pm}$ are linear combinations of the $D^0$ and 
$\overline{D^0}$ states: $D^0_{\pm}= (D^0 \pm \overline{D^0})/\sqrt{2}$,
both branches lead to the same final states $B^- \to D^0_{\pm} K^-$. So, the
condition of CP interferometry is fulfilled. The
decays $B^- \to D^0_{\pm} K^-$ are described by the amplitudes:
\be
A(B^- \to D_{\pm}^0 K^-) = \frac{1}{\sqrt{2}}\,\left [ A(B^- \to D^0 K^-) \pm 
A(B^- \to \overline{D^0} K^-) \right ]\,.
\ee
Since, the weak phase of the $b \to u$ transition is 
$\gamma$ but the $b \to c$ 
transition has no phase, a measurement of the CP asymmetry through the 
interference of these two amplitudes yields $\gamma$.
The four equations that will be used to extract $\gamma$ are:
\bea
 R_{\pm}&\equiv& \frac{ {\cal B}(B^- \to D_{\pm}^0 K^-) + {\cal B}(B^+ \to 
D_{\pm}^0 K^+)}{{\cal B}(B^- \to D^0 K^-) + {\cal B}(B^+ \to D^0 K^+)}
= 1 + r_{\rm DK}^2 \pm 2 r_{\rm DK} \cos \delta_{\rm DK} \cos \gamma\,, 
\nonumber\\[-1mm]
\label{Lyuba} \\[-1mm]
A_{\pm} &\equiv& \frac{ {\cal B}(B^- \to D_{\pm}^0 K^-) - {\cal B}(B^+ \to 
D_{\pm}^0 K^+)}{ {\cal B}(B^- \to D_{\pm}^0 K^-) + {\cal B}(B^+ \to D_{\pm}^0 K^+)}
= \frac{\pm 2 r_{\rm DK} \sin \delta_{\rm DK} \sin \gamma}{1 +r_{\rm DK}^2 \pm 
2 r_{\rm DK} \cos \delta_{\rm DK} \cos \gamma}\,.
\nonumber
\label{dkasym}
\eea
With three unknowns ($r_{\rm DK}, \delta_{\rm DK}, \gamma)$, but four
quantities which will be measured $R_{\pm}$ and $A_{\pm}$, one has, in 
principle, an over constrained system. Here, $r_{\rm DK}$ is the ratio of
the two tree amplitudes~\cite{Gronau:2002mu} 
$r_{\rm DK}\equiv \vert T_1/T_2 \vert  \sim (0.1 - 0.2)$, 
with $T_1$ and $T_2$ being the CKM suppressed $(b \to u)$ and CKM 
allowed $(b \to c)$ amplitudes, respectively, and 
$\delta_{\rm DK}$ is the relative strong phase between them. The construction 
of the final states involves flavour and CP-tagging of the various $D^0$ 
states, which can be done, for example, through the decays 
$D^0_{+} \to \pi^+ \pi^-$,
$D^0_{-} \to K_S \pi^0$, and $D^0 \to K^- \pi^+$. 
Also, more decay modes can be added to increase the data sample.

Experimentally, the quantities $R_{\pm}$ are measured through the ratios:
\bea
R(K/\pi) &\equiv & \frac{{\cal B}(B^- \to D^0 K^-)}
{{\cal B}(B^- \to D^0 \pi^-)}\,; 
\qquad
R(K/\pi)_{\pm} \equiv \frac{{\cal B}(B^\pm \to D^0_{\pm} K^\pm)}
{{\cal B}(B^\pm \to D^0_{\pm} \pi^\pm)}\,.
\label{rkpipm}
\eea

With all three quantities $(R(K/\pi) and R(K/\pi)_{\pm}$ measured, 
one can determine  
$R_{\pm} = R(K/\pi)_{\pm}/R(K/\pi)$. Present measurements in the $B \to 
DK$ and $B \to D \pi$ decays have been recently summarized by 
Jawaherey~\cite{Jawahery:2003}: 
\bea
R_{+} &=& 1.09 \pm 0.16\,, \qquad 
A_{+}= 0.07 \pm 0.13 ~~{\rm [BELLE, BABAR]}\,, 
\nonumber\\
R_{-} &=& 1.30 \pm 0.25\,, \qquad 
A_{-}= -0.19 \pm 0.18 ~~{\rm [BELLE]}\,, 
\label{gammadk} \\
&\Longrightarrow& ~~ r_{\rm DK} =0.44^{+0.14}_{-0.24},
\qquad \langle A_{\rm CP}(DK) 
\rangle =0.11 \pm 0.11\,.
\nonumber
\eea
Thus, CP asymmetry in the $B \to DK$ modes is consistent with zero and 
the ratios $R_{\pm}$ are both in excess of 1. Hence, no useful constraint 
on $\gamma$ can be derived from these data at present.
One needs more precise measurements of $R_{\pm}$ and $A_{\pm}$ to 
determine $\gamma$ from this method. More useful decay modes to construct 
the $B \to DK$ triangle will have to 
be identified to reduce the statistical errors. In a recent paper, Atwood and
Soni~\cite{Atwood:2003jb} have advocated
to also include the decays of the vector states in the analysis, 
such as $B^- \to K^{*-} D^0$, $B^- \to K^- D^{*0}$, and 
$B^- \to K^{*-} D^{*0}$, making use of the $D^{*0} \to D^0
\gamma$ and $D^{*0} \to D^0 \pi^0$ modes. 
This is a promising approach but requires 
$O(10^9)$ $B\bar{B}$ pairs and reconstruction of many decay modes of 
the $D^{(*)} K^{(*)}$ final states to allow a 
significant measurement of~$\gamma$.

A variant of the $B \to DK$ method of measuring $\gamma$ is to use the decays
$B^\pm \to DK^\pm$ followed by  multibody decays of the $D$-meson, such as
$D^0 \to K_S \pi ^- \pi^+$, $D^0 \to K_S K^- K^+$ and $D^0 \to K_S \pi^- \pi^+ 
\pi^0$.
This was suggested some time ago by Atwood, 
Dunietz and Soni~\cite{Atwood:2000ck},
and was revived more recently by Giri {\it et al.}~\cite{Giri:2003ty}, 
in which a binned Dalitz plot analysis  of the decays 
$D^0/\overline{D^0} \to K_S \pi^- \pi^+$ 
was proposed. Assuming no CP asymmetry in $D^0$ decays, the amplitude of the
$B^+ \to D^0 K^+ \to (K_S \pi^+ \pi^-) K^+$ can be written as
\be
M_{+}= f(m_+^2,m_-^2) + r_{DK} {\rm e}^{i( \gamma + \delta_{DK})} f(m_-^2, m_+^2)\,,
\label{dalitz1}
\ee
where $m_+^2$ and $m_-^2$ are the squared invariant masses of the $K_S\pi^+$ and 
$K_S \pi^-$ combinations in the $D^0$ decay, and $f$ is the complex amplitude of
the decay $D^0 \to K_S \pi^+ \pi^-$. The quantities $r_{DK}$ and $\delta_{DK}$ are 
the relative magnitudes and  strong phases of the two amplitudes, already 
discussed earlier. The amplitude for the charge conjugate $B^-$ decay is
\be
M_{-}= f(m_-^2,m_+^2) + r_{DK} {\rm e}^{i(- \gamma + \delta_{DK})} f(m_+^2, m_-^2)\,.
\label{dalitz2}
\ee
Once the functional form of $f$ is fixed by a choice of a model 
for $D^0 \to K_S \pi^+ \pi^-$ decay, the Dalitz distribution for 
$B^+$ and $B^-$ decays can be fitted simultaneously by the expressions 
for $M_+$ and $M_-$, with $r_{DK}$, $\delta_{DK}$
and $\gamma$ (or $\phi_3$) as free parameters. The model-dependence could 
be removed by a binned Dalitz~distribution~\cite{Giri:2003ty}.

This method has been used by the BELLE collaboration~\cite{Abe:2003cn}
to measure the angle $\phi_3$. As the binned Dalitz distribution at this 
stage is
limited by statistics and some technical issues involving backgrounds and 
reconstruction efficiencies have to be resolved, an unbinned model-dependent 
analysis of the $B^\pm \to D^0 K^\pm$ decays followed by the decay 
$D^0 \to K_S \pi^+ \pi^-$  has been performed. The model for the function 
$f$ is based on a coherent sum of $N$ two-body plus one non-resonant 
decay amplitudes, and the $N=7$ resonances used are: 
$K^{*+} \pi^-$, $K_S \rho^0$, $K^{*-}\pi^+$, $K_S\omega$,
$K_S f_0(980)$, $K_S f_0(1430)$ and $K_0^{*} (1430)^+ \pi^-$:
\be
f(m_+^2,m_-^2) = \sum_{j=1}^{N} a_j {\rm e}^{i \delta_j} A_j (m_+^2,m_-^2) + b {\rm 
e}^{i \delta_0}\,,
\label{belleffunc}
\ee
where $A_j (m_+^2,m_-^2)$, $a_j$ and $\delta_j$ are the matrix element, amplitude
and strong phase, respectively, for the $j$-th resonance, and $b$ and $\delta_0$
are the amplitude and phase for the non-resonant component. Further details can be 
seen in the BELLE paper~\cite{Abe:2003cn}. The result based on 140 fb$^{-1}$ data
has yielded
\be
r_{DK}=0.33 \pm 0.10\,, \qquad 
\delta_{DK}= (165 ^{+17}_{-19})^\circ\,, \qquad 
\phi_3=(92 
^{+19}_{-17})^\circ\,.
\label{bellephi3}
\ee
As the errors are quite non-parabolic, they do not represent the accuracy of
the measurement. Rather, the constraint plots on the pair of parameters 
$(\phi_3,\delta_{DK})$ and $(r_{DK}, \phi_3)$, which can be seen in the BELLE 
publication~\cite{Abe:2003cn}, are used to get the information on
these parameters. The resulting 90\% C.L. ranges~are~\cite{Abe:2003cn}: 
\be
0.15 < r_{DK} < 0.50\,, \qquad 
104^\circ < \delta_{DK} < 214^\circ\,,  \qquad 
61^\circ < \phi_3 < 
142^\circ\,,
\label{bellephi3fits}
\ee
and the significance of direct CP violation effect (
including systematics) is 2.4
standard deviations. This measurement is in agreement with the SM expectations
$43^\circ \leq \phi_3 \leq 86^\circ$, though lot less precise. 

As the final topic of this review, we discuss the $B \to K \pi$ decays,
which in the SM are dominated by QCD penguins. Their 
branching ratios, averaged over the charge conjugate states, and the present 
measurements of the CP asymmetries are summarized in Table \ref{brasym}. The 
branching ratios listed in this table (except for the 
$B^0 \to \pi^0 \pi^0$ mode)  are in 
agreement with the theoretical predictions based on the
QCD factorization~\cite{Beneke:1999br} and pQCD~\cite{Keum:2003qi} approaches.
These approaches  will be put to more stringent tests with precise
measurements of the CP asymmetries $A_{\rm CP}(K\pi)$ 
in various decay modes, and
also through the ratios of the branching ratios, which provide a better focus 
on the relative strong phases and magnitudes of the QCD penguin, tree and 
electroweak penguin 
contributions~\cite{Neubert:1998jq,Buras:2000gc,Gronau:2003kj,Beneke:2001ev}.

Concerning the determination of $\gamma$ from these decays, 
whose present status we discuss 
below, we note that the current data on $B \to K \pi$ decays has some 
puzzling features,
which have to be understood before we determine $\gamma$ from this data. 
In particular, the following two ratios with 
their currently measured~values:
\bea
R_n &\equiv& \frac{\overline{\cal B}(K^+ \pi^-)}
{2 \, \overline{\cal B}(K^0 \pi^0)}
=0.76 \pm 0.10\,, \nonumber\\[-1.5mm] 
\label{rcrndef} \\[-1.5mm] 
R_c &\equiv& \frac{\overline{\cal B}(K^+ \pi^0)}{\overline{\cal B}(K^0 \pi^+)}
=1.17 \pm 0.13\,,
\nonumber
\eea
have received some attention lately~\cite{Buras:2003yc,Yoshikawa:2003hb} as  
possible harbingers of new physics. It has been argued that these data require a  
much enhanced electroweak penguin contribution than is assumed or inferred
from the $B \to \pi \pi$ data and approximate SU(3) symmetry.  
 Taking the current data on the face value,
a range  $0.3 \leq r_{\rm EW}\equiv\vert P^\prime_{\rm EW}\vert/\vert P^\prime \vert \leq 
0.5$, and a 
significant strong phase difference,
$\delta_{\rm EW} - \delta_T$ between the electroweak penguin and tree amplitudes  are 
needed to explain the data. Typical estimates of $r_{\rm EW}$ are, however, in the 
ball park of $r_{\rm EW} \sim 0.15$~\cite{Beneke:2001ev}, with a negligible phase 
difference.   
We have mentioned earlier that also the measurement of $C_{\pi \pi}$
is hinting at a significant strong phase; leaving this phase as a free parameter the 
allowed range of $\alpha$ from the analysis of the time-dependent CP asymmetry is in 
good agreement  with the SM-based indirect estimates of the same.

 The presence of significant strong phase differences in $B \to  \pi \pi$ and/or $B \to K  
\pi$ decays implies that the strong interaction effects in these and related decays are 
not quantitatively described by perturbative methods such as QCD factorization. 
It is, however, less clear if a value of $r_{\rm EW} \sim 0.3 - 0.5$ can also be 
attained in an improved  theoretical framework within the SM. So, the current $B \to K \pi$ 
data are somewhat puzzling. However, judging from the  difference $R_c-R_n=0.41 \pm 0.16$,
and the estimates giving $R_c-R_n \sim 0.1$, this appears to be about a $2 \sigma$ problem.
In view of the fact that current data on $R_c$ and $R_n$ are not quite understood, it is 
advisable not to use these ratios to constrain $\gamma$, as both $R_c$ and $R_n$ involve 
the poorly understood electroweak penguin contribution.

To constrain $\gamma$ from $B \to K \pi$ data, the so-called mixed 
ratio $R_0$, defined below, and advocated some time ago by Fleischer and 
Mannel~\cite{Fleischer:1997um} is potentially useful. The amplitude for the process
$B^+ \to K^0 \pi^+$ is written in (\ref{pipikpirel}). Using isospin symmetry, the
decay
amplitude for $B^0 \to K^+ \pi^-$ can be written as
\be
A(B^0 \to K^+ \pi^-)=\vert P^\prime \vert {\rm e}^{i \delta} - \vert T^\prime \vert {\rm 
e}^{i \gamma}\,.
\ee
There is also a small color-suppressed electroweak penguin contribution which can be safely
neglected. Denoting $r\equiv \vert T^\prime \vert / \vert P^\prime \vert$, one has the 
following relations:
\bea
 R_0 &\equiv& \frac{\bar{\Gamma}(K^\pm \pi^\mp)}{\bar{\Gamma}(K^0 \pi^\pm)} = 1-2 r \cos
\delta
\cos \gamma +r^2\,, \nonumber\\
{\cal A}_{\rm CP}(K^+ \pi^-) &\equiv& \frac{\Gamma(K^- \pi^+) - \Gamma(K^+ \pi^-)}{\Gamma(K^-
\pi^+) + \Gamma(K^+ \pi^-)} =-2r \sin \delta \sin \gamma/R_0\,,
\eea
which are both functions of $r$, $\delta$ and $\gamma$. Fleischer and
 Mannel~\cite{Fleischer:1997um} have shown that $R_0>\sin^2 \gamma$ for any $r$ and $\delta$.
So, if $R_0<1$, one has an  interesting bound on $\gamma$. The current value $R_0=0.898 \pm 
0.071$
is consistent with one at about $90\%$ C.L., and hence no useful bound emerges on $\gamma$.  
However, if
${\cal A}_{\rm CP}(K^+ \pi^-)$ is measured precisely, then the two equations for $R_0$ and 
$R_0{\cal A}_{\rm CP}(K^+ \pi^-)$ can be used to eliminate $\delta$ and, in principle, a 
useful constraint on $\gamma$ emerges. Current measurements yield
${\cal A}_{\rm CP}(K^+\pi^-)=-0.095 \pm 0.029$, which at $3 \sigma$ is probably the 
only significant direct CP-violation observed so far in $B$-decays.

 To extract a value of $\gamma$ from these
measurements (or to put a useful bound),  one has to assume a value for $r$, or extract it 
from data under some assumptions. Using 
arguments based on factorization and  SU(3)-breaking, Gronau and
Rosner~\cite{Gronau:2003br} 
estimate $r$ from the average of the CLEO, BELLE and BABAR data, getting $r=0.142 
^{+0.024}_{-0.012}$ and a bound $\gamma < 80^\circ$ at $1\sigma$. However, there is no 
bound on $\gamma$ at 95\% C.L. to be compared with the corresponding indirect bounds
from unitarity. More precise measurements of $R_0$ and ${\cal A}_{\rm CP}(K^+\pi^-)$ are
required to get useful constraints on $\gamma$.

In conclusion, CP asymmetry has been observed in $B \to D K$ decays
using the Dalitz distributions at 2.4$\sigma$ level. Likewise, direct CP asymmetry
is seen at 3 $\sigma$ level in ${\cal A}_{\rm CP} (K^+ \pi^-)$. However, the current 
significance of the CP asymmetry and the  
model-dependence of the resonant structure in the decays of the $D$-meson in the former, 
and imprecise
knowledge of $R_0$ (as well as of ${\cal A}_{\rm CP} (K^+ \pi^-)$) in the latter, hinder 
at present in drawing quantitative and model-independent conclusions on $\gamma$.  
This situation may change with more precise measurements of the various ratios and CP 
asymmetries in the $B \to K \pi$ and $B \to DK$ decays. Both the KEK and SLAC 
$B$-factories are now collecting data at record luminosities and we trust that 
improved determinations of $\gamma$ (or $\phi_3$) will not take very long to come.

\section{Summary and Concluding Remarks}

We have reviewed the salient features of the CKM phenomenology and $B$-meson physics,
with emphasis on new experimental results and related theoretical developments.
The data discussed here are spread over an energy scale of over three orders of magnitude, 
ranging from  muon decay, determining $G_F$, to the top-quark decays, determining $\vert 
V_{tb} \vert$, and have been obtained from  diverse experimental
facilities. Their interpretation has required the 
development of a number of theoretical tools, with the Lattice-QCD, QCD sum rules, chiral 
perturbation theory, and heavy quark effective theory at the forefront.
We reviewed representative applications of each of them. Progress in computational
technology has enabled a quantitative determination of all the CKM matrix elements. 
While the
precision on some of them can be greatly improved, all currently available  
measurements are compatible with the assumption that the CKM matrix is the 
only source of flavour changing transitions in the hadronic sector. In fact, there is 
currently no compelling experimental evidence suggesting deviations from the CKM theory.

However, there are some aspects of the data which are puzzling and deserve further 
research. Those under current experimental scrutiny are summarized below. 
\begin{itemize}
\item Test of unitarity in the first row of the CKM matrix yields $\Delta_1=(3.3 \pm 
1.3) \times 10^{-3}$, which is $2.5$ standard deviations away from zero.
Further experimental and theoretical work, yielding robust
evaluations of the low energy constants of chiral perturbation theory, will have an
impact on this issue. Precise measurements of $K_{\ell 3}$ decays are being done 
at DA$\Phi$NE and elsewhere. They will yield a determination of $\vert V_{us} \vert$
at significantly better than a per cent level. Analysis of  $\tau$-decays
from the $B$-meson factories will also help. Resolution of the current
inconsistencies in the determination of $g_A/g_V$ in polarized neutron $\beta$-decay 
experiments will improve the precision on $\vert V_{ud} \vert$ from this 
method. Together, they will enable a precise determination of $\Delta_1$.

\item Experiments at LEP have measured the decays $W^\pm \to q^\prime \bar{q} (g)$,  
enabling a quantitative test of the
unitarity involving the first two rows of the CKM matrix. The result $\sum 
\vert V_{ij} \vert^2 -2=0.039 \pm 0.025$ is consistent with being zero at $1.6$ standard 
deviations. Experiments at  CLEO-C and BES-III, but also the $B$-factory experiments BABAR 
and BELLE, will measure the matrix elements $\vert V_{cs} \vert$ and $\vert V_{cd} \vert$ at 
about $1\%$ accuracy, allowing an improved test of the CKM unitarity in the second row. 
Progress in Lattice-QCD technology will be required to have precise knowledge of  
the $D \to (K,K^*,\pi,\rho)$ form factors.

\item The difference $S_{\phi K_S} -S_{J/\psi K_S}$ involving the time-dependent CP 
asymmetries in the decays $B \to \phi K_S$ and $B \to J/\psi K_S$, which vanishes in 
the first approximation in the SM, is currently found to 
deviate from zero.  With $S_{\phi K_S}= -0.14 \pm 0.33$ (not including 
the scale factor) and $S_{\phi K_S}= -0.14 \pm 0.69$ (including the scale 
factor), the BELLE and BABAR measurements are 1.3 (2.7) standard deviations away from
$S_{J/\psi K_S}=0.736 \pm 0.049$ with (without) the scale factor. Including all the $b \to s 
\bar{s} s$ penguin-dominated final states measured so far gives  $\sin 2 
\beta  - \langle  \sin2 \beta_{\rm eff} \rangle = 0.50 \pm 0.25$, with the scale factor, 
which differs from 0 by $2$ standard deviations. The presence of a large scale factor 
(typically 2) in the determination of $S_{\phi K_S}$ implies that improved measurements
of this (and related) quantities are needed to settle the present inconsistency.
They will be undertaken at the current and planned $B$ factories, and also at LHC-B and B-TeV.     

\item Measurements of the ratios $R_c$ and $R_n$ involving the $B \to K \pi$ 
decays hint at the electroweak penguin contributions in these decays   
at significantly larger strength than their estimates in the SM. However, 
judged from the current measurements, $R_c-R_n=0.41 \pm 0.16$, and 
theoretical estimates yielding $R_c-R_n \simeq  0.1$ based on  the default value $r_{\rm EW}= \simeq 
0.15$, the mismatch with the SM has a significance of about 2 standard 
deviations. Improved measurements
at the $B$ factories and theoretical progress in understanding non-leptonic $B$-meson
decays will clarify the current puzzle.

\item Enhanced electroweak penguin contributions would also influence semileptonic 
rare $B$- and $K$-decays, providing important consistency checks.   
First round of experiments of the electroweak penguins in 
$B \to (K,K^*,X_s) \ell^+ \ell^-$ have been reported by the BABAR and BELLE 
collaborations. Current data are summarized~\cite{Nakao:2003gc} 
in~Table~\ref{bsllbabelle} together with the SM-based 
estimates~\cite{Ali:2002jg} of the same. A comparison shows that the experimental 
measurements are well accounted for in the SM. As the exclusive decays have larger
theoretical errors due to the uncertain form factors for which the QCD sum rule
estimates~\cite{Ali:1999mm} have been used, we quantify a possible mismatch
using the theoretically cleaner inclusive decay $B \to X_s \ell^+ \ell^-$.
Noting that the average of the BELLE and BABAR measurements is~\cite{Nakao:2003gc} 
${\cal B}\,(B \to X_s \ell^+ \ell^-)=(6.2~\pm 1.1^{+1.6}_{-1.3})\,\times 10^{-6}$, 
the difference ${\cal B}(X_s \ell^+ \ell^-)_{\rm exp} - 
{\cal B}(X_s\ell^+ \ell^-)_{\rm SM}=(2.0 \pm  2.0) \times 10^{-6}$ 
amounts to 1 standard deviation. Note that the errors are dominantly
experimental. From this we infer that there is no 
evidence of any abnormal electroweak penguin contribution
in the reliably calculable semileptonic rare $B$-decays. 
Experiments at the  $B$ factories and the hadron 
colliders  will greatly improve the precision on the decays 
$B \to (K,K^*,X_s)  \ell^+ \ell^-$, measuring also various distributions sensitive to 
physics beyond the SM~\cite{Ali:2002jg,Ali:1999mm}.
\begin{table}[htbp] 
\caption{$B \to K^{(*)} \ell^+ \ell^-$ and $B \to X_s \ell^+ \ell^-$
branching ratios in current experiments~\protect\cite{Nakao:2003gc} and comparison with 
the SM-estimates~\protect\cite{Ali:2002jg}.
\label{bsllbabelle}}

\vspace*{-2mm}

\renewcommand{\arraystretch}{1.3} 
\begin{center}

\begin{tabular}{l c c c}
\hline \hline\\[-4mm]
Mode & BELLE & BABAR & Theory (SM) \\[1.5mm] \hline \\[-4mm]
${\cal B}$($B \to K \ell^+ \ell^-$) ($\times$~10$^{-7}$) &
4.8$^{+1.0}_{-0.9}$~$\pm$~0.3~$\pm$~0.1 & 
6.9$^{+1.5}_{-1.3}$~$\pm$~0.6 &
3.5~$\pm$~1.2 \\
${\cal B}$($B \to K^* \ell^+ \ell^-$) ($\times$~10$^{-7}$) $\hspace*{5mm}$ &
$\hspace*{2mm}$ 11.5$^{+2.6}_{-2.4}$~$\pm$~0.7~$\pm$~0.4 $\hspace*{2mm}$ &
$\hspace*{2mm}$ 8.9$^{+3.4}_{-2.9}$~$\pm$~1.1 $\hspace*{2mm}$ &
11.9~$\pm$~3.9 \\
${\cal B}$($B \to X_s \ell^+ \ell^-$) ($\times$~10$^{-6}$) &
6.1~$\pm$~1.4$^{+1.4}_{-1.1}$ &
6.3~$\pm$~1.6$^{+1.8}_{-1.5}$ &
4.2~$\pm$~0.7
\\[2mm] 
\hline \hline
\end{tabular}
\end{center}
\end{table}

\end{itemize}

From the theoretical point of view, it is very likely that the current 
deviations from the CKM unitarity involving the first row will not stand 
the force of improved measurements. The anomalies in the penguin-dominated 
$B$-decays are also likely to find an experimental resolution, though  
mapping out the QCD and electroweak penguins in  
$B$-decay experiments is crucial in reaching a definitive conclusion.

 Finally, it should be underlined that a number of 
benchmark measurements in the $B$- and $K$-meson sectors still remain to be done.
On the list of the future experimental milestones are the following:
\begin{itemize}
\item  Precise determinations of the weak phases $\alpha$ (or $\phi_2$) and $\gamma$ (or 
$\phi_3$).
\item  Measurement of the $B_s^0$ - $\overline{B_s^0}$ mass difference $\Delta M_{B_s}$
\item  Measurement of the branching ratio ${\cal B}(B_s^0 \to \mu^+ \mu^-)$
\item Precise measurements of the dilepton invariant mass spectra in $B \to (X_s,K,K^*) \ell^+ \ell^-$ and the 
forward-backward asymmetries in the decays $B \to (X_s,K^*) \ell^+ \ell^-$.
\item Measurements of the CKM-suppressed  radiative and semileptonic rare decays $b \to d 
\gamma$ and $b \to d \ell^+ \ell^-$ in the inclusive modes, and some exclusive decays
such as $B \to (\rho, \omega) \gamma$ and $B \to (\pi,\rho,\omega) \ell^+ \ell^-$.
\item Measurements of the rare decays $B \to (X_s, K,K^*) \nu \bar{\nu}$ and $K \to \pi 
\nu \bar{\nu}$.
\item Last, but not least, is the challenging measurement of $\arg (\Delta M_{B_s})$, also 
called $\delta \gamma$, having a value $\delta \gamma =-\lambda^2\eta\simeq -2^\circ$ in 
the SM. While it appears to be a formidable task to attain  the experimental sensitivity to
probe the SM in $\delta \gamma$, searches for 
beyond-the-SM physics will be undertaken at the hadron collider experiments   
through the CP asymmetry ${\cal A}_{\rm CP}(B_s \to J/\psi \phi)$.
\end{itemize}
There measurements  will test the CKM theory in not so well 
charted sectors where new physics may find it easier to intervene. 

 It is time to reminisce. Retrospectively, some forty 
years ago, Cabibbo rotation solved the problem of the apparent non-universality of the
Fermi weak interactions. The GIM mechanism and the KM proposal
were crucial steps in the quest of understanding the FCNC processes and CP violation in 
the framework of universal weak interactions. These theoretical developments  
brought in their wake an entire new world of flavour physics. In the meanwhile, all the 
building blocks predicted by these theories are in place. Thanks to dedicated experiments and 
sustained  progress in theory, the field of flavour physics has developed into a precision 
science. All current measurements within errors are compatible with the CKM theory. While
 there is every 
reason to celebrate and rejoice this great synthesis, this does not necessarily imply that we have 
reached the end of the great saga of discoveries having their roots in flavour physics. 
Looking to  the future, one conclusion can already be drawn: As experiments and 
theoretical techniques become more refined, new aspects of physics will come under 
experimental scrutiny and the known ones will be measured with unprecedented precision. It is
conceivable that also at the end of the next round of experiments, CKM theory will continue to 
prevail. It is, however, also conceivable that a consistent description of experiments in flavour 
physics may require the intervention of new particles and forces. Perhaps, the current 
experimental deviations from the CKM theory, while not statistically significant, are 
shadows being cast by the coming events. If we are lucky, the hints in the data with renewed 
efforts could turn into irrefutable solid evidence of new physics, or perhaps some of the 
crucial experiments listed above would force us to seek for explanations which go beyond the 
CKM theory. This remains to be seen. Only future experiments can tell if we have reached the shores of a new
world or whether the new shores that we have reached are still governed by the standard model 
physics. In any case, the outcome of the ongoing and planned experiments in flavour physics will be a vastly 
improved knowledge about the laws of Natute.
\section{Acknowledgement}
The material presented here is based on a number of lectures and invited 
talks given at various meetings. These include: Four-Seas 
Conference, Thessaloniki, Greece (2002); 
2nd International Workshop on B Physics and CP Violation, 
National Taiwan  University, Taiwan (2002); International Workshop on Quark Mixing and CKM 
Unitarity,  Heidelberg, Germany (2002); and the International Meeting on Fundamental 
Physics, Soto de Cangas (Asturias), Spain (2003). These topics were also part of  
a Mini-Course of lectures on $B$ Physics and the CKM 
Phenomenology, given at LAPP-Annecy, France, in September 2003, and  the 
Academic Lectures on $B$ Physics, in progress at the
High Energy Accelerator Research Organization KEK, Tsukuba, Japan. This 
paper will appear in the proceedings of the conference IMFP, Soto de Cangas, Spain. I thank 
Alberto Ruiz for his generous acceptance of the manuscript despite its being oversized, 
and Fernando Barreiro for his hospitality in Madrid and for the pleasant journey to the 
Cangas. 
Many thanks to Damir Becirevic, Shoji Hashimoto, Masashi Hazumi, Enrico Lunghi, Matthew 
Moulson, Mikihiko Nakao, Alexander Parkhomenko and Yoshi Sakai  for  
providing valuable inputs, having helpful discussions, and reading all or parts of this 
manuscript. I thank Alexander and Enrico  also for updating several figures presented 
here which are based on research work done in collaboration with them.
I am grateful to Yuji Totsuka San, Yasuhiro Okada and members of the Theory 
Group for their warm hospitality at KEK, where this manuscript was written.
This work has been supported by the KEK Directorate under a grant from 
the Japanese Ministry of Education, Culture, Sports, Science and Technology. Domo 
Arigato Gozaimas!

\end{document}